\documentclass[12pt,letterpaper,onecolumn]{article}

\usepackage[dvipsnames]{xcolor}

\usepackage{amsfonts}
\usepackage{amsmath}
\usepackage{amssymb}
\usepackage{amsthm}
\usepackage[appendix=inline]{apxproof}
\usepackage{array}
\usepackage[blocks]{authblk}

\usepackage{booktabs}
\usepackage{bm}

\usepackage[disableredefinitions]{complexity}
\usepackage{csquotes}
\usepackage{enumitem}
\usepackage[T1]{fontenc}
\usepackage{geometry}
\usepackage{graphicx}

\usepackage{hyperref}
\usepackage[capitalise]{cleveref}

\usepackage{import}
\usepackage[utf8]{inputenc}

\usepackage{lineno}

\usepackage{mathtools}
\usepackage{mathrsfs}

	\usepackage{needspace}
	\usepackage{relsize}
	\usepackage{refcount}
	\usepackage{reptheorem}
\usepackage{setspace}
\usepackage{soul}
\usepackage{subcaption}
\usepackage{thm-restate}
\usepackage{thmtools}
\usepackage{tikz}
\usetikzlibrary{decorations.pathreplacing}
\usepackage{tikz-cd}
\usepackage{titling}
\usepackage{todonotes}

\usepackage{xspace}
\newcommand{\barv}{\bar{v}}

\newcommand{\bbB}{\mathbb{B}}

\newcommand{\bbN}{\mathbb{N}}

\newcommand{\bbR}{\mathbb{R}}

\newcommand{\bbZ}{\mathbb{Z}}

\newcommand{\bfA}{\mathbf{A}}

\newcommand{\bfD}{\mathbf{D}}
\newcommand{\bfE}{\mathbf{E}}

\newcommand{\bfJ}{\mathbf{J}}

\newcommand{\bfL}{\mathbf{L}}
\newcommand{\bfM}{\mathbf{M}}

\newcommand{\bfQ}{\mathbf{Q}}

\newcommand{\bfV}{\mathbf{V}}

\newcommand{\bfX}{\mathbf{X}}

\newcommand{\calA}{\mathcal{A}}
\newcommand{\calB}{\mathcal{B}}
\newcommand{\calC}{\mathcal{C}}

\newcommand{\calF}{\mathcal{F}}
\newcommand{\calG}{\mathcal{G}}

\newcommand{\calL}{\mathcal{L}}

\newcommand{\calQ}{\mathcal{Q}}

\newcommand{\rmP}{\mathrm{P}}

\newcommand{\rmR}{\mathrm{R}}

\newcommand\sg[1]{}
\newcommand\sgchanged[1]{}
\newcommand\am[1]{}
\newcommand\amchanged[1]{}

\newcommand{\reduce}{\leq}
\newcommand{\depth}{\mathit{depth}}

\renewcommand{\SIZE}{\text{\normalfont\scshape Size}}
\newcommand{\SIZEDEPTH}{\text{\normalfont\scshape SizeDepth}}
\newcommand{\WIRESIZEDEPTH}{\text{\normalfont\scshape WSizeDepth}}

\newcommand{\scrC}{\mathscr{C}}

\newcommand{\N}{\mathbb{N}}

\newcommand{\QDA}{\bfQ\bfD\bfA}
\newcommand{\bfone}{\mathbf{1}}

\newcommand{\QESL}{\bfQ\bfE\bfJ_{\bfone}}

\newcommand{\rmFO}{\mathrm{FO}}

\newcommand{\resp}{\ensuremath{\textit{resp.}}}

\DeclareMathOperator{\cont}{cont} 

\newcommand{\problemx}[3]{
\par\noindent\underline{\sc#1}\par\nobreak\vskip.2\baselineskip
\begingroup\clubpenalty10000\widowpenalty10000
\setbox0\hbox{\bf INPUT:\ }\setbox1\hbox{\bf QUESTION:\ }
\dimen0=\wd0\ifnum\wd1>\dimen0\dimen0=\wd1\fi
\vskip-\parskip\noindent
\hbox to\dimen0{\box0\hfil}\hangindent\dimen0\hangafter1\ignorespaces#2\par
\vskip-\parskip\noindent
\hbox to\dimen0{\box1\hfil}\hangindent\dimen0\hangafter1\ignorespaces#3\par
\endgroup}

\theoremstyle{plain}
\newtheorem{theorem}{Theorem}[section]
\newtheorem{problem}{Problem}[section]

\newtheorem{lemma}[theorem]{Lemma}
\newtheoremrep{lemma}[theorem]{Lemma}

\newtheorem{fact}[theorem]{Fact}

\newtheorem{corollary}[theorem]{Corollary}
\newtheoremrep{corollary}[theorem]{Corollary}

\newtheorem{remark}[theorem]{Remark}

\newtheorem{definition}[theorem]{Definition}
\newtheorem{proposition}[theorem]{Proposition}
\newtheoremrep{proposition}[theorem]{Proposition}

\theoremstyle{definition}
\newtheorem{example}[theorem]{Example}
\Crefname{lemma}{Lemma}{Lemmata}
\Crefname{equation}{Line}{Lines}\crefrangelabelformat{equation}{(#3#1#4)--(#5#2#6)}
\Crefname{inequality}{Inequality}{Inequalities}\creflabelformat{inequality}{(#2#1#3)}
\Crefname{equivalence}{Equivalence}{Equivalences}\creflabelformat{equivalence}{(#2#1#3)}
\Crefname{property}{Property}{Properties}\creflabelformat{property}{(#2#1#3)}

\newcommand{\defeq}{\stackrel{\text{def}}{=}}

	\newcommand{\stab}{\text{stab}}

\newif\ifdraft
\drafttrue
\draftfalse
\ifdraft
\usepackage{showlabels}
\renewcommand\sg[1]{\todo[inline,size=\scriptsize,backgroundcolor=Yellow]{#1 - \textbf{Stefan}}}
\renewcommand\sgchanged[1]{{\color{red}{#1}}}
\renewcommand\am[1]{\todo[inline,size=\scriptsize,backgroundcolor=Magenta]{#1 - \textbf{Amal}}}
\renewcommand\amchanged[1]{{\color{blue}{#1}}}
\fi

\newcommand{\ACO}{\mathbf{AC^0}}
\newcommand{\WLACO}{\mathbf{WLAC^0}}

\newcommand{\pre}{\mathit{pre}}

\newcommand{\congC}{\mathbin{\mathscr{C}}}

\DeclareMathOperator{\opc}{c}

\setlist[enumerate]{label=(\arabic*),font=\normalfont}

\renewcommand{\PSPACE}{\mathbf{PSPACE}}
\newcommand{\wires}{{\mathit{wires}}
}
\newcommand{\OmegaHL}{\Omega}
\newcommand{\eval}{\textrm{eval}}
\newcommand{\fanin}{\textit{fan-in}}

\renewcommand{\themainthm}{\getrefnumber{mainthm:number}}

\usetikzlibrary{positioning}

\title{A Circuit Complexity Trichotomy For Regular Languages}

\newif\ifanonym
\anonymtrue
\anonymfalse
\ifanonym
\author{}
\else
\author{
    Stefan G\"oller\\
\small{\url{stefan.goeller@uni-kassel.de}}}
\affil{School of Electrical Engineering and Computer Science, Universit\"at Kassel, Germany}
\author{Amaldev Manuel\\\small{\url{amal@iitgoa.ac.in}}}
\affil{School of Mathematics and Computer Science, Indian Institute of Technology Goa, India}
\fi

\begin{document}

\maketitle
\begin{abstract}
	We prove the following trichotomy for the 
	non-uniform unbounded-fan-in circuit complexity of regular languages:
	the circuit complexity of a regular language is
	either $O(1)$, $\OmegaHL(\log n)\cap O(\log n \log \log n)$ 
	or $\OmegaHL(n)\cap O(n)$.\footnote{Here, 
	$\OmegaHL$ refers to Hardy-Littlewood $\Omega$, formally
	$f\in\OmegaHL(g)$
	if there exists a constant $c>0$ such that
	$f(n)\geq c\cdot g(n)$ for infinitely many $n$.}

Our main technical contribution is a characterization of the sublinear class in terms of algebra, word congruences, regular expressions, and logic. 
	We show that the regular languages of sublinear circuit complexity are precisely those computable by circuits of size $O(\log^2 n)$ and constant depth,
	as well as by circuits of size $O(\log n\log\log n)$ and depth 
	$O((\log\log n)^2)$.
	As a consequence, every regular language of sublinear circuit complexity 
	belongs to $\rmFO^2[\Sigma,<,\mathrm{mod}]$, which in turn
	is contained in wire-linear $\ACO$.

	We further prove that deciding membership in each of the three classes is
	$\PSPACE$-complete for given nondeterministic finite automata.
	We similarly classify the circuit complexity of regular languages with
	a neutral letter.
\end{abstract}

{\small
{\bf 
\noindent
Disclosure of AI assistance:} 
The proof of the
$\SIZEDEPTH\bigl(O(\log n\log\log n), O((\log\log n)^2)\bigr)$
upper bound presented in Section~\ref{S loglogn} 
(which involves Theorem~\ref{T Spielman} derived
from Spielman's encoding result~\cite{Spielman96}, 
Lemma~\ref{L Spielman} and
Theorem~\ref{T upper loglog}) was found by
ChatGPT-6 Astra. 
The proof of Lemma~\ref{L Spielman} was incorporated with minor adjustments
to conform to our notation. The proof of
Theorem~\ref{T upper loglog} was substantially rewritten by the authors.
All proofs were verified by the authors.
ChatGPT-6 Astra was also used for proofreading, spellchecking,
and editing. 

All other definitions, proofs, and mathematical results were
obtained by the authors.
}
\newpage
\tableofcontents
\newpage

\setcounter{tocdepth}{4}
\newif\ifshowproofs
\showproofsfalse
\showproofstrue

\ifshowproofs
\else
  \RenewEnviron{proof}[1][]{}
\fi

%!TEX root = main.tex

\section{Introduction}
In this work we prove a classification of the circuit complexity, i.e., the asymptotic size of the smallest circuit using unbounded fan-in Boolean gates, of regular languages. 
While this is an established problem \cite{Koucky2009,Kouckyhandbook}, there
are three motivations for this study that also tell the nature of it. A first is to understand the parallel complexity of regular languages. Unbounded fan-in Boolean circuits 
are a foundational model of parallel computation, just like finite state automata for sequential computation. 
A classification linking the two sheds light into the interplay between the two paradigms which is relevant for developing hybrid models and complexity measures.  

A second reason is that regular languages play a key role in the study of complexity classes within $\mathbf{NC}^1$, i.e., the class of languages computed by polynomial-sized Boolean circuits with bounded fan-in and logarithmic depth. For instance, they separate $\mathbf{AC}^0$ and $\mathbf{NC}^1$~\cite{FurstSaxeSipser1984} and any separation of $\mathbf{ACC}^0$ or $\mathbf{TC}^0$ from $\mathbf{NC}^1$ can be witnessed by an 
$\mathbf{NC}^1$-complete regular language~\cite{Koucky2009}.
While the circuits in this paper use Boolean gates, circuit complexity with threshold and modular gates is an interesting direction to pursue. 

A third reason and a particularly interesting one from an automata-theoretic viewpoint is to develop a catalogue of algebraic classes corresponding to circuit complexity of regular languages akin to the pseudovarieties of finite semigroups in the case of finite state automata. The latter theory has been useful in many ways, as a vocabulary for articulating results and goals, in unifying concepts across diverse paradigms, and as a combinatorial tool for developing algorithms. 
The theorems of Sch\"utzenberger~\cite{Schutz65} and McNaughton-Papert~\cite{McNPap71} identify the star-free languages with 
those having aperiodic syntactic monoids and those 
definable in first-order logic with letter predicates and
the order predicate $<$, that is, in $\mathrm{FO}[\Sigma,<]$,
and can be seen as a prime example of this.
Their circuit analogue, namely the Barrington-Compton-Straubing-Th\'erien theorem \cite{BCST92}, relates regular languages in $\mathbf{AC}^0$, quasi-aperiodic stamps and first-order logic over arbitrary numerical predicates. 

In a previous work \cite{GM26arxiva} we developed a framework to carry out such a classification. 
We use the $\calC$-varieties of Straubing~\cite{Straubing02,StraubingPin10}, i.e., in particular length-multiplying pseudovarieties of surjective morphisms onto finite semigroups (lm-pseudovariety of stamps) 
for our task.     

\subsection{Our contribution}
\subsubsection{The Trichotomy Theorem}
Let $\opc(L)$ denote the circuit complexity of a language $L$. 
We consider nonuniform circuits with unbounded fan-in AND
and OR gates, and count only non-input gates. We study the 
circuit model of \cite{Kouckyhandbook} where the input follows a subset encoding, that is to say, for each input length $n>0$, there is an input gate for each pair $(i, \Gamma)$, where $i\in [0,n-1]$ and $\Gamma\subseteq \Sigma$, that is true if the input letter at position $i$ is in $\Gamma$. We refer to \S\,\ref{S Circuit Complexity} for more details.

For classes $\calF$ and $\calG$ of functions from $\N_{>0}$ to $\N_{>0}$, the complexity class
$\SIZEDEPTH(\calF,\calG)$ is the set of languages recognized by a circuit family whose size is in $\calF$  and depth is in $\calG$.

Our main contribution is the following theorem.

\begin{samepage}
\begin{theorem}[Circuit Complexity Trichotomy for Regular Languages]\label{T Trichotomy}
Let $L\subseteq\Sigma^+$ be a regular language.
Then precisely one of the following holds, where $\Omega$ refers to Hardy-Littlewood
	notation (see Definition~\ref{D Hardy} for more details):
\begin{enumerate}
  \item \label{case:constant}
    \textbf{\upshape Constant:}
    $\opc(L)\in O(1)$.

  \item \label{case:sublinear}
	  \textbf{\upshape Almost Logarithmic:}
    $\opc(L)\in
      \Omega(\log n)\cap O(\log n\,\log\log n)$.
	\begingroup
\renewcommand{\thefootnote}{\fnsymbol{footnote}}\!\!\footnote{In fact, we show
		$L\in\SIZEDEPTH(O(\log^2 n),O(1))\cap
		\SIZEDEPTH(O(\log n\,\log\log n),O((\log\log n)^2))$.}
\endgroup

  \item \label{case:linear}
    \textbf{\upshape Linear:}
    $\opc(L)\in\Omega(n)\cap O(n)$.
\end{enumerate}
\end{theorem}
\end{samepage}

It is well-known that regular languages admit circuit families of linear size and
logarithmic depth. Therefore, Theorem~\ref{T Trichotomy} follows from the constant-size and sublinear-size characterizations below.

In \cite{GM26arxiva}, the class of regular languages computed by constant-size circuit families was  characterized in terms of syntactic stamps, congruences, regular expressions and logic.

Let $\varphi:\Sigma^+\to S$ be a morphism onto a finite semigroup $S$. The smallest number $s\geq 1$ satisfying $\varphi(\Sigma^s)=\varphi(\Sigma^{2s})$ is called the {\em stability index of $\varphi$}; the {\em stable semigroup of $\varphi$} is 
$\stab(\varphi)=\varphi(\Sigma^s)$.

\begin{theorem}[\cite{GM26arxiva}, $O(1)$-Characterization Theorem]
\label{T const main}
Let $L\subseteq\Sigma^+$ be a regular language. Then
	the following statements are equivalent:
	\begin{enumerate}
		\item $\opc(L)\in O(1)$.
		\item $\opc(L)\in o(\log n)$.
		\item The syntactic morphism $\eta_L$ belongs to the length-multiplying pseudovariety of stamps $\QESL$, i.e., 
			the stable semigroup $\stab(\eta_L)$ of $\eta_L$ satisfies the identities
			$epxqf=epxxqf$ and $epxyqf=epyxqf$, 
			where $e,f$ range over idempotents 
			and where $p,q,x,y$ range over all elements of $\stab(\eta_L)$.
		\item $L$ is a union of $\approx_k$-classes for some
			$k\in\bbN_{>0}$, where
			$\approx_k$ denotes the following congruence on $\Sigma^+$:
			$u\approx_k v$ if, and only if, $u=v$ or
			\begin{itemize}
				\item $|u|,|v|>2k$,
				\item $u=u'xu''$ and $v=v'yv''$ have the same 
					$k$-length prefix $u'=v'$ and 
			the same $k$-length suffix $u''=v''$, 
		\item $|x|\equiv|y|\text{ mod }k$, and
		\item $x$ and $y$ agree on the set of letters appearing
			in positions that are equivalent mod $k$.
			\end{itemize}
		\item $L$ is a finite Boolean combination of finite languages and languages
			 $u(\Gamma_0\dots \Gamma_{k-1})^+v$, where
			$k\in\bbN_{>0}$, $\emptyset\not=\Gamma_0,\dots,\Gamma_{k-1}\subseteq\Sigma$
			and $u,v\in\Sigma^+$.
		\item $L$ is definable in 
			$\rmFO^1[\Sigma, \mathrm{arb}]$,
		more precisely in the one-variable fragment of first-order
			logic with arbitrary numerical predicates.
			\end{enumerate}
\end{theorem}

Note that the above theorem implies that if a regular language $L$ does not have
a constant-size circuit family, then $c(L)\in \Omega(\log n)$.  
In the present work we prove an analogous result for regular languages recognized by sublinear-size circuit families. 

The objects used in the sublinear-size characterization theorem are refinements or generalizations of the ones used in the $O(1)$-characterization theorem. Their full definitions are involved and therefore we summarize them informally before stating the theorem.

The congruence $\simeq_k$  (Definition \ref{D diag}) is a refinement of $\approx_k$: 
two words $u$ and $v$ are $\simeq_k$ equivalent if they have length at most $2k$ and are equal,
or their length is larger, they are $\approx_k$-equivalent, and their infixes obtained by dropping the prefix and suffix of length $k$ satisfy the following property: in each residue class mod $k$, the frequency of each letter is the same up to threshold $k$. 
Moreover, the sequence of rare positions
(positions whose letter occurs fewer than $k$ times in its respective residue class mod $k$) 
in the middle factor of $u$ and $v$ must agree on their labels as well as 
all pairwise distances among these positions and the middle-factor endpoints
mod $k$ and up to threshold $k$.

The length-multiplying pseudovariety of semigroup stamps $\bfX$ (Definition \ref{D X}) is defined by the infinite set of identities given below, for each $m\in \bbN$ and $i\in \{0,\ldots, m-1\}$
\begin{align}
&(x_0 \cdots x_{m-1})^\omega
  p_0 \cdots p_{i-1} x_i p_{i+1} \cdots p_{m-1}
  (y_0 \cdots y_{m-1})^\omega \notag\\
=~&(x_0 \cdots x_{m-1})^\omega
  p_0 \cdots p_{i-1} y_i p_{i+1} \cdots p_{m-1}
	(y_0 \cdots y_{m-1})^\omega.\label{E X identities}
\end{align}
Variables in such identities are replaced
by nonempty words of a common length, and $\omega$
denotes the idempotent power of the resulting image.
It allows the exchange of letters in idempotent contexts provided the letters appear in congruent positions within the idempotents. The pseudovariety $\bfX$ is not characterizable in terms of stable semigroups (Proposition \ref{P isomorphic}).

The Lindstr\"om extension of first-order logic allows the use of generalized quantifiers whose denotations are specified explicitly. We define the one-variable fragment of the Lindstr\"om extension 
(Definition~\ref{def:lind-one}) analogous to the one-variable fragment of FO. We use this fragment with bounded quantifiers (Definition~\ref{def:bounded-q}) in the sublinear characterization theorem. Intuitively a quantifier $Q$ is bounded, if 
every relation in its denotation is a Cartesian product of sets of uniformly bounded cardinality. The typical example is the quantifier $\exists^{<t}$ for $t \in \bbN$ whose denotation is precisely all subsets of the domain of cardinality $<t$.  
Below $\mathcal B_\tau$ denotes bounded quantifiers defined by
fixed quantifier-free formulas over $\tau$.
The notation $\mathrm{num}[O(f(b)),O(g(b))]$ specifies numerical
predicates computable with the indicated size and depth on
arguments of bit length $b$; $+\omega$ denotes all fixed-offset
predicates $+j$.

A disjoint monomial is a language of the form $\Gamma^*a_0\Gamma^*\ldots a_{n-1}\Gamma^*$ such that $\Gamma\subseteq \Sigma$ and $a_0,\ldots, a_{n-1}\in \Sigma\setminus\Gamma$. 
Straubing showed that Boolean combinations of disjoint monomials characterize the languages recognized by monoids in the pseudovariety generated by finite nilpotent monoids~\cite{mnil}.
We define a modular generalization of them parameterized by a 
modulus $k\geq 1$ (Definition \ref{def:mod-disjoint}).
For $k,t >0$, $i\in \{0,\ldots, k-1\}$ and $a\in \Sigma$, the threshold language $\mathrm{Thr}_{i,k}(a,t)$ is the set of all nonempty words that have at least $t$ occurrences of the letter $a$ in positions $i$ modulo $k$.

\refstepcounter{theorem}
\label{mainthm:number}
\repthm{T log main}

The nontrivial directions are $(3) \implies (4) \implies (5)$, $(9)\implies (1)$ and $(10)\implies (2)$. The direction $(3) \implies (4)$ is proved contra-positively.

For the direction $(3) \implies (4)$ we prove that if the syntactic morphism $\eta_L$ 
of a language $L$ is not in $\bfX$ then any circuit computing $L$ has size $\Omega(n)$. 
Our linear lower bound is proved by a gate-counting argument.
Starting from a violation of the defining identities of $\bfX$,
we reduce a suitable fibre $\eta_L^{-1}(m)$ to $L$.
For a fixed length $k$ and every $n\geq 2$, we construct $n-1$
input pairs of words $(u_1,v_1)\ldots, (u_{n-1},v_{n-1})$
with both words in each pair of length $k(n+1)$
such that $u_i \in \eta_L^{-1}(m)$ and $v_i \not \in\eta_L^{-1}(m)$. 
We say a non-input gate separates such a pair $(u_i,v_i)$ if the 
gate evaluates to $1$ on the first word and $0$ on the second. 
Clearly, in a circuit computing $\eta_L^{-1}(m)$ there must be a 
non-input gate that separates each pair in the sequence. 
The nontrivial part of the lower bound is to prove that 
any non-input gate in a circuit can separate at most one pair
that is not already separated by any of its non-input predecessors.
This yields that any circuit computing $\eta_{L}^{-1}(m)$ has $\Omega(n)$ gates. Thus, if $\eta_L \not \in \bfX$ then $c(L)$
is $\Omega(n)$. 

 To prove, $(4) \implies (5)$, one proves that given a language $L$
 for which $\eta_L\in \bfX$ there is a $k>0$ such that $L$ is a finite union of equivalence classes of $\simeq_k$, i.e., if two words are $\simeq_k$-equivalent then they have the same image under $\eta_L$. 
 To prove the claim, a large enough $k$ is chosen such that for any two words $u,v$ such that $u \simeq_k v$ there exists a word $w$ such that $w$ and $u$ as well as $w$ and $v$ have the same image under the syntactic morphism of $L$. To construct such a $w$, we pump suitable factors in $u$ and $v$ to obtain idempotent contexts containing the frequent letters, and then apply the identities of $\bfX$ to rename the frequent letters to make them syntactically identical as required.  

 The directions $(9)\implies (1)$ and $(10) \implies (2)$ use the following consequence of Friedman's construction~\cite{Friedman86}: for each fixed $t\in \bbN$ there exists a circuit family of logarithmic size 
 and constant depth that computes whether the input contains at least $t$ occurrences of the letter $a$. Combining the tests for thresholds $t$ and $t+1$, we can therefore
check within the same bounds whether the input contains exactly
$t$ occurrences of a given letter $a$.
 Assume $w$ is a word containing precisely $t$ occurrences of $a$. For the translations from the logic to circuits, one needs to compute the distances between the $a$'s up to a fixed threshold. We first compute the positions of these occurrences in binary. Once these positions are available, comparisons of their
order and distances up to a fixed threshold require only logarithmically many gates and constant depth.
The more general numerical predicates allowed in (9) and (10) can likewise be evaluated within
the respective target size and depth bounds.
Thus, the complexity of the resulting circuit is bounded by the circuit complexity of computing indices of a fixed number of $a$'s. 
 In \S\,\ref{S upper} we show that this can be achieved with constant-depth circuits of size $O(\log^2 n)$ (Theorem \ref{thm:occ-log2n}) as well as with circuits of size $O(\log n\log\log n)$ and depth 
 $O((\log\log n)^2)$ (Theorem \ref{T upper loglog}). The $O(\log^2 n)$ upper bound uses Friedman's construction 
 to index each occurrence of $a$ uniquely and then computes its position in binary. 

To obtain the $O(\log n\log\log n)$ upper bound, we use Spielman's
error-correcting codes with linear-size, logarithmic-depth
encoders~\cite{Spielman96}. Combining these encoders with a sampling
argument, we obtain codes over a fixed alphabet with sufficiently
large relative distance. 
We reconstruct the occurrence positions by recursively dividing their binary representations into
blocks of size $b\in O(\log n)$, each of which is subdivided into two blocks of equal length. 
The before-mentioned codes yield a test of size $O(b)$ and depth
$O(\log(b+1))$ for whether a candidate $b$-bit block occurs
among the marked position indices. At each merge, we use
this test to filter the constantly many concatenations
of blocks recovered from the two children.
Since the complete
indices have $O(\log n)$ bits, the recursion has
$O(\log\log n)$ levels, each using $O(\log n)$ gates in total
and adding at most $O(\log\log n)$ depth. This yields circuits
of size $O(\log n\log\log n)$ and depth
$O((\log\log n)^2)$ for computing the occurrence positions, and hence the desired upper bound.

\subsubsection{Effectiveness of the Trichotomy}

 As is often the case with algebraic characterizations of classes of regular languages,
 we obtain decidability of membership:

\refstepcounter{theorem}
\label{intro:decidability}
\repthm{T intro decidability}

It is shown in \cite{GM26arxiva} (Theorem 5.1) that given a regular language $L$ as a nondeterministic finite automaton, the problem of deciding if
$\opc(L)\in O(1)$ is $\mathbf{PSPACE}$-complete. We show that for a given language membership of its syntactic morphism in the pseudovariety $\bfX$ is in $\mathbf{PSPACE}$. This is done in two steps. In the first step it is shown that to check if a syntactic morphism is in $\bfX$ it suffices to check polynomially many identities in the size of the syntactic semigroup. In the second step, it is shown how to compute on the fly whether the identities are satisfied via suitably computed transition matrices of the nondeterministic finite automaton.
To show the lower bound one reduces from the universality problem of NFA in both cases.
\subsubsection{Corollaries of the Characterization Results}
A language has a neutral letter if there exists a fixed letter whose insertion or deletion preserves membership. In~\cite{GM26arxiva}, the neutral letter languages of constant circuit complexity were characterized (Corollary 4.16). In this work we show an analogous result for sublinear circuit complexity.
Here $\congC_1^{t,0}$ records each letter count up to threshold $t$ and the ordered sequence of occurrences of letters whose total frequency is less than $t$. Moreover
$\bfM\mathbf{Nil}$ denotes the class of finite aperiodic monoids whose idempotents
		  commute with every element.
		  
\refstepcounter{theorem}
\label{intro:neutral}
\repthm{T Neutral Letter}

The pseudovarieties of finite semigroups $\bfD\bfA$ and $\bfL\bfD\bfA$ are defined, respectively, by the identities 
$(xy)^\omega x(xy)^\omega=
(xy)^\omega$ and 
$
(z^\omega x z^\omega y z^\omega)^\omega
  \,x\,
  (z^\omega x z^\omega y z^\omega)^\omega
  =
  (z^\omega x z^\omega y z^\omega)^\omega
$. We show that a stamp is in $\QDA$ whenever it satisfies the set of 
identities from (\ref{E X identities})
with the restriction that each $p_j$ is either $x_j$ or $y_j$.
Hence $X\subseteq\QDA$. Our following lemma in fact shows
that this inclusion is strict.

\refstepcounter{theorem}
\label{intro:QDA}
\repthm{L X in QDA}

\refstepcounter{theorem}
\label{intro:Sigma}
\repthm{C polylog QDA}

\refstepcounter{theorem}
\label{intro:WLAC}
\repthm{C polylog WLAC0}

\subsection{Related work}
\label{subsec:related}

\paragraph{Results on Circuit Size}
Hromkovič proved in~\cite{Hromkovic85a} that various explicit Boolean 
functions require 
a linear number of gates in the unbounded fan-in Boolean circuit 
model of unrestricted depth.
Extending an earlier result of
Chandra, Fortune, and Lipton~\cite{ChandraFortuneLipton1985} for
group-free monoids, Kouck\'y~\cite{Koucky2009} showed
that every regular language $L\in\ACO$ can be
recognized by $\ACO$ circuits with
$O(n f^{-1}(n))$ wires, for every unbounded,
nondecreasing primitive recursive function $f$,
where the circuit depth may depend on $f$.

\paragraph{Length-Multiplying Pseudovarieties}

The notion of $\calC$-variety was introduced by Straubing in~\cite{Straubing02}. \cite{ActionCvariety,StraubingPin10} extended Eilenberg's variety theory from finite monoids to general
$\mathcal{C}$-varieties. An analogous equational framework based on 
generalized implicit operations was developed by Kunc~\cite{Kunc10} yielding an equational characterization of these pseudovarieties and, in particular, of the one corresponding to $\mathbf{AC}^0$.

Several key results already use or could be stated in terms of lm-varieties. 
The characterization of regular languages in $\ACO$ by
Barrington, Compton, Th{é}rien and Straubing~\cite{BCST92} can be stated in the following way: the class of regular languages in $\ACO$ corresponds to the lm-pseudovariety $\bfQ\bfA$ of quasi-aperiodic stamps. Likewise, the regular languages in 
 $\mathbf{WLAC}^0$, i.e.~those computed by constant-depth circuit families 
  with a linear number of  wires,
 have recently been characterized
 by Paperman and Cadilhac~\cite{CadilhacP22} as those languages whose
 syntactic morphism lies in the lm-pseudovariety of stamps $\mathbf{QLDA}$, and equivalently those definable in $\rmFO^2[\Sigma, <,+1, \mathrm{mod}]$.

Algebraic characterizations are also known for several fragments
of first-order logic with modular predicates. These include
$\Sigma_1[<,\mathrm{mod}]$ and its Boolean closure
$\calB\Sigma_1[<,\mathrm{mod}]$~\cite{Chaubard} as well as $\rmFO^2[<,\mathrm{mod}]$~\cite{Papermanfo2,KW15} which coincides with the 
$\Sigma_2\cap\Pi_2$ fragment of the logic {$\rmFO[<,\mathrm{mod}]$~\cite{KW15}}.

\paragraph{Programs over Monoids}
The surveys~\cite{BridgesTessonTherien, LogicMeetsAlgebra} and the monograph \cite{Straubingbook} 
discuss in detail the semigroup theory related to circuits. 
Following Barrington's theorem~\cite{Barrington1989}, there has been considerable interest in programs over monoids~\cite{BarringtonTherien1988,McKenzieNC1,PeladeauTCS} and in particular obtaining lower bounds using them. 
Regular languages defined by polynomial-sized programs over monoid varieties 
constitute lm-varieties~\cite{McKenzieNC1}.
In particular, the lm-pseudovariety of stamps $\QESL$ that is used to characterize  
$O(1)$ circuit complexity in \cite{GM26arxiva} has been studied in the context of programs over monoids~\cite{NathanJ,GrosshansMS22}.

\paragraph*{Classification and Trichotomy Results}
Any regular language whose syntactic monoid contains a non-solvable group is $\mathbf{NC}^1$-hard ~\cite{BCST92}. If moreover the syntactic monoid of a regular language $L$ does not contain
any non-solvable group, then $L$ is in $\mathbf{ACC}^0$~\cite{BarringtonTherien1988}. Thus, every regular language is either in $\mathbf{ACC}^0$, or $\mathbf{NC}^1$-complete.

There exist other trichotomy complexity results for regular languages.
Tesson and Th\'erien classify the two-party deterministic
communication complexity of regular languages, maximized over
partitions of the input positions, into constant, logarithmic,
and linear~\cite{TessonTherien2005}.
Amarilli, Labbé, and Paperman study out-of-order membership and prove a constant, logarithmic, 
and linear space trichotomy for the associated finite-monoid evaluation problem~\cite{ALP26} .
Ganardi, Hucke, Lohrey, Mamouras, and Starikovskaya
completely characterize the deterministic space 
complexity of recognizing regular languages in the sliding-window streaming 
model, proving that every regular language requires either constant, logarithmic, 
or linear space~\cite{Ganardi2025SlidingWindow}.
Aaronson, Grier, and Schaeffer gave a complete 
classification of the quantum query complexity of regular 
languages~\cite{AaronsonGS19}.

\subsection{Organization of this paper}

In \S\,\ref{S Preliminaries} we introduce general mathematical notation,
in particular words and languages, recognition by finite semigroups,
pseudovarieties, circuit complexity, and first-order logic over words.

In \S\,\ref{S sublinear} we state our main technical result (Theorem~\ref{T log main}) 
and derive its corollaries; before that we introduce the necessary
notions involving the word congruence, the Lindström one-variable fragment
of first-order logic, and modular disjoint monomials.
Some of the implications appearing in Theorem~\ref{T log main} are proven in
\S\,\ref{S sublinear}, some of them are shown in subsequent sections.

Our circuit complexity lower bound is shown in \S\,\ref{S lower}: it
is proven that regular languages whose syntactic
morphism is not in $\bfX$ have circuit complexity $\Omega(n)$.
In \S\,\ref{S AlgCong} we show that for every regular language $L$
with $\eta_L\in\mathbf X$, there exists $k\geq 1$ such that
$\simeq_k$-equivalent words have the same image under $\eta_L$.
\S\,\ref{S upper} provides the circuit constructions 
for computing the $t$-threshold function
and for extracting the positions of a fixed number of
occurrences of a given letter.

\S\,\ref{S Decidability} establishes
$\textbf{PSPACE}$-completeness of membership in each of the three classes.

A conclusion and open problems are given in \S\,\ref{S Conclusios}.

%!TEX root = main.tex
\newcommand{\Rare}{\mathrm{Rare}}
\newcommand{\Occ}{\mathrm{Occ}}

\section{Preliminaries}\label{S Preliminaries}

By $\N=\{0,1,\dots\}$ we denote the non-negative integers and by 
$\N_{>0}=\{1,2,\dots\}$ the positive integers.
By $[i,j]$ we denote the interval $\{i,i+1,\dots,j\}$.
For each set $X$ we denote by $2^X$ the powerset of $X$.
For each $k\in\bbN_{>0}$ 
let $\equiv_k$ denote the congruence modulo $k$ on $\bbN$
and let $k\N=\{n\in\bbN\mid n\equiv_k0\}$
denote the set of all multiples of $k$. 
For each $t\in\bbN$ 
let $\equiv^t$ denote the threshold congruence on $\bbN$ defined as 
    $m \equiv^t n$ if $m = n$  or  $m,n\geq t$. 
We define
$m \equiv_{k}^t n$ if $m=n$, or $m,n\geq t$ and $m\equiv_k n$.

\subsection{Words and Languages}
For a finite alphabet~$\Sigma$, let $\Sigma^*$ (\resp~$\Sigma^+)$ denote the set of all 
finite words (\resp~non-empty finite words) over~$\Sigma$.
Let $\varepsilon$ denote the {\em empty word} and let
$\cdot$ denote concatenation of words (that we mostly drop). 
We also use $\prod$ in the context of concatenation of words.
The length of a word~$w$ is denoted by~$|w|$.
For words $u,v\in \Sigma^+$ such that $|u|=|v|$ we define the Hamming difference set 
$H(u,v)$ between $u$ and $v$ to be $H(u,v)=\{i \in [0,|u|-1] \mid (u)_i\neq (v)_i\}$. 

By $\Sigma^{s\N}$ (\resp~$\Sigma^{s\N_{>0}}$) we denote the set 
of (\resp~non-empty) words whose length is a multiple of $s$.
For each letter $a\in\Sigma$ we denote by $|w|_a$ the number of
occurrences of the letter $a$ in $w$.
For each $k\in\bbN$ let $\Sigma^{>k}$ denote $\{w\in\Sigma^*:|w|>k\}$
and let $\Sigma^{<k}$ denote $\{w\in\Sigma^*:|w|<k\}$.

Throughout the paper we write words $w\in\Sigma^n$ as $w=w_0\dots w_{n-1}$,
where 
$0,\dots,n-1$ are {\em the positions of $w$} and
$w_0,\dots, w_{n-1}\in\Sigma$ are letters.
We also use the notation $(w)_i=w_i$ to denote the letter at position $i$ of $w$.
By $w^R=w_{n-1}\dots w_0$ we denote the {\em reversal of $w$}.
For a position $i\in[0,n-1]$ and a letter $a\in\Sigma$ we denote
by $w[i\leftarrow a]=w_0\dots w_{i-1}aw_{i+1}\dots w_{n-1}$ the word
that is obtained from $w$ by replacing the letter at position $i$ by $a$.
The \emph{alphabetic content} of a word $w\in\Sigma^*$ is defined
as $\cont(w)=\{(w)_i\in\Sigma\mid i\in[0,|w|-1]\}$.

We say $u\in\Sigma^*$ is a \emph{prefix} (\resp~\emph{suffix}) of $v\in\Sigma^*$, 
if there exists $w\in \Sigma^*$ such that 
$v=uw$ (\resp~$v=wu$). More generally, $u$ is a \emph{factor} of $v$ if there 
exist $w,w'\in\Sigma^*$ such 
that $v=wuw'$.

If $u$ is a prefix of $v$, then the \emph{left quotient} of $v$ by $u$, denoted by~$u^{-1}v$, is the 
word~$w\in\Sigma^*$ such that $v=uw$.
Analogously the \emph{right quotient} of $v=wu$ by $u$ is the 
word~$vu^{-1}=w$. 

A {\em language} is a subset of $\Sigma^*$ and it is {\em positive} if
it is a subset of~$\Sigma^+$. 
The {\em positive left-quotient} of a language $L\subseteq\Sigma^+$ by a word $x\in\Sigma^*$
is the positive language $x^{-1}L = \{w \in \Sigma^+ \mid xw \in L\}$.
The {\em positive right-quotient} is defined analogously.
Unless stated otherwise we will be working with {\em positive} languages in the rest of this
paper.

\subsection{Recognition by Finite Semigroups}\label{S Semigroups}

A semigroup $(S,\cdot)$ is a set $S$ with a binary operation 
$\cdot : S \times S \rightarrow S$ that is associative.  
We often denote the semigroup operation by juxtaposition
and often write $S$ for $(S,\cdot)$.
An {\em idempotent} of a semigroup $S$ is an element $e\in S$ satisfying $e\cdot e=e$;
by $E(S)$ we denote the set of all idempotents of $S$.
In a finite semigroup $S$ every element $s$ has a unique idempotent
power $s^\omega\in E(S)$.
Moreover, there exists $n\geq 1$ such that 
$s^n$ is an idempotent for all $s\in S$, for instance
$n=|S|!$.
Any such $n$ is called an {\em exponent} of $S$.

Let $(S,\cdot)$ and $(T,+)$ be semigroups.  
A {\em (semigroup) morphism} from $S$ to $T$ is a map $\varphi:S\rightarrow T$ such that 
$\varphi(s_1\cdot s_2)=\varphi(s_1)+\varphi(s_2)
\quad \text{for all } s_1,s_2\in S$.
In the context of a morphism $\varphi:\Sigma^+\to S$ we call a word $u\in\Sigma^+$
{\em idempotent} if $\varphi(u)$ is idempotent.
Let~$\Sigma$ and~$\Gamma$ be finite alphabets. 
Any morphism 
$\varphi : \Sigma^+ \to \Gamma^+$ is completely specified by 
its restriction~$\varphi\restriction \Sigma : \Sigma \to \Gamma^+$. 
It is \emph{length-multiplying} if there exists $\ell\in\bbN_{>0}$ 
such that $\varphi(a) \in \Gamma^\ell$ for each~$a \in \Sigma$. 
We often write $u=_\varphi v$ to denote $\varphi(u)=\varphi(v)$.

A {\em semigroup stamp} is a {\em surjective} semigroup morphism of
the form $\varphi:\Sigma^+\to S$, where
$\Sigma$ is a finite alphabet and $S$ is finite.
Since we are only concerned with semigroup
stamps we henceforth call them {\em stamps}.

A subset $T\subseteq S$ is a \emph{subsemigroup} of $S$ 
if $T$ is closed under the semigroup operation. 
An equivalence relation $\sim$ on $S$ is a \emph{congruence} 
if $x \sim y$ implies $xs \sim ys$ and $sx \sim sy$ for all $x,y,s\in S$.  
The {\em quotient} of $S$ by $\sim$, denoted by $S/{\sim}$, 
is the semigroup $\{[s] \mid s \in S\}$ formed by equivalence classes of $S$
with the multiplication $[s_1][s_2]=[s_1s_2]$.
A semigroup $T$ {\em divides} a semigroup $S$, denoted as $T \preceq S$, 
if $T$ is a quotient of a subsemigroup of $S$.  

A language $L \subseteq \Sigma^+$ is \emph{recognized} by a finite semigroup $S$
if there exists a morphism 
$\varphi : \Sigma^+ \to S $
and a subset $P \subseteq S$ such that 
$L = \varphi^{-1}(P)$.
A language $L \subseteq \Sigma^+$ is said to be \emph{recognizable} 
if it is recognized by some finite semigroup. 

Recognizable languages are closed under Boolean operations, quotients by words and images 
under morphisms and inverse morphisms. 
Given a recognizable language $L\subseteq\Sigma^+$ the 
{\em syntactic congruence} $\sim_L$ on $\Sigma^+$ is given by
$u\sim_L v$ if
$xuy\in L\Leftrightarrow xvy\in L$ for each $x,y\in \Sigma^*$.
The corresponding semigroup $\Sigma^+/{\sim_L}$ is called the {\em syntactic semigroup}, 
denoted by $S(L)$ and the canonical morphism $\eta_L: \Sigma^+ \to S(L)$ is called the {\em syntactic morphism of $L$}. 
Note that $\eta_L$ is a stamp and we therefore refer to $\eta_L$ as
the {\em syntactic stamp of $L$} also.

It is well-known that the class of recognizable languages coincides
with the class of languages accepted by finite-state automata or equivalently
by regular expressions, namely the class of {\em regular languages}.

Let $\varphi:\Sigma^+\to S$ be a morphism onto a finite semigroup $S$.
The smallest number $s\geq 1$ satisfying $\varphi(\Sigma^s)=\varphi(\Sigma^{2s})$ is called the {\em stability index of $\varphi$};
the {\em stable semigroup of $\varphi$} is 
$\stab(\varphi)=\varphi(\Sigma^s)$.
We remark that $s$ is (at most) exponential in $|S|$.

\subsection{Pseudovarieties}\label{S Pseudo}

A \emph{pseudovariety of finite semigroups} is a class $\bfV$ 
of finite semigroups closed under finite direct 
products and 
division, i.e., if $S\in\bfV$ and $T$ divides $S$, then $T$ is in $\bfV$.  

The \emph{diagonal direct product} of the stamps $\varphi_1: \Sigma^+ \to S_1$ and $\varphi_2: \Sigma^+ \to S_2$ is the stamp 
$\varphi:w \mapsto (\varphi_1(w),\varphi_2(w))$ from $\Sigma^+$ onto the 
subsemigroup $\varphi(\Sigma^+) \subseteq S_1\times S_2$. 
The stamp $\varphi_1 : \Sigma^+ \rightarrow S_1$ {\em lm-divides} 
the stamp  $\varphi_2 : \Gamma^+ \rightarrow S_2$ 
if there exists a length-multiplying morphism 
$\alpha:\Sigma^+ \rightarrow \Gamma^+$ and a surjective morphism $\beta: \operatorname{Im}\,(\varphi_2 \circ \alpha) \rightarrow S_1$ 
such that $\varphi_1 = \beta \circ \varphi_2 \circ \alpha$. 
A class of stamps $\bfV$ is an
{\em lm-pseudovariety of stamps} if it is closed under finite diagonal direct products and lm-divisions.

If $\bfV$ is a pseudovariety of finite semigroups, 
then the collection of all stamps whose stable semigroup is in $\bfV$ forms an 
lm-pseudovariety of stamps, denoted by $\bfQ\bfV$ \cite{Straubing02}.

\subsection{Circuit Complexity}\label{S Circuit Complexity}

\subsubsection{Asymptotics}

Consider two functions $f,g:\N_{>0}\to\bbR_{\geq 0}$.
As usual, we write $f\in O(g)$ if there exists
$c\in\bbR_{>0}$ such that 
 $f(n)\leq c\cdot g(n)$ for all but finitely many $n\in\bbN_{>0}$.
 We write $f\in\Theta(g)$ if $f\in O(g)$ and $g\in O(f)$.
Recall that $f\in o(g)$ if for all
$c\in\bbR_{>0}$ 
we have $f(n) < c\cdot g(n)$ for all but finitely many $n\in\bbN_{>0}$.

For lower bounds the following weaker notion is more
suitable for circuit complexity lower bounds.
We refer to~\cite{VitanyiM84} for more details.

\begin{definition}[Hardy-Littlewood $\Omega$ notation]\label{D Hardy}
	For two functions $f,g:\N_{>0}\to\bbR_{\geq 0}$ we write 
	$f\in\OmegaHL(g)$ if 
	there exists $c\in\bbR_{>0}$ such that $f(n) \geq c\cdot g(n)$
	for infinitely many $n\in\N_{>0}$.
\end{definition}

Note that $f\in\OmegaHL(g)$ if, and only if, $f\not\in o(g)$.

\subsubsection{The Circuit Model}
Next we introduce the particular circuit model used in this paper.
The presentation of the inputs employs a subset encoding following
Kouck\'y~\cite{Kouckyhandbook}.
With respect to succinctness it is asymptotically equivalent to the
one-hot encoding model
under the promise that the input is well-formed, as precisely formulated
in Proposition 6.16 in~\cite{GM26arxiva}.

\begin{samepage}
	\begin{definition}\label{D Circuit}
	For each $n\in\bbN_{>0}$ an {\em $n$-circuit} 
over a finite non-empty alphabet $\Sigma$ 
is a tuple $C=(G,\prec,\lambda,out)$, where the following hold:
\begin{itemize}
	\item  $(G,\prec)$ is a finite
		directed acyclic graph,
		i.e. $\prec^+$ is a strict partial order.
		The set $G$ is a non-empty set of {\em gates} and 
		$\prec\ \subseteq G\times G$ is a set of {\em wires}. 
		The set of gates $G_0\subseteq G$ of 
		in-degree $0$ are the {\em input gates}. 
		The set of gates $G_1=G\setminus G_0$
		of in-degree $>0$ are the {\em non-input gates}.
\item $\lambda:G\rightarrow\{\wedge,\vee\}\cup \left([0,n-1]\times 2^\Sigma\right)$
assigns each gate a type, 
		where moreover we have $\lambda(G_0)= [0,n-1]\times 2^\Sigma$
		and $\lambda(G_1)\subseteq\{\wedge,\vee\}$.
	\item By $G_{out}=\{g\in G_1\mid g\text{ has out-degree 0}\}$
		we denote the set of {\em output gates}.
		Moreover $out:[0,m-1]\to G_{out}$ is a bijection, where
		$m\in\bbN_{>0}$.
		\end{itemize}
\end{definition}
\end{samepage}
An {\em input} to $C$ is a word from $\Sigma^n$.
Let us define the function $\eval:\Sigma^n\times G\to\{0,1\}$
by induction on $\prec$ for all $w=w_0\dots w_{n-1}$.
For the induction base, i.e., when $g$ is of in-degree $0$
and $\lambda(g)=(i,\Gamma)$
we define $\eval(w,g)=1$ if $w_i\in\Gamma$ and $\eval(w,g)=0$ otherwise.
For the induction step, in case $g$ is an $\wedge$-gate 
(\resp~$\vee$-gate)
we have $\eval(w,g)=\bigwedge_{h\prec g}\eval(w,h)$
(\resp~$\eval(w,g)=\bigvee_{h\prec g}\eval(w,h)$).
The circuit $C$ {\em computes the function $f_C:\Sigma^n\to\{0,1\}^m$}
given by $w_0\dots w_{n-1}\mapsto\prod_{i=0}^{m-1}\eval(w,out(i))$ for all
$w\in\Sigma^n$, where $\prod$ denotes concatenation.
In case $m=1$ we also define {\em the language computed by the circuit $C$} as
$L(C)=\{w\in \Sigma^n\mid f_C(w)=1\}$.

The {\em size} of $C$ is $|C|=|G_1|$, i.e. 
the number of non-input gates. Note that since we require that 
$|G_{out}|\geq 1$ we always have $|C|\geq 1$.
The {\em wire size} of $C$ is $\wires(C)=|\!\!\prec\!\!|$.
The {\em depth} of $C$, denoted by $\depth(C)$, is the length 
(i.e.,~the number of wires) of a longest path from 
some input gate to some output gate in $(G,\prec)$.
The {\em fan-in} of $C$ is defined as
$\fanin(C)=\max\{|\{h\in G\mid h\prec g\}|: g\in G_1\}$.

A {\em circuit family over the alphabet $\Sigma$} is an infinite tuple 
$\calC=(C_n)_{n\in\bbN_{>0}}$, 
where each $C_n$ is an $n$-circuit over $\Sigma$.
The circuit family $\calC=(C_n)_{n\geq 1}$ over
$\Sigma$ {\em computes
the function} $f_\calC:\Sigma^+\to\{0,1\}^+$ given by 
$w\mapsto f_{C_{|w|}}(w)$.
In case each $C_n$ has precisely one output gate
we say that $\calC$ {\em computes the language} 
$L(\calC)=\bigcup\{L(C_n)\mid n\geq 1\}$.
The {\em size} of the family $\calC$, denoted 
by $|\calC|:\N_{>0}\to\bbN_{>0}$, is defined to be the function $n\mapsto |C_n|$.
The {\em depth} (\resp~{\em fan-in}) of the family $\calC$, denoted 
by $\depth(\calC)$ (\resp~{\em $\fanin(\calC)$}) 
is defined to be the function $n\mapsto\depth(C_n)$
(\resp~$n \mapsto\fanin(C_n)$).
The {\em wire size} of the family, denoted by $\wires(\calC):\N_{>0}\to\bbN_{>0}$,
is defined to be the function $n\mapsto \wires(C_n)$.
We remark that our circuit families have unbounded fan-in and
that they are nonuniform:
we impose no condition on our circuit families, other than that 
they satisfy the required size bounds.
Note that given any language $L\subseteq \Sigma^+$ and any input length 
$n\in\bbN_{>0}$
there exists an $n$-circuit $C_n$ such that $L\cap \Sigma^n=L(C_n)$.
Note that for every language $L\subseteq\Sigma^+$ there exists a circuit
family $\calC$ computing $L$, where moreover $|\calC|$ is exponential.
The {\em circuit complexity of a language $L\subseteq\Sigma^+$} is defined
to be the function $\opc(L):\bbN_{>0}\to\bbN_{>0}$, where 
\begin{align}
	\opc(L): n\mapsto
	\min\{|C|:\text{$C$ is an $n$-circuit computing $L\cap\Sigma^n$}\}.
\end{align}

\subsubsection{Reductions}

In this paper we work with the following notion of reduction~\cite{GM26arxiva}.
Let $L_1\subseteq \Sigma_1^+$ and $L_2\subseteq \Sigma_2^+$ be languages.
We write
$L_1\reduce L_2$ if 
there exists a length-multiplying morphism 
$\varphi:\Sigma_1^+\rightarrow \Sigma_2^+$ such that
$L_1$ is a finite Boolean combination over
a finite set of languages all of the form $\varphi^{-1}(u^{-1}L_2v^{-1})$,
where $u$ and $v$ range over $\Sigma_2^*$.

\begin{lemma}[{\cite[Lemma 3.8]{GM26arxiva}}]\label{L reduce}
	Let $L$ be a regular language.
	Then for each $x\in S(L)$ we have $\eta_L^{-1}(x)\reduce L$.
\end{lemma}

Let $f:\bbN_{>0}\to\bbR_{\geq 0}$ be a function. 
We call $f$ {\em non-decreasing} if 
$f(n)\leq f(n+1)$ for all $n\in\bbN_{>0}$.
We call $f$ {\em eventually positive} if $f(n)>0$ for all but finitely 
many $n\in\bbN_{>0}$.

Following~\cite{GM26arxiva} a 
	function $f:\bbN_{>0}\to\bbR_{\geq 0}$ is {\em mild} if 
	$f$ is non-decreasing, eventually positive, and satisfies 
	$f(n)\in\Theta(f(\lceil\varepsilon n\rceil))$ for every $\varepsilon\in\bbR_{>0}$.
	Note that the eventually positive constant functions are mild
	and every non-decreasing and eventually positive polynomial is mild.
	Moreover, $n\mapsto\log^k n$ is mild for all $k\in\N_{>0}$.
 The function $n\mapsto 2^n$ is not mild. 

\begin{samepage}
	\begin{theorem}[{\cite[Theorem 3.7]{GM26arxiva}}]\label{T asymptotics}
	Let $f:\bbN_{>0}\to\bbR_{\geq 0}$ be a mild function.
	Let $L_1$ and $L_2$ be languages such that $L_1\leq L_2$.
	\begin{enumerate}
		\item If $\opc(L_2)\in O(f)$, then $\opc(L_1)\in O(f)$.
		\item If $\opc(L_1)\in\OmegaHL(f)$, then $\opc(L_2)\in \OmegaHL(f)$.
	\end{enumerate}
\end{theorem}
\end{samepage}

\subsubsection{Circuit Complexity Classes}

Next, we define circuit complexity classes. 
For a class of functions $\calF$ from $\N_{>0}$ to $\N_{>0}$ define the size complexity class
$\SIZE(\calF)$ as follows
\begin{align}
	\SIZE(\calF)=\{L \subseteq \Sigma^+ \mid \exists 
	\text{fam.} \calC: L(\calC)=L, |\calC|\in \calF\}.
\end{align}
Analogously, for classes $\calF$ and $\calG$ of functions, define
\begin{align}
\SIZEDEPTH(\calF,\calG)=\{ L \subseteq \Sigma^+ \mid \exists
\text{fam.}\ \calC: 
	L(\calC)=L, |\calC|\in \calF,\depth(\calC)\in\calG\}.\label{D SIZEDEPTH}
\end{align}
Finally, we define the class 
\begin{align}
\WIRESIZEDEPTH(\calF,\calG)=
\{ L \subseteq \Sigma^+ \mid \exists
\text{fam.}\ \calC: 
	L(\calC)=L, \wires(\calC)\in \calF,\depth(\calC)\in\calG\}\label{D WSIZEDEPTH}
\end{align}
and $\WLACO=\WIRESIZEDEPTH(O(n),O(1))$.

Assume that $L$ is a regular language. For an element $s\in S(L)$, it is 
straightforward to construct a circuit family of linear size and logarithmic depth 
that computes the set of words whose image under the syntactic morphism 
$\eta_L$ is $s$.  As a consequence, we obtain

\begin{samepage}
\begin{fact}
\label{prop:reg-lin} For each regular language $L$ we have
	$L\in\SIZEDEPTH(O(n),O(\log n))$.
\end{fact}
\end{samepage}

\subsection{First-Order Logic over Words}\label{S FO}

A word $w=a_0\dots a_{n-1}\in\Sigma^n$ is represented as a first-order structure over the domain $\{0,\dots,n-1\}$ along with the unary letter predicates $P_a$, for each $a\in \Sigma$, where $P_a(i)$ holds if, and only if, $a_i=a$. We use $\Sigma$ to denote the set of unary letter predicates in the vocabulary. 

The constant symbols $\min$ and $\max$ denote the \emph{first} and the \emph{last} positions of the word, respectively.
The binary predicate $<$ denotes the order on the positions. For $k\geq 0$, the binary relation $+k$, written as $x+k=y$, holds if and only if $y$ is the $k$-th \emph{successor} of $x$.
The collection of all such successor relations is denoted in the vocabulary by $+\omega$. For $r,k\in \bbN$ with $k\geq 1$,
the unary modular predicate $x \equiv r \bmod k$  is true at a position $j$ if $j\equiv r \bmod k$. The collection of modular predicates is denoted by $\mathrm{mod}$ in the vocabulary.

Let $\tau$ be a relational vocabulary on words. The set $\rmFO[\tau]$ of first-order $\tau$-formulas is the smallest set containing the atomic formulas $R(t_0,\ldots,t_{r-1})$, for each $r$-ary relation symbol $R\in\tau$, and $t=t'$, where $t_0,\ldots, t_{r-1},t,t'$ are constants or variables, and closed under Boolean connectives and existential and universal quantification.

A predicate or constant is \emph{numerical} if its interpretation only depends on the domain, i.e., its interpretation is common for all words of the same length. 
By $\mathrm{arb}$ we denote the class of all \emph{arbitrary} numerical predicates and constants. 

Given a word $w$ an assignment $I$ maps variables to $[0,|w|-1]$. Formulas are interpreted in $w$ under assignments, with relation symbols interpreted according to
$w$, equality interpreted as equality on $[0,|w|-1]$, Boolean connectives interpreted as usual, and quantifiers ranging over $[0,|w|-1]$. We write $w,I\models\varphi$ when $\varphi$ is satisfied
under $I$.
For a formula $\varphi\in\rmFO[\tau]$ and variables
$x_0,\ldots,x_{k-1}$, we write $\varphi^{w,I}\subseteq [0,|w|-1]^k$
for the relation
\begin{align}
\varphi^{w,I} &= \left\{ (a_0,\ldots,a_{k-1})\in [0,|w|-1]^k :  w,I[x_0\mapsto a_0, \ldots,x_{k-1}\mapsto a_{k-1}]
\models\varphi \right\},
\end{align}
where $I[x_0\mapsto a_0,\ldots,x_{k-1}\mapsto a_{k-1}]$
denotes the assignment obtained by updating $I$ on the specified
variables, leaving all other variable assignments unchanged.

For $m\in \bbN_{>0}$, the counting quantifier $\exists^{< m} x$
is true of a formula $\varphi(x)$ if the number of elements satisfying $\varphi$ is $<m$, that is,
\begin{align}
w,I \models \exists^{<m} x\, \varphi(x) \Longleftrightarrow \big| \varphi^{w,I}\big|< m.
\end{align}
The quantifier $\exists^{\geq m} x$ is defined as the negation of $\exists^{< m}$, that is
\begin{align}
    \exists^{\geq m} x\, \varphi(x) \defeq \neg \exists^{< m} x\, \varphi(x).
\end{align}

We write $C[\tau]$  for the extension of $\rmFO[\tau]$ obtained by adding counting quantifiers.
For each $k\geq 0$, the logic $\rmFO^k[\tau]$ is the fragment of $\rmFO[\tau]$ consisting of formulas that use at most $k$ variables.
Likewise $C^k[\tau]$ is the $k$-variable fragment of $C[\tau]$. In particular, $\rmFO^1[\tau]$ is the fragment using only the variable $x$ and $\rmFO^2[\tau]$ is the fragment using only the variables $x$ and $y$.

A sentence is a formula with no free variable.
The language $L(\varphi)$ of a sentence $\varphi$ is the set of all words satisfying $\varphi$. A language $L\subseteq \Sigma^+$ is defined by the sentence $\varphi$ if $L(\varphi)=L$.

\begin{example}
\label{example:logic-const}
Let $w=a_0\ldots a_{n-1}\in \Sigma^+$. The singleton set $\{w\}$ is defined by the following formula in $\rmFO^1[\,\Sigma, \min, \max, +\omega\,]$.
  \begin{align}
    \varphi_w &\defeq \min+(n-1)=\max \,\wedge \bigwedge_{0\leq i \leq n-1}\exists x\, (\min+i = x \wedge P_{a_{i}}(x)).
    \end{align}
The language $(ab)^+$ is defined by the formula $\varphi_{(ab)^+} \in \rmFO^1[\,\Sigma, \min, \max, \mathrm{mod}\,]$.
\begin{align}
 \varphi_{(ab)^+} &\defeq \max \equiv 1 \bmod 2\, \wedge\forall x\, (( x \equiv 0 \bmod 2 \rightarrow P_a(x) ) \wedge (x \equiv 1 \bmod 2 \rightarrow P_b(x))).
\end{align}
The language $\mathit{b^*ab^*ab^*} \subseteq \{a,b\}^*$ is defined by the formula $\varphi_{\mathit{b^*ab^*ab^*}} \in  C^1[\,\Sigma\,]$.
\begin{align}
\varphi_{\mathit{b^*ab^*ab^*}} &\defeq \exists^{\geq 2}x\, P_a(x) \wedge \exists^{< 3}x\, P_a(x).
\end{align}
\end{example}

%!TEX root = main.tex

\section{Regular Languages of Sublinear Circuit Complexity}
\label{S sublinear}
This section is devoted to stating our main result.
It is organized as follows.
In \S\,\ref{S Congruence} we introduce ordered modular threshold
congruence.
In \S\,\ref{S X} we introduce the pseudovariety $\bfX$ and its examples.
In \S\,\ref{S LFO} we introduce the Lindstr\"om extension of first-order logic and its one-variable fragment.
Disjoint (modular) monomials are the subject of \S\,\ref{S Disjoint}.
In \S\,\ref{S log main} we state our main result.
Corollaries of our main result are discussed in \S\,\ref{S Corollaries}.

\subsection{Ordered Modular Threshold Congruence}\label{S Congruence}

We start with some properties of the congruences $\equiv^t$ and $\equiv_k^t$.
The proofs follow directly from definition and are therefore omitted.
\begin{samepage}
	\begin{lemma}
\label{L equiv}
	Let $s,k\in\bbN_{>0}$ and $t,t'\in\bbN$. 
		Then the following properties hold:
	\begin{enumerate} 
		\item $\equiv_{k}^t$ is a congruence on $\bbN$ with respect to addition.
		\item If $t\geq t'$, then $\equiv^t$ refines $\equiv^{t'}$
			and $\equiv_k^t$ refines $\equiv_k^{t'}$.
		\item If $s|k$, then $\equiv_k^t$ refines $\equiv_s^{t}$.
		\item If $s|k$, then $\equiv_k^k$ refines $\equiv_s^s$.
	\end{enumerate}
\end{lemma}
\end{samepage}
Let $k\in\bbN_{>0}$. Assume that $w=a_0\dots a_{n-1}\in \Sigma^*$. 
The {\em content} of $w$ is $\cont(w) = \{a_i \mid i \in [0,n-1]\}$.  
For each $i\in\bbZ$ we define
\begin{align}
	\pi_{i,k}(w) = \prod_{j\in [0,n-1],\atop
	j\equiv_k i}a_j,
\end{align}
where concatenation is done in increasing order.
The content of the modular class $i\in \bbZ$ is given by  
$\cont_{i,k}=\cont \mathbin{\circ} \pi_{i,k}$.

    Given a threshold $t\in\bbN_{>0}$, a comparison operation 
${\bowtie}\in\{=,<,>,\leq,\geq\}$,
and a word $w\in \Sigma^*$ let 
$\cont^{\bowtie\, t}(w)=\{a\in \Sigma:|w|_a\bowtie t\}$.
We define
$\cont^{\bowtie\, t}_{i,k} = \cont^{\bowtie\, t} \mathbin{\circ} \pi_{i,k}$.

The set of occurrences of letter $a$ in positions of $w$ that are in residue class $i\bmod{k}$ is denoted by 
\begin{align}
    \rmP_{a,i,k}(w)=\{j\in [0,n-1] : a_j=a, j \equiv_k i\}.
\end{align}

The letters in $\cont_{i,k}^{< t}(w)$ are the \emph{rare} letters in the 
residue class $i$. The set of rare positions in modular class $i$ is 
\begin{align}
\rmR_{i,k}^t(w) &= \bigcup_{a \in \cont_{i,k}^{< t}(w)} \rmP_{a,i,k}(w).
\end{align}
Finally, the set of rare positions in $w$ is 
\begin{align}
\rmR_{k}^t(w)&=\bigcup_{i\in[0,k-1]}
			\rmR_{i,k}^t(w).
\end{align}

\begin{definition}[The Modular Threshold-Counting Congruence $\scrC_k^t$ \cite{GM26arxiva}]
Let $k,t\in \bbN_{>0}$.  The congruence $\congC_k^t$ on $\Sigma^+$ is defined as follows: $u \congC_k^t v$ if
\begin{align}
  |u|\equiv_k |v| \text{ and }   \left |\pi_{i,k}(u)\right |_a \equiv^t \left |\pi_{i,k}(v)\right |_a 
\end{align}
	for all $a\in \Sigma$ and $i\in[0,k-1]$.
\end{definition}

Clearly, each non-empty word in $\Sigma^{< k}$ forms its own equivalence class. 
Also, $u \congC_k^1 v$ if $|u|\equiv_k|v|$ and $\cont_{i,k}(u)=\cont_{i,k}(v)$ 
for all $i\in\bbN$.

\begin{remark}\label{R calR}
	If $u \mathbin{\congC_k^t} v$ then $\left |\pi_{i,k}(u)\right |_a \equiv^t \left |\pi_{i,k}(v)\right |_a$  
	for all $a\in \Sigma$ and $i\in[0,k-1]$, and thus $|\rmR_{k}^t(u)|  = |\rmR_{k}^{t}(v)|$.
\end{remark}
Next we define a refinement of $\scrC_k^t$.

\begin{samepage}
\begin{definition}[The Ordered Modular Threshold Congruence $\scrC_k^{t,d}$]
	\label{D ordered}
Let $k,t\in \bbN_{>0}$ and $d\in\bbN$.
	The relation $\congC_k^{t,d}$ on $\Sigma^+$ is defined as follows: $u \congC_k^{t,d} v$ if
\begin{enumerate}
	\item $u \mathbin{\congC_k^t} v$, and
\item there is an order-preserving bijection $\beta: \rmR_k^t(u) \cup \{0,|u|-1\} \to  
\rmR_k^t(v) \cup \{0,|v|-1\}$ such that
		\begin{enumerate}[label=(a)]
\item[(a)] $(u)_{i}=(v)_{\beta(i)}$ for all $i\in\rmR_k^t(u)$, and
\item[(b)] $j-i\equiv^d_k \beta(j)-\beta(i)$ for 
	all $i,j\in\rmR_k^t(u)\cup\{0,|u|-1\}$ such that $i<j$.
\end{enumerate}
\end{enumerate}
\end{definition}
\end{samepage}

\begin{remark}\label{R beta}
	Suppose $u\congC_k^{t,d} v$ and let $\beta$ witness
	Point (2) of Definition~\ref{D ordered}.
	Then $\beta$ satisfies the following:
	\begin{enumerate}
		\item $\beta(i)\equiv_k i$ for all $i\in\rmR_k^t(u)\cup\{0,m-1\}$, where $m=|u|$.
		\item The restriction of $\beta$ to $\rmR_k^t(u)$ is a bijection from $\rmR_k^t(u)$ to $\rmR_k^t(v)$.
	\end{enumerate}
\end{remark}

The remark below follows from the definition.
\begin{remark}\label{R cong equiv}
	The relation $\congC_k^{t,d}$ is an equivalence relation.
	Moreover, $u \congC_k^{t,d} v$ if, and only if, 
	$u^R \congC_k^{t,d} v^R$.
\end{remark}

	\begin{samepage}
		\begin{lemmarep}\label{L congC}
Let $k,t,t',s \in \bbN_{>0}$ and $d,d'\in\bbN$.
\begin{enumerate}
	\item  $\congC_{k}^{t,d}$ is a congruence.
    \item  $\congC_{k}^{t,d}$ refines $\congC_{s}^{t,d}$ whenever $s|k$.
    \item  $\congC_{k}^{t,d}$ refines $\congC_{k}^{t',d'}$ whenever $t\geq t'$ and $d\geq d'$.
\end{enumerate}
\end{lemmarep}
	\end{samepage}
\begin{appendixproof}
	(1):\quad
	By Remark~\ref{R cong equiv} it suffices to show
	that $u\mathbin{\congC_k^{t,d}} v$ 
	implies $ua\mathbin{\congC_k^{t,d}} va$ 
	for all $u,v\in\Sigma^+$ and all $a\in\Sigma$.

	Let us assume $u\mathbin{\congC_k^{t,d}} v$. Since $\congC_k^t$ is a 
	congruence by Lemma~4.4 of~\cite{GM26arxiva},
	we have $ua\mathbin{\congC_k^t}va$, which shows Point (1) of 
	Definition~\ref{D ordered}.

	It remains to prove Point (2) of Definition~\ref{D ordered}.
	Let $m=|u|$ and $n=|v|$.
	By assumption there exists an order-preserving bijection 
	$\beta:\rmR_k^t(u)\cup\{0,m-1\}\to\rmR_k^t(v)\cup\{0,n-1\}$ 
	satisfying 
	\begin{align}
		(u)_{i}=(v)_{\beta(i)}&\quad\text{ for all $i\in\rmR_k^t(u)$, and}\label{E uivi}\\
		j-i\equiv^d_k \beta(j)-\beta(i)&\quad\text{ for 
		all $i,j\in\rmR_k^t(u)\cup\{0,m-1\}$ such that $i<j$.}\label{E betaij}
	\end{align}

We need to show that there exists an order-preserving bijection 
	$\beta':\rmR_k^t(ua)\cup\{0,m\}\to\rmR_k^t(va)\cup\{0,n\}$ 
	satisfying 
	\begin{align}
		(ua)_{i}=(va)_{\beta'(i)}&\text{ for all $i\in\rmR_k^t(ua)$, and}\label{E uauv}\\
		j-i\equiv^d_k \beta'(j)-\beta'(i)&\text{ for 
		all $i,j\in\rmR_k^t(ua)\cup\{0,m\}$ such that $i<j$.}\label{E beta}
	\end{align}

Note that $\rmR_k^t(ua)\cap[0,m-1]\subseteq\rmR_k^t(u)$.
Let $\beta'$ be the map from $\rmR_k^t(ua)\cup\{0,m\}$ to 
	$\rmR_k^t(va)\cup\{0,n\}$ 
	such that $\beta'(i)=\beta(i)$ for all  
	$i \in (\rmR_k^t(ua)\cup\{0\})\cap[0,m-1]$ and $\beta'(m)=n$.

Let us first prove (\ref{E uauv}).
In the following, let $\ell=|u|\bmod{ k}=|v|\bmod{ k}$.

Let us first consider the case when $m\in\rmR_k^t(ua)$,
	hence $|\pi_{\ell,k}(ua)|_a<t$.
	Thus, $|\pi_{\ell,k}(u)|_a< t-1$. Since $u\mathbin{\congC_k^t}v$
	it follows $|\pi_{\ell,k}(u)|_a=|\pi_{\ell,k}(v)|_a< t-1$.
	Hence $\rmR_k^t(ua)=\rmR_k^t(u)\uplus\{m\}$, $n\in\rmR_k^t(va)$, so
	$\rmR_k^t(va)=\rmR_k^t(v)\uplus\{n\}$. 
It follows that $\beta'$ is an order-preserving bijection
	from $\rmR_k^t(ua)\cup\{0,m\}$ to $\rmR_k^t(va)\cup\{0,n\}$
and that $(ua)_m=a=(va)_n=(va)_{\beta'(m)}$.
Combining the latter with
	(\ref{E uivi}), we have that (\ref{E uauv}) follows.

	Let us now consider the case when $m\not\in\rmR_k^t(ua)$.
	Hence we have
	\begin{align}
		\rmR_k^t(ua)=\rmR_k^t(u)\setminus\{i\in[0,m-1]\cap (k\bbN+\ell) \mid (u)_i=a\}.
	\end{align}
	Since $u\congC_k^{t}v$ we have both 
	\begin{align}
		\rmR_k^t(va)=\rmR_k^t(v)\setminus\{i\in[0,n-1]\cap (k\bbN+\ell)\mid (v)_i=a\}
	\end{align}
	and $|\rmR_k^t(ua)|=|\rmR_k^t(va)|$.
	We claim that $\beta(i)\equiv_k i$ for all $i\in\rmR_k^t(ua)$:
	in case $i=0$ we have $\beta(0)=0$ since $\beta$ is order-preserving,
	in case $i>0$ we have $\beta(i)=\beta(i)-0\equiv_k i-0=i$ by
	 (\ref{E betaij}).
	Therefore $\beta'(i)\in\rmR_k^t(va)$ for all $i\in\rmR_k^t(ua)$.

Thus, the mapping $\beta'$ is an order-preserving bijection from 
$\rmR_k^t(ua)\cup\{0,m\}$ to $\rmR_k^t(va)\cup\{0,n\}$. By (\ref{E uivi})
	we can conclude (\ref{E uauv}).

It remains to prove~\eqref{E beta}.
Let $i<j$ belong to $R_k^t(ua)\cup\{0,m\}$.
If $j<m$, the required relation follows directly
from~(\ref{E betaij}), since $\beta'$ agrees with $\beta$
below $m$.

If $j=m$, recall that $\beta(m-1)=n-1$.
For $i<m-1$, equation~\eqref{E betaij} gives
\begin{align}
(m-1)-i \equiv_k^d (n-1)-\beta'(i).
\end{align}
For $i=m-1$, the same relation holds because both sides
are zero. Adding one by Point (1) of Lemma~\ref{L equiv} yields
\begin{align}
m-i \equiv_k^d n-\beta'(i)
    = \beta'(m)-\beta'(i).
\end{align}
This proves~\eqref{E beta}.

	\medskip

	\noindent
(2):\quad Let $s|k$ and $u\mathbin{\congC_k^{t,d}} v$. We need to prove
$u\mathbin{\congC_s^{t,d}} v$.
We have $u\mathbin{\congC_s^t}v$ by Point (2) of Lemma~4.4 of~\cite{GM26arxiva}.
Thus Point (1) of Definition~\ref{D ordered} for $\congC_s^{t,d}$ is established.

It remains to prove Point (2).
	By assumption there exists an order-preserving bijection 
	$\beta:\rmR_k^t(u)\cup\{0,m-1\}\to\rmR_k^t(v)\cup\{0,n-1\}$ 
	satisfying 
	\begin{align}
		(u)_{i}=(v)_{\beta(i)}&\quad\text{ for all $i\in\rmR_k^t(u)$, and}\label{E uiviX}\\
		j-i\equiv^d_k \beta(j)-\beta(i)&\quad\text{ for 
		all $i,j\in\rmR_k^t(u)\cup\{0,m-1\}$ such that $i<j$.}\label{E betaijX}
	\end{align}

Since $s|k$ note that $\rmR_s^t(u)\subseteq\rmR_k^t(u)$
and $\rmR_s^t(v)\subseteq\rmR_k^t(v)$.
We claim that $\beta(p)\in\rmR_s^t(v)$ for all $p\in\rmR_s^t(u)$.
Note that there is a letter $a\in\Sigma$ such that
$(u)_p=a=(v)_{\beta(p)}$ by (\ref{E uiviX}).
Secondly, we have
$\beta(p)=\beta(p)-0\equiv_s p-0=p$ by 
(\ref{E betaijX}) whenever $p>0$ and directly from the definition
of $\beta$ when $p=0$.
Letting $\ell=p\bmod{ s}=\beta(p)\bmod{ s}$ 
it follows 
\begin{align}
	p\in\rmR_s^t(u) &\quad\Longleftrightarrow\quad a\in\cont_{\ell,s}^{<t}(u)
	\label{E p in Rstu1}\\
	&\quad\Longleftrightarrow\quad a\in\cont_{\ell,s}^{< t}(v)& 
	\text{(Since $u\mathbin{\congC_s^t}v$)}\label{E p in Rstu2}\\
	&\quad\Longleftrightarrow\quad \beta(p)\in\rmR_s^t(v).\label{E p in Rstu3}
\end{align}

Since $u\mathbin{\congC_s^t} v$ we have $|\rmR_s^t(u)|=|\rmR_s^t(v)|$ by Remark~\ref{R calR}.
Hence the restriction $\beta'$ of $\beta$ to the domain
$\rmR_s^t(u)\cup\{0,m-1\}$ is an
order-preserving bijection from $\rmR_s^t(u)\cup\{0,m-1\}$ to 
$\rmR_s^t(v)\cup\{0,n-1\}$.
From (\ref{E uiviX}) we immediately obtain $(u)_i=(v)_{\beta'(i)}$
for all $i\in\rmR_s^t(u)$.
From (\ref{E betaijX}) and Point (3) of Lemma~\ref{L equiv} we obtain
$j-i\equiv^d_s \beta'(j)-\beta'(i)$ for all $i,j\in\rmR_s^t(u)\cup\{0,m-1\}$ such that $i<j$.
This proves Point (2) of Definition~\ref{D ordered} for $\congC_s^{t,d}$.

\medskip

	\noindent
(3):\quad
Let $t\geq t'$, $d\geq d'$ and $u\mathbin{\congC_k^{t,d}} v$. We need to prove
$u\mathbin{\congC_k^{t',d'}} v$.
Since $u\mathbin{\congC_k^{t}} v$ and $\equiv^{t}$ is a refinement of $\equiv^{t'}$
by Point (2) of Lemma~\ref{L equiv} we have $u\mathbin{\congC_k^{t'}} v$.
This shows Point (1) of Definition~\ref{D ordered} for $\congC_k^{t',d'}$.

It remains to prove Point (2) of Definition~\ref{D ordered} for $\congC_k^{t',d'}$.
	By assumption there exists an order-preserving bijection 
	$\beta:\rmR_k^t(u)\cup\{0,m-1\}\to\rmR_k^t(v)\cup\{0,n-1\}$ 
	satisfying 
	\begin{align}
		(u)_{i}=(v)_{\beta(i)}&\quad\text{ for all $i\in\rmR_k^t(u)$ and}\label{E uiviY}\\
		j-i\equiv^d_k \beta(j)-\beta(i)&\quad\text{ for 
		all $i,j\in\rmR_k^t(u)\cup\{0,m-1\}$ such that $i<j$.}\label{E betaijY}
	\end{align}
	Analogously to Case (2) one proves that $\beta(p)\in\rmR_k^{t'}(v)$
	for all $p\in\rmR_k^{t'}(u)$
	and that $|\rmR_{k}^{t'}(u)|=|\rmR_{k}^{t'}(v)|$.
	Hence we define $\beta'$ to be the restriction of $\beta$ to the 
	domain $\rmR_k^{t'}(u)\cup\{0,m-1\}$,
	which is obviously an order-preserving bijection from 
$\rmR_{k}^{t'}(u)\cup\{0,m-1\}$ to $\rmR_{k}^{t'}(v)\cup\{0,n-1\}$.
Point (2) (a) of Definition~\ref{D ordered} for $\congC_k^{t',d'}$ is immediate.
	Since $\equiv_k^{d}$ is a refinement of $\equiv_k^{d'}$ by Point (2) 
	of Lemma~\ref{L equiv}
	and $\equiv_k^{d'}$ is a congruence by Point (1) of
	Lemma~\ref{L equiv}, by (\ref{E betaijY}) we have
			$j-i\equiv^{d'}_k \beta'(j)-\beta'(i)$ for 
		all $i,j\in\rmR_k^{t'}(u)\cup\{0,m-1\}$ such that $i<j$.
This shows Point (2) (b) of Definition~\ref{D ordered} for $\congC_k^{t',d'}$.
\end{appendixproof}

	Let $k\in\bbN$ and let $u\in\Sigma^*$ be a word with $|u|\geq k$.
	As in Definition 4.5 in~\cite{GM26arxiva} 
	by $\alpha_{k}(u)$ (\resp~$\omega_{k}(u)$) we denote
	the unique prefix (\resp~suffix) of $u$ of length $k$. 
In case $|u|>2k$ let $\kappa_{k}(u)$ denote the 
unique factor of $u$ such that $u = \alpha_{k}(u)\cdot \kappa_{k}(u) \cdot \omega_{k}(u)$.

\begin{samepage}
\begin{definition} 
\label{def:simeq}
	For $k,t\in\bbN_{>0}$ and $d\in\N$ 
	we define the binary relation $u\simeq_{k}^{t,d}v$ over
	$\Sigma^+$ as follows:
	\begin{itemize}
		\item either $\min(|u|,|v|) \leq 2k$ and $u=v$, or
     \item $|u|,|v| > 2k$ and
        \begin{enumerate}
		     \item $\alpha_k(u)=\alpha_k(v)$, 
		     \item $\omega_k(u)=\omega_k(v)$, 
		     \item $\kappa_k(u) \mathbin{\congC_k^{t,d}} \kappa_k(v)$.
        \end{enumerate}
	\end{itemize}
\end{definition}
\end{samepage}

\begin{remark}\label{R reverse}
	$\simeq_k^{t,d}$ is an equivalence relation. Moreover,  
	$u\simeq_k^{t,d} v$ if, and only if, $u^R\simeq_k^{t,d} v^R$.
\end{remark}

\begin{samepage}
\begin{lemmarep}
Let $k,s,t\in \bbN_{>0}$ and $d\in\bbN$.
\begin{enumerate}
	\item  $\simeq_{k}^{t,d}$ is a congruence.
	\item  $\simeq_k^{t,d}$ refines $\simeq_s^{t,d}$ whenever $s|k$.
\end{enumerate}
\label{L simeq}
\end{lemmarep}
\end{samepage}
\begin{appendixproof}
(1):\quad 
	To prove $\simeq_k^{t,d}$ is an equivalence relation, 
	it suffices to prove that $\simeq_k^{t,d}$ is an equivalence relation
	on $\Sigma^{> 2k}$. 
	On $\Sigma^{> 2k}$ we have that equality under $\alpha_k$,
	$\omega_k$ and $\congC_k^{t,d}$-equivalence under $\kappa_k$ are equivalence relations, 
	respectively.
	Hence their intersection defines an equivalence relation.

	If $\min(|u|,|v|)\leq 2k$, then $u=v$ and $ua=va$ and we are done.

	Let us argue that for all $u,v\in\Sigma^{>2k}$ and all
	$a\in\Sigma$ we have that $u\simeq_k^{t,d} v$ implies $ua\simeq_k^{t,d} va$.
		Since $\alpha_k(u)=\alpha_k(v)$, trivially we have
	$\alpha_k(ua)=\alpha_k(u)=\alpha_k(v)=\alpha_k(va)$.
	Since $\omega_k(u)=\omega_k(v)$ we have 
	$\omega_k(ua)=\omega_{k-1}(u)a=\omega_{k-1}(v)a=\omega_{k}(va)$.
Since $|u|\equiv_k |v|$, 
	it follows $|ua|\equiv_k |va|$.
	To finish the proof 
	observe that $\kappa_k(ua) =  \kappa_k(u) \cdot a'$ and 
	$\kappa_k(va) =  \kappa_k(v) \cdot a'$,
	where $a'=(\omega_{k}(u))_{0}=(\omega_{k}(v))_{0}$. 
	Since $\kappa_k(u) \congC_k^{t,d}  \kappa_k(v)$, 
	it follows by Point (1) of Lemma~\ref{L congC} 
	that $\kappa_k(ua) \congC_k^{t,d}  \kappa_k(va)$. 
	Thus $ua \simeq_k^{t,d} va$. By a symmetric argument 
	$au \simeq_k^{t,d} av$. Thus $\simeq_k^{t,d}$ is a congruence.\\
    \newline
	\noindent (2):\quad Assume that $s|k$ and 
	$u\simeq_k^{t,d} v$.
	Without loss of generality let us assume $|u|,|v|>2k$.
	Since $s|k$, equality under $\alpha_k$ implies
	equality under $\alpha_s$, equality under $\omega_k$
	implies equality under $\omega_s$. Finally, we have
    $\kappa_s(u) = 
		x\kappa_k(u)y$ and 
$\kappa_s(v) = 
		x\kappa_k(v)y$, where  $x=
\omega_{k-s}(\alpha_k(u))$ and $y=\alpha_{k-s}(\omega_k(u))$. 
	Since $\kappa_k(u)\mathbin{\congC_k^{t,d}} \kappa_k(v)$, we have 
	$\kappa_k(u)\mathbin{\congC_s^{t,d}} \kappa_k(v)$ 
	by Point (2) of Lemma~\ref{L congC}. 
	Also, $x\kappa_k(u)y\mathbin{\congC_s^{t,d}}
x\kappa_k(v)y$ since 
	$\congC_s^{t,d}$ is a congruence by Point (1) of Lemma~\ref{L congC}. 
	Thus $\kappa_s(u) \congC_s^{t,d} \kappa_s(v)$.
\end{appendixproof}

\begin{definition}[The Diagonal Congruence $\simeq_k$]\label{D diag}
	For all $k\in\bbN_{>0}$ we let $\simeq_k\mathbin{=}\simeq_{k}^{k,k}$.
\end{definition}

\subsection{The lm-Pseudovariety $\bfX$}\label{S X}

An \emph{identity} is an equation $t=t'$, where
$t$ and $t'$ are finite $\omega$-terms built from a set of variables
using multiplication and $\omega$-power. 
For example $xy=yx$, $x = x^\omega$, and $(xy^\omega z)^\omega=xy^\omega$ are all identities. 
A {\em semigroup $S$ satisfies} the identity $t=t'$ if the two sides evaluate to the same element under every assignment of the variables to elements of $S$.
For instance, the first two identities are satisfied by $S$ if, and only if,
it is commutative and every element of $S$ is idempotent. By convention, we use variables $e,f,\dots$ as shorthand for idempotent elements. 

Given a map $\sigma:V\to\Sigma^\ell$ and a stamp $\varphi:\Sigma^+\to S$
we define $\varphi(\sigma(t))$ for each finite $\omega$-term $t$
by induction as follows: $\varphi(\sigma(t))=\varphi(\sigma(x))$ in case $t=x$,
$\varphi(\sigma(t_1t_2))=\varphi(\sigma(t_1))\varphi(\sigma(t_2))$ and
finally $\varphi(\sigma(t^\omega))=\varphi(\sigma(t))^\omega$.
We sometimes implicitly extend $\sigma$ to a morphism from $V^+$ to $\Sigma^+$.
A {\em stamp $\varphi:\Sigma^+\to S$ satisfies} the identity $t=t'$ over
the set of variables $V$ if
$\varphi(\sigma(t))=\varphi(\sigma(t'))$ for every $\ell\in\bbN_{>0}$ and every
morphism defined by a map $\sigma:V\to\Sigma^\ell$.

In this section we will introduce our lm-pseudovariety of stamps $\bfX$.
Its name is motivated by e{\textbf{X}}changing a frequent 
letter (resp. infix) for another frequent letter (resp. infix).

\begin{samepage}
	\begin{definition}[The lm-pseudovariety of stamps $\bfX$]\label{D X}
	Let $\{p_i,x_i,y_i\mid i\in\N\} \subseteq V$ be a set of variables.
		For each $k\in\bbN_{>0}$ let $\bfX_k$ denote the 
		lm-pseudovariety of stamps satisfying the following identities,
		where $m\in[1,k]$ and $i\in[0,m-1]$:
\begin{align}
&(x_0 \cdots x_{m-1})^\omega
  p_0 \cdots p_{i-1} x_i p_{i+1} \cdots p_{m-1}
  (y_0 \cdots y_{m-1})^\omega \notag\\
={}&(x_0 \cdots x_{m-1})^\omega
  p_0 \cdots p_{i-1} y_i p_{i+1} \cdots p_{m-1}
  (y_0 \cdots y_{m-1})^\omega
	\label{eq:Yeq3}
\end{align}
Let $\bfX = \bigcap_{k \geq 1}\bfX_k$.
\end{definition}
\end{samepage}

We remark that $\bfX_1\supseteq\bfX_2\supseteq\bfX_3\supseteq\dots$.
     
Pin and Straubing proved in (\cite{StraubingPin10}, Theorem 2.11) that a class 
of stamps on monoids forms an lm-pseudovariety if, and only if, 
it is defined by a set of profinite identities. An analogous statement holds 
for stamps on semigroups. Since $\bfX$ is defined by a set of identities, it is closed under diagonal direct products and lm-divisions. Hence, $\bfX$ is an lm-pseudovariety of semigroup stamps. 
We do not make use of this property in the rest of the paper.

Next we will give two seemingly similar looking regular languages
with different circuit complexities.

\begin{samepage}
	\begin{example}\label{E in X}
		Consider the language $L_1= (a+b)^*c (a+b) c (a+b)^*$
		over the alphabet $\Sigma=\{a,b,c\}$.
		Note that $\eta_{L_1}(a)=\eta_{L_1}(b)$ and that
		$S(L_1)$ contains only two idempotents,
		namely $\eta_{L_1}(aa)$ and $\eta_{L_1}(cc)=0$;
		i.e. $0\cdot x=x\cdot 0=0$ for all $x\in S(L_1)$.
		We claim that $\eta_{L_1}$ is in $\bfX$, or 
		equivalently that $\eta_{L_1}$ satisfies 
		all identities of the form (\ref{eq:Yeq3}).
		Recall Definition~\ref{D X} and consider
		the corresponding identity (\ref{eq:Yeq3})
		for some $k\in\bbN_{>0}$, $m\in[1,k]$, and $i\in[0,m-1]$.
		Moreover fix any map 
		$\sigma:\{x_j,p_j,y_j\mid j\in[0,k-1]\}\to\Sigma^\ell$,
		where $\ell\in\bbN_{>0}$.
		One can distinguish two
		cases: either at least one of the words 
		$\sigma(x_0\dots x_{m-1})$ and
		$\sigma(y_0\dots y_{m-1})$
		contains the letter $c$, or not.
		In the former case, when say $\sigma(y_0\dots y_{m-1})$ 
		contains a $c$, we have
		that $\eta_{L_1}$ evaluates both sides of identity~(\ref{eq:Yeq3})
		to $0$.
In the latter case, $\sigma(x_i)$ and $\sigma(y_i)$ belong to $\{a,b\}^\ell$. Since $\eta_{L_1}(a)=\eta_{L_1}(b)$, their syntactic images coincide. Hence the two middle factors have the same image under $\eta_{L_1}$.
		Hence both the left-hand side
		and the right-hand side of (\ref{eq:Yeq3}) 
		are evaluated to the same element.
		\end{example}
	\end{samepage}

	\begin{example}\label{E not in X}
	Consider the language $L_2= (a+b)^*c a c (a+b)^*$
		over the alphabet $\Sigma=\{a,b,c\}$.
		We claim that $\eta_{L_2}$ is not in $\bfX_3$ and hence not in $\bfX$.
		It is worth noting that $S(L_2)$ contains only two idempotents,
		namely $\eta_{L_2}(b)$ and $\eta_{L_2}(cc)$. 
		We remark that $\eta_{L_2}(cc)=0$ is the
		zero of $S(L_2)$.
		Also note that $\eta_{L_2}(a)$ is not an idempotent, in particular
		$\eta_{L_2}(a)\not=\eta_{L_2}(b)$, yet 
		$\eta_{L_2}(a^i)=\eta_{L_2}(b^j)$
		for all $i\geq 2$ and all $j\geq 1$.
		We claim that $\eta_{L_2}$ does not satisfy the following identity
		of $\bfX_3$:
\begin{align}
(x_0x_1x_2)^\omega p_0x_1p_2 (y_0y_1y_2)^\omega&=
	(x_0x_1x_2)^\omega p_0y_1p_2 (y_0y_1y_2)^\omega\label{E X3 id}
		\end{align}
		To show this, let $V=\{p_j,x_j,y_j\mid j\in[0,2]\}$ and
		consider the map $\sigma:V\to\Sigma$,
		where $\sigma(x_0)=\sigma(x_1)=\sigma(x_2)=a$,
		$\sigma(y_0)=\sigma(y_1)=\sigma(y_2)=b$ and 
		$\sigma(p_0)=\sigma(p_1)=\sigma(p_2)=c$.
		Hence the left-hand side of the identity (\ref{E X3 id})
		is evaluated to $\eta_{L_2}(aaa cac bbb)\in\eta_{L_2}(L_2)$
		and the right-hand side 
		to $\eta_{L_2}(aaa cbc bbb)\not\in\eta_{L_2}(L_2)$.
		Thus $\eta_{L_2}$ is not in $\bfX_3$
		and therefore not in $\bfX$.
	\end{example}

\begin{proposition}\label{P isomorphic}
	Consider the languages $L_1$ and $L_2$ from Example~\ref{E in X}
	and Example~\ref{E not in X} respectively.
Then
$\operatorname{stab}(\eta_{L_1}) \cong
\operatorname{stab}(\eta_{L_2})$.
\end{proposition}
\begin{proof}
	We employed the software Semigroup Explorer~\cite{khoulani2026semigroups}.
	The stability index of both $\eta_{L_1}$ and $\eta_{L_2}$ is four.
	The stable semigroup of $\eta_{L_1}$ is given by the 
	multiplication table in Figure \ref{fig:stable-semigroup-isomorphism}.

	\begin{figure}
  \centering
  \small
  \setlength{\arraycolsep}{3pt}
  \renewcommand{\arraystretch}{1.15}
  \[
  \begin{array}{
    c|ccccccc
    @{\qquad}
    c|ccccccc
    @{\qquad}
    c@{\;\mapsto\;}c
  }
    \multicolumn{8}{c}{\operatorname{stab}(\eta_{L_1})}
    &
    \multicolumn{8}{c}{\operatorname{stab}(\eta_{L_2})}
    &
    \multicolumn{2}{c}{\varphi}
    \\[4pt]
    \cdot & aa   & aac & aaca & acaa & caa & cac & cc
    &
    \cdot & b   & bc & bca & acb & cb  & cac & cc
    &
    \multicolumn{2}{c}{}
	  \\ \cmidrule(r{1em}){1-8}\cmidrule(l{1em}){9-16}
    aa    & aa   & aac & aaca & cc   & cc  & cac & cc
    &
    b     & b   & bc & bca & cc  & cc  & cac & cc
    &
    aa & b
    \\
    aac   & cc   & cc  & cc   & cac  & cc  & cc  & cc
    &
    bc    & cc  & cc & cc  & cac & cc  & cc  & cc
    &
    aac & bc
    \\
    aaca  & cc   & cc  & cc   & cc   & cac & cc  & cc
    &
    bca   & cc  & cc & cc  & cc  & cac & cc  & cc
    &
    aaca & bca
    \\
    acaa  & acaa & cc  & cc   & cc   & cc  & cc  & cc
    &
    acb   & acb & cc & cc  & cc  & cc  & cc  & cc
    &
    acaa & acb
    \\
    caa   & caa  & cc  & cc   & cc   & cc  & cc  & cc
    &
    cb    & cb  & cc & cc  & cc  & cc  & cc  & cc
    &
    caa & cb
    \\
    cac   & cac  & cc  & cc   & cc   & cc  & cc  & cc
    &
    cac   & cac & cc & cc  & cc  & cc  & cc  & cc
    &
    cac & cac
    \\
    cc    & cc   & cc  & cc   & cc   & cc  & cc  & cc
    &
    cc    & cc  & cc & cc  & cc  & cc  & cc  & cc
    &
    cc & cc
  \end{array}
  \]
  \caption{The stable semigroups of $\eta_{L_1}$ and $\eta_{L_2}$
    and an isomorphism $\varphi$ between them.
    Corresponding rows and columns appear in matching order.}
  \label{fig:stable-semigroup-isomorphism}
\end{figure}
\end{proof}

By Points (2) and (4) of Theorem~\ref{T log main} the circuit complexity
of the regular language $L_1=(a+b)^*c (a+b) c (a+b)^*$ is $O(\log n\log\log n)$
(its syntactic morphism is in $\bfX$ as argued in Example~\ref{E in X}), 
whereas by Points (3) and (4) of Theorem~\ref{T log main} 
the circuit complexity of the
language $L_2=(a+b)^*c a c (a+b)^*$ is $\OmegaHL(n)$ (its
syntactic morphism is not in $\bfX$ as argued in Example~\ref{E not in X}).

\begin{samepage}
\begin{lemma}[Exchange Principle]
\label{lem:exchange-principle}
Let $\varphi: \Sigma^+ \to S$ be a morphism in $\bfX$ with stability index 
	$s\in\bbN_{>0}$. 
	Let $u,v,w\in \Sigma^{s\bbN_{>0}}$ be words and in particular let 
	$u$ and $w$ be idempotents. 
	Assume that $u=u_0\ldots u_{|u|-1}$, $v=v_0\ldots v_{|v|-1}$ and 
	$w=w_0\ldots w_{|w|-1}$. 
Then for all $i \in [0,|u|-1]$, $j \in [0,|v|-1]$, $k \in [0,|w|-1]$ 
	such that $i\equiv_s j \equiv_s k$, we have
\begin{align}
	u  v[j\leftarrow u_i] w =_{\varphi} 	u  v[j\leftarrow w_k]w.
\end{align}
\end{lemma}  
\end{samepage}
\begin{proof}
	Assume that $u,v,w\in \Sigma^{s\bbN_{>0}}$ and 
	$i \in [0,|u|-1], j \in [0,|v|-1]$, $k \in [0,|w|-1]$. 
	Note that for each $y\in \Sigma^{s\bbN_{>0}}$
	we have 
	$\varphi(y) \in \stab(\varphi)=\varphi(\Sigma^s)$. 
	Thus for each $y\in \Sigma^{s\bbN_{>0}}$, let $\widehat{y}\in \Sigma^s$ denote an arbitrarily chosen word whose image is $\varphi(y)$.

Let $\ell_x=|x|/s$ for each $x\in \{u,v,w\}$. 
Let $r\in [0,s-1]$ be such that $r\equiv_s i \equiv_s j \equiv_s k$. 
For each $x\in \{i,j,k\}$ define $q_x=\lfloor x/s\rfloor$, i.e.:
\begin{align}
i &= q_is+r\\
j &= q_js+r\\
k &= q_ks+r
\end{align}
Let $\ell=\ell_u+\ell_v+\ell_w$ and $q=q_i+q_j+q_k$.   
	Define $u',v',w'\in\Sigma^{s\bbN_{>0}}$ as follows:
\begin{align}
u'&=\widehat{u}^{(q-q_i)}u\widehat{u}^{(\ell-\ell_u+q_i)}\\
v'&=\widehat{u}^{(q-q_j)}v\widehat{w}^{(\ell-\ell_v+q_j)}\\
w'&=\widehat{w}^{(q-q_k)}w\widehat{w}^{(\ell-\ell_w+q_k)}
\end{align}
It is easy to verify that $|u'|=|v'|=|w'|=(q_i+q_j+q_k)s+(\ell_u+\ell_v+\ell_w)s=
	(q+\ell)s$.
	Let $n=(q_i+q_j+q_k)s+r$.
Note that $n=i+(q-q_i)s=j+(q-q_j)s=k+(q-q_k)s$. 
	Thus, $(u')_n=u_i$, $(v')_n=v_j$, and $(w')_n=w_k$.
		Note that, if $x,y,z\in \Sigma^+$ are such that $|x|=|y|=|z|$ and $x,z$ are idempotent, then for all $i\in[0,|x|-1]$ and exponents $m$ of $S$ we have
\begin{align}
	xy[i\leftarrow (x)_i]z &=_\varphi x^m \cdot y[i\leftarrow(x)_i]z^{m-1}\cdot z^m\\
&=_\varphi x^m \cdot y[i\leftarrow (z)_i] z^{m-1}\cdot z^m &&(\text{Applying the identity of $\bfX$})\\
&=_\varphi xy[i\leftarrow (z)_i]z.\label{eq:exchange-X}
\end{align}

Bearing in mind that $u$ and $w$ are idempotents
we have:
\begin{align}
uv[j\leftarrow u_i]w &=_\varphi u' \widehat{u}^{(q-q_j)} v[j\leftarrow u_i] \widehat{w}^{(\ell-\ell_v+q_j)} w' &&(\text{As $u'=_\varphi \widehat{u}=_\varphi u$, $w'=_\varphi \widehat{w}=_\varphi w$})\\
&=_\varphi u'v'[n\leftarrow u_i]w'&&(\text{Since $(v')_n=v_j$})\\
&=_\varphi u'v'[n\leftarrow (u')_n]w' && (\text{Since $(u')_n=u_i$}) \\
	&=_\varphi u'v'[n\leftarrow(w')_n]w' &&(\text{By (\ref{eq:exchange-X})})\\
&=_\varphi u'v'[n\leftarrow w_k]w'&& (\text{Since $(w')_n=w_k$}) \\
&=_\varphi uv[j\leftarrow w_k]w.&&(\text{As $u'=_\varphi \widehat{u}=_\varphi u$, $w'=_\varphi \widehat{w}=_\varphi w$})
\end{align}
\end{proof}

\subsection{Lindstr\"om One-Variable Fragment} \label{S LFO} \label{S Lindstrom} 
Let $k\in \bbN_{>0}$. A \emph{$k$-ary relational Lindstr\"om quantifier} $Q$ on words is defined by its denotation $Q^{w,I} \subseteq 2^{[0,n-1]^k}$ on each word $w$ with length $n$ and an interpretation $I$ of variables. In other words $Q$ can be viewed as a mapping from each pair $w,I$ to a set of $k$-ary relations over $[0,|w|-1]$.
A quantification by $Q$ of $k$ formulas $\phi_0(x_0),\ldots, \phi_{k-1}(x_{k-1})$ with distinct free variables $x_0,\ldots, x_{k-1}$ respectively is denoted as
\begin{align}
    \varphi &= Qx_0\ldots x_{k-1}\, [\phi_0(x_0),\ldots, \phi_{k-1}(x_{k-1})].
\end{align}
On a pair $w,I$ the satisfaction of the formula $\varphi$ is defined as
\begin{align}
 w,I &\models \varphi  \quad \mathrm{ iff }\quad \prod_{i=0}^{k-1} \phi_i^{w,I} \in Q^{w,I}.
\end{align}

Let $t \in \bbN_{>0}$. Observe that $\exists^{<t}x\, \varphi(x) \equiv Q_{\exists^{< t}} x \,[\varphi(x)]$ where the unary quantifier $Q_{\exists^{< t}}$ has the denotation 
\begin{align}
    Q_{\exists^{< t}}^{w, I} &= \{ X \in 2^{[0,|w|-1]} : |X| < t \}
\end{align}
on the pair $w,I$. Since $\forall x\, \varphi(x) \equiv \exists^{<1}x\, \neg \varphi(x)$
and $\exists x\, \varphi(x) \equiv \neg \forall x\, \neg \varphi(x)$ the quantifiers $\forall$ and $\exists$ are also expressed by Lindstr\"om quantifiers using negation.

Next we introduce the Lindstr\"om one-variable fragment.
\begin{definition}[Lindstr\"om One-Variable Fragment]
\label{def:lind-one}
Let $\calQ$ be a set of Lindstr\"om quantifiers. The first-order logic over the vocabulary $\tau$ with the Lindstr\"om quantifiers in $\calQ$ is denoted by $\calL[\calQ,\tau]$. The \emph{one-variable} fragment of $\calL[\calQ,\tau]$, denoted by $\calL^{1}[\calQ,\tau]$, is the smallest set of formulas satisfying the following conditions:
\begin{itemize}
    \item Every atomic formula over $\tau$ using at most one variable belongs to $\calL^{1}[\calQ,\tau]$.
    \item Let $\{x_0,\ldots, x_{k-1}\}$ be a set of  pairwise distinct variables and let $\phi_0,\ldots, \phi_{k-1}\in\calL^{\mathrm{1}}[\mathcal Q,\tau]$ be such that the set of free variables of  $\phi_i$ is a subset of $\{x_i\}$ for $i \in [0,k-1]$. If $Q\in \calQ$ is a $k$-ary quantifier then $Qx_0\ldots x_{k-1}
[\phi_0,\ldots,\phi_{k-1}] \in 
\calL^{1}[\mathcal Q,\tau]$.
\item If $\phi, \psi$ are in $\calL^{1}[\calQ,\tau]$ and the resulting formula has at most one free variable, then $\neg \phi$, $\phi \vee \psi$ and $\phi \wedge \psi$ are in $\calL^{1}[\calQ,\tau]$. 
\end{itemize}
\end{definition}

Thus every formula in $\calL^{1}[\mathcal Q,\tau]$ has at most one free variable, while every formula obtained by applying a numerical or Lindstr\"om quantifier is a sentence. Therefore we limit the denotation $Q^{w,I}$ of a quantifier $Q$ to depend only on the word $w$, not on the interpretation $I$. Henceforth we drop $I$ from the denotation of quantifiers used in the one-variable fragment.

\begin{remark}
Let  
$\varphi = Qx_0\ldots x_{k-1}\, [\phi_0(x_0),\ldots, \phi_{k-1}(x_{k-1})]$
be a formula in $\calL^{1}[\mathcal Q,\tau]$. Assume that $\psi$ is a  subformula of $\varphi$ with no free variable. The formula $\varphi$ is equivalent to the formula $(\psi \wedge \varphi[\psi/\top]) \vee (\neg \psi \wedge \varphi[\psi/\bot])$ where $\varphi[\psi/\top]$ (\resp~$\varphi[\psi/\bot]$) is obtained by replacing every occurrence of $\psi$ in $\varphi$ by $\top$ (\resp~$\bot$). Iterating this decomposition expresses every sentence as a Boolean combination of atomic sentences and formulas $Qx_0\ldots x_{k-1}\, [\phi_0(x_0),\ldots, \phi_{k-1}(x_{k-1})]$ whose arguments are quantifier-free and contain no nonconstant sentence as a subformula.
\label{remark:sentence-free}
\end{remark}

A Lindstr\"om quantifier $Q$ is \emph{numerical} if its denotation $Q^w$  on a word $w$ depends only on the length of the word $w$. For each $n\in\bbN_{>0}$, we write $Q^n$ for the denotation of $Q$ on words of length $n$. All Lindstr\"om quantifiers we consider henceforth are numerical.

\begin{definition}[Bounded Quantifiers $\calB$]
\label{def:bounded-q}
    Let $k,t\in \bbN_{>0}$. A $k$-ary numerical Lindstr\"om quantifier $Q$ over words is \emph{$t$-bounded} if for every length $n \in \bbN_{>0}$ and every relation $R\in Q^{n}$, there exist subsets $P_0^R,\ldots, P_{k-1}^R \subseteq [0,n-1]$, satisfying $R = P_0^R \times \ldots \times P_{k-1}^R$ and $|P_i^R|< t$ for each $i \in [0,k-1]$. The bounded numerical Lindstr\"om quantifiers are denoted by $\calB$.  
    
    Let $\tau$ be a vocabulary of numerical predicates. The $t$-bounded numerical quantifier $Q$ is $\tau$-definable if there exists a \emph{quantifier-free formula} $\varphi(x_0,\ldots, x_{k-1})$ over the vocabulary $\tau$ 
    such that 
\begin{align}
    Q^{n} &= \left\{ \prod_{i=0}^{k-1} P_i : P_i \subseteq [0,n-1], |P_i| < t \text{ for all $i\in [0,k-1]$}, \prod_{i=0}^{k-1} P_i \subseteq \varphi^{n} \right\}.
\end{align}

The class of bounded numerical Lindstr\"om quantifiers that are $\tau$-definable is denoted by $\calB_\tau$.
\end{definition}

\begin{example}Below we define some examples of Lindstr\"om quantifiers. 

\smallskip
\noindent 1.\qquad The quantifier $Q_{\exists^{<t}}$, where $t \in \bbN_{>0}$, is numerical and $t$-bounded  and is definable in the empty vocabulary by the formula $\top$. Consequently, the quantifiers $\{\exists, \forall, \exists^{<t}, \exists^{=t}, \exists^{\geq t}\}$ are expressible using Boolean combinations of numerical bounded quantifiers definable in the empty vocabulary. We use this fact
implicitly in the remainder of this section.

\smallskip
\noindent 2.\qquad The language $L_2=c^*ac^*bc^* \subseteq \{a,b,c\}^+$ is defined by the formula
    \begin{align}
   \varphi_2 &= \exists x P_a(x) \wedge \exists x P_b(x) \wedge Q_{2}xy [P_a(x), P_b(y)]
    \end{align}
    where the quantifier $Q_2$ has the denotation 
    \begin{align}
        Q_2^n&= \{\emptyset \} \cup \{\{i\} \times \{j\} : i,j \in [0,n-1], i<j\}
    \end{align} for each length $n\in \bbN_{>0}$. The formula $\varphi_2$ is 
true on a word $w$ if and only if $P_a^w \times P_b^w =\{(i,j) \}$ for some $(i,j) \in [0,n-1]^2, i<j$. Therefore $L(\varphi_2)=L_2$. The quantifier $Q_2$ is $2$-bounded and it is defined by the formula $x<y$. Hence $Q_2\in \calB_{[<]}$ and $L_2 \in \calL^1[\calB_{[<]},\Sigma]$.

\smallskip
\noindent 3.\qquad The language $L_3=c^*ac^*bc^*ac^* \subseteq \{a,b,c\}^+$ is defined by the formula
    \begin{align}
 \varphi_3 &=\exists^{=2} x P_a(x) \wedge \exists^{=1} x P_b(x) \wedge Q_{3}xyz [P_a(x), P_b(y), P_a(z)]
    \end{align}
      where for each length $n\in \bbN_{>0}$ the quantifier $Q_3$ has the denotation 
    \begin{align}
          Q_3^n&=\left\{\prod_{i=0}^2 P_i : P_i \subseteq [0,n-1], |P_i| < 3, \text{ for each $i\in [0,2]$}, \prod_{i=0}^2 P_i \subseteq \chi^n \right\}
        \end{align} 
where $\chi(x,y,z)$ is the formula
    \begin{align}
           \chi(x,y,z) &= x<y<z \vee z<y<x \vee x=z
      \end{align} 
The formula $\varphi_3$ is 
true on a word $w$ if and only if $P_a^w \times P_b^w \times P_a^w  =\{i,k\} \times \{j\} \times \{i,k\}$ for some $i,j,k \in [0,n-1], i<j<k$. Hence $L(\varphi_3)=L_3$.   The quantifier $Q_3$ is $3$-bounded and is $[<]$-definable. Therefore  $L_3 \in \calL^1[\calB_{[<]},\Sigma]$.

\smallskip
\noindent 4.\qquad The language $L_4=a^+b^+ \subseteq \{a,b\}^+$ is defined by the formula
 \begin{align}
 \varphi_4 &=Q_{4}xy [P_a(x), P_b(y)]
    \end{align}
    where the quantifier $Q_4$ has the denotation
    \begin{align}
          Q_4^n&=\{ [0,\ell] \times [\ell+1, n-1]   \mid \ell \in [0,n-2]\}
      \end{align} 
      for each length $n\in \bbN_{>0}$. Clearly $\varphi_4$ is true if and only if $P_a^w \times P_b^w$ is of the form $[0,\ell] \times [\ell+1, n-1]$ for some $\ell \in [0,n-2]$, and hence $L(\varphi_4)=L_4$. 
    It is easy to see that the quantifier $Q_4$ is not bounded.

\smallskip
\noindent 5.\qquad The language $L_5=(a+b)^*c(a+b)c(a+b)^* \subseteq \{a,b,c\}^+$ is defined by the formula
 \begin{align}
 \varphi_5 &=\exists^{=2}x\,P_c(x) \wedge Q_{5}xy [P_c(x), P_c(y)]
    \end{align}
    where $Q_5$ is the $3$-bounded quantifier   defined by the formula $\chi=x=y \vee x=y+2 \vee y=x+2$. That is to say,
    \begin{align}
          Q_5^n&=  \left\{ P_1 \times P_2   \mid P_1, P_2 \subseteq [0,n-1], |P_1|,|P_2| < 3, (x,y) \models \chi \text{ for all $(x,y)\in P_1\times P_2$ }\right\}
      \end{align} 
      for each length $n\in \bbN_{>0}$. 
      Note that $\emptyset \in Q_5^n$ and
\begin{align}
     \{\{\ell,  \ell+2\} \times \{  \ell,\ell+2\}   \mid \ell \in [0,n-3]\} \subseteq Q_5^n.
\end{align}
    
      If $w$ satisfies $\varphi_5$ then $P_c^w =\{ \ell, \ell+2\}$ for some $\ell \in [0,|w|-3]$ and hence $L(\varphi_5)=L_5$. Therefore  $L_5 \in \calL^1[\calB_{[+2]},\Sigma]$.

\smallskip
\noindent 6.\qquad The language $L_6=(a+b)^*cac(a+b)^* \subseteq \{a,b,c\}^+$ is defined by the formula

\begin{align}
 \varphi_6 &=Q_{6}xyz [P_c(x), P_a(y), P_c(z)]
    \end{align}
    where the quantifier $Q_6$ has the denotation
    \begin{align}
          Q_6^n&=\{ \{\ell,\ell+2\} \times (P \cup \{\ell+1\})  \times \{\ell,\ell+2\}  : \ell \in [0,n-3], P \subseteq [0,n-1]\setminus\{\ell,\ell+2\}\}
      \end{align} 
      for each length $n\in \bbN_{>0}$. Clearly $w\in L(\varphi_6)$ if and only if $P_c^w \times P_a^w \times P_c^w = \{\ell,\ell+2\} \times (P \cup \{\ell+1\})  \times \{\ell,\ell+2\}$ for some $\ell \in [0,n-3], P \subseteq [0,n-1]\setminus\{\ell,\ell+2\}$ if and only if $w \in L_6$. Clearly the quantifier $Q_6$ is not bounded.
\end{example} 

\begin{definition}For $x\in[0,2^n-1]$, let $\mathrm{bin}_n(x)$ denote the
$n$-bit binary representation of $x$, padded with leading zeros.

A $k$-ary numerical predicate $R\subseteq\bbN^k$ is
\emph{computable by a circuit family of size $O(f(n))$ and
depth $O(g(n))$} if there exists a family of circuits
$(C_n)_{n\in\bbN_{>0}}$, where $C_n:(\{0,1\}^n)^k\longrightarrow\{0,1\}$, such that $|C_n|=O(f(n))$, $\depth(C_n)=O(g(n))$, and,
for every $n\in\bbN_{>0}$ and every
$(v_1,\ldots,v_k)\in[0,2^n-1]^k$,
\begin{align}
(v_1,\ldots,v_k)\in R
\quad\Longleftrightarrow\quad
C_n\bigl(\mathrm{bin}_n(v_1),\ldots,
\mathrm{bin}_n(v_k)\bigr)=1.
\end{align}
We denote by $\mathrm{num}[O(f),O(g)]$ the class of all
numerical predicates computable by circuit families of
size $O(f(n))$ and depth $O(g(n))$.
\end{definition}

\subsection{Disjoint and Modular Disjoint Monomials}\label{S Disjoint}
We sometimes identify a letter $a\in\Sigma$ with the singleton
language $\{a\}$, as is customary in regular expressions, whenever
no confusion can arise.

A {\em disjoint monomial}~\cite{mnil} over the alphabet $\Sigma$ 
is a language of the form
$\Gamma^*a_0\Gamma^*\dots a_{n-1}\Gamma^*$, where
$a_0,\dots, a_{n-1}\in\Sigma\setminus\Gamma$ and $\Gamma\subseteq\Sigma$.

We define the modular variant of disjoint monomials.
\begin{samepage}
\begin{definition}[Modular Disjoint Monomials]
	Let $k \in \bbN_{>0}$. A language over the alphabet $\Sigma$ is a {\em $k$-modular disjoint monomial} if
it is of the form
	\begin{align}
			 (\Gamma_0\dots\Gamma_{k-1})^*U_0  (\Gamma_0\dots\Gamma_{k-1})^*\dots	U_{n-1}
				(\Gamma_0\dots\Gamma_{k-1})^*
				,
	\end{align} where $n\in\bbN$,
			$\Gamma_0,\dots,\Gamma_{k-1}\subseteq\Sigma$ and  for each $i\in [0,n-1]$ the set of words $U_i$ is a product $U_i=U_{i,0}\ldots U_{i,q_ik-1}$, where $q_i\in \bbN_{>0}$ and $U_{i,j} = \Gamma_{j \bmod k} \text{ or } U_{i,j} \subseteq \Sigma \setminus \Gamma_{j \bmod k}$ for each $j\in [0,q_ik-1]$.
            
	    \label{def:mod-disjoint}
\end{definition}
\end{samepage}

\begin{proposition}
\label{lem:dis-mono-logic}
    For each disjoint monomial $L$ there exists a formula $ \varphi \in \calL^1[\calB_{[<]}, \Sigma]$ such that $L(\varphi)=L \cap \Sigma^+$.
\end{proposition}
\begin{proof}
Let $L$ be the disjoint monomial $\Gamma^*a_0\dots\Gamma^*a_{n-1}\Gamma^*$, where
$a_0,\dots, a_{n-1}\in\Sigma\setminus\Gamma$ and $\Gamma\subseteq\Sigma$. If  $n=0$, then $L=\Gamma^*$ and $L\cap \Sigma^+=\Gamma^+$ is defined by the formula $\forall x \vee_{a\in \Gamma} P_a(x) \in\calL^1[\calB_{[<]}, \Sigma]$. 

Otherwise, let $u=a_0\ldots a_{n-1} \in \Sigma^+$ and let $t=1+\max \{|u|_a : a\in \cont(u)\}$. Let $G_u$ be the set of all permutations $\pi:[0,n-1]\to [0,n-1]$ on the positions of $u$ that leave $u$ intact, i.e., $u_{\pi(0)}\ldots u_{\pi(n-1)} = u$. Let $\chi(x_0,\ldots, x_{n-1})$ be the formula
\begin{align}
    \chi(x_0,\ldots, x_{n-1}) \defeq \bigvee_{i,j \in [0,n-1] \atop i\neq j} x_i = x_j \vee \bigvee_{\pi\in G_u} x_{\pi(0)} < x_{\pi(1)} \cdots < x_{\pi(n-1)}.
\end{align}
where $a<b<c$ is short for $a<b \wedge b <c$. Let $Q$ be the $t$-bounded quantifier defined by $\chi$. 

Let $\varphi_{\cont}$ be the formula
\begin{align}
    \varphi_{\cont} &=  \forall x \bigvee_{a \in \cont(u) \cup \Gamma} P_a(x) \wedge  \bigwedge_{a\in \cont(u)} \exists^{=|u|_a}x P_a(x).
\end{align}
We claim that the formula $\varphi$ where 
\begin{align}
\varphi = \varphi_{\cont} \wedge Qx_0\ldots x_{n-1}[P_{a_0}(x_0),\ldots, P_{a_{n-1}}(x_{n-1})] 
\end{align}
defines the language $L$. 

We show that $L(\varphi)=L$. First, suppose that $w\in L$. Then
$w=v_0a_0v_1a_1\cdots v_{n-1}a_{n-1}v_n$, where $v_i\in\Gamma^*$ for every $i\in[0,n]$. The occurrences of letters from $\Sigma\setminus\Gamma$ in $w$ are precisely the occurrences of $(a_0,\ldots,a_{n-1})$ in $w$. Consequently, $w$ satisfies $\varphi_{\mathrm{cont}}$.

Let $i_j\in[0,|w|-1]$ be the position of the occurrence of $a_j$ in $w$, for each $j\in[0,n-1]$. Then
\begin{align}
i_0<i_1<\cdots<i_{n-1}.
\end{align}

Moreover, since each $a\in\cont(u)$ occurs exactly $|u|_a$ times in $w$, we have
$P_{a_j}^w=\{i_\ell\mid \ell \in [0,n-1], a_\ell=a_j\}$
for every $j\in[0,n-1]$. Every tuple $(v_0,\ldots, v_{n-1}) \in P_{a_0}^w\times\cdots\times P_{a_{n-1}}^w$ is either of the form $(i_{\pi(0)},\ldots,i_{\pi(n-1)})$, where $\pi\in G_u$, or 
there exists $\ell, \ell' \in [0,n-1]$ such that $\ell\neq \ell'$,
$a_\ell = a_{\ell'}$ and $v_\ell = v_{\ell'}$.
By the definition of $\chi$, every such tuple satisfies $\chi$. Hence
$P_{a_0}^w\times\cdots\times P_{a_{n-1}}^w
\subseteq\chi^{|w|}$. Therefore $w\models\varphi$.

Conversely, suppose that $w\models\varphi$. By $\varphi_{\mathrm{cont}}$,
every letter occurring in $w$ belongs to $\cont(u)\cup\Gamma$, and
each $a\in\cont(u)$ occurs exactly $|u|_a$ times in $w$. Since
$\cont(u)\cap\Gamma=\emptyset$, the occurrences of letters from
$\Sigma\setminus\Gamma$ in $w$, read from left to right, form a word
$u'$ such that
$\cont(u')=\cont(u)$ and $|u'|_a=|u|_a$ for every
$a\in\cont(u)$. It suffices to prove that $u=u'$. 
Since $w\models\varphi$, we have
$P_{a_0}^w\times\cdots\times P_{a_{n-1}}^w
\subseteq\chi^{|w|}$. For each $i\in[0,n-1]$, let $v_i$ be the position in $w$ of the
$(r_i+1)$-st occurrence of $a_i$ among the letters of $u'$, where
\begin{align}
r_i=|\{j<i:a_j=a_i\}|.
\end{align}
This occurrence exists because $|u'|_{a_i}=|u|_{a_i}$. By
construction, the positions $v_0,\ldots,v_{n-1}$ are pairwise
distinct, and $(v_0,\ldots,v_{n-1})\in\chi^{|w|}$.

Since the coordinates are pairwise distinct, the first disjunct of
$\chi$ is false. Hence, the tuple satisfies an ordering disjunct
indexed by some permutation $\pi\in G_u$. Consequently, the letters
of $u'$ occur in the order $a_{\pi(0)},\ldots,a_{\pi(n-1)}$.
Since every permutation in $G_u$ preserves $u$, it follows that
$u'=u$. Thus, the letters of $w$ outside $\Gamma$, read in their original
order, form $u$, and all remaining letters belong to $\Gamma$. Therefore, $w\in L$.

We conclude that $L(\varphi)=L$. Since $Q$ is $t$-bounded and is definable in $[<]$, it follows that
$L\in\mathcal L^1[\mathcal B_{[<]},\Sigma].$
\end{proof}

\begin{proposition}
\label{lem:mod-dis-mono-logic}
    For every $k$-modular disjoint monomial $L$, there exists a formula $ \varphi \in \calL^1[\calB_{[<,+\omega]}, \Sigma, <, +\omega, \mathrm{mod},\min,\max]$ such that  $L(\varphi)=L\cap \Sigma^+$.
\end{proposition}
\begin{proof}
    Let $k\in \bbN_{>0}$ and $B= \Gamma_0\dots\Gamma_{k-1}$ where $\Gamma_i \subseteq \Sigma$ for each $i\in [0,k-1]$.
    If $L$ is of the form $B^*$ then it is definable by a formula in $\rmFO^1[\Sigma,<,+\omega,\mathrm{mod},\min,\max]$. Otherwise, let  \begin{align}
        L = B^*U_0 B^*\dots U_{n-1}B^*
    \end{align}
    be a $k$-modular disjoint monomial,
	where $n\in \bbN_{>0}$ and
    \begin{align}
        U_i = U_{i,0}\ldots U_{i,q_ik-1},
    \end{align} $q_i\in \bbN_{>0}$ for each $i\in [0,n-1]$, and $U_{i,j} = \Gamma_{j \bmod k}$ or $U_{i,j} \subseteq \Sigma \setminus \Gamma_{j \bmod k}$ for each $j\in [0, q_ik-1]$. 
    Let $z=(q_0 + \cdots + q_{n-1})k$. For $i \in \bbN$, let $\Gamma_i = \Gamma_{i \bmod k}$. 
    Since the logic $\calL^1[\calB_{[<,+\omega]}, \Sigma, <, +\omega, \mathrm{mod},\min,\max]$ is closed under Boolean operations, we may assume without loss of generality that for each $i\in [0,n-1]$ and $j \in [0, q_ik-1]$ either $U_{i,j} = \Gamma_{j}$, or it is a singleton $U_{i,j} = \{a\}$ where $a \in \Sigma \setminus \Gamma_{j}$.
    
    We introduce some abbreviations. Let $\Gamma\subseteq \Sigma$ and $t \in \bbN_{0}$.
    \begin{align}
    \Gamma(x) &\defeq \vee_{a\in \Gamma} P_a(x)\\
    x \leq y &\defeq x < y \vee x=y\\
        x +t \leq y &\defeq x \leq y \wedge \bigwedge_{0 \leq t' < t} x +t' \neq y
    \end{align}
Note that the last abbreviation also applies when $t=0$.
    
Define $\varphi_{\mathit{len}}$ by
\begin{align}
    \varphi_{\mathit{len}} = \max \equiv k-1 \bmod k \wedge  \min + (z-1) \leq \max.
\end{align}
It is satisfied by a word $w$ if and only if $|w| \bmod k = 0$ and $|w| \geq z$. 

 Let $R$ be the set of positions
    \begin{align}
        R = \{(i,j) \mid i \in [0,n-1], j\in [0, q_ik-1], U_{i,j} \neq  \Gamma_{j}\}.
    \end{align}
If $R=\emptyset$, then $L=B^*B^{q_0 + \cdots + q_{n-1}}$. Then, $L$ is defined by the formula
    \begin{align}
        \varphi=\varphi_{\mathit{len}} \wedge \bigwedge_{i\in [0,k-1]} \forall x(x \equiv i \bmod k \rightarrow \Gamma_i(x)).
    \end{align}
    Otherwise, assume that $|R|=h>0$ and let 
    \begin{align}
         R=\{(i_0,j_0),\ldots,(i_{h-1},j_{h-1})\}
    \end{align}
   in the lexicographic order. For each $r \in [0,h-1]$ let $\{a_{r}\}= U_{i_r,j_r}$. 
   For $j \in [0,k-1]$ and $a\in \Sigma$, let
    \begin{align}
        \Delta_{j,a} &= \{ r \in [0,h-1] \mid j_r \equiv j \bmod k, a_r = a\},\\
         \Delta_{j} &= \{ a \in \Sigma \mid \Delta_{j,a} \neq \emptyset\}.
    \end{align}
    
    Define $\varphi_{\cont}$ by
    \begin{align}
        \varphi_{\cont} = \varphi_{\mathit{len}} &\wedge \bigwedge_{i\in [0,k-1]} \forall x(x \equiv i \bmod k \rightarrow \Gamma_i(x) \vee \Delta_i(x)) \\
        &\wedge \bigwedge_{j\in [0,k-1]} \bigwedge_{a\in \Delta_{j}} \exists^{=\left |\Delta_{j,a}\right |} x\,(x \equiv j \bmod k \wedge P_a(x)).
    \end{align}

 For each $r\in[0,h-1]$, let
\begin{align}
    \lambda_r = k\sum_{d=0}^{i_r-1} q_d + j_r
\end{align}
be the position of the letter $a_r$
in the product $U_0\cdots U_{n-1}$. Clearly
$0\leq\lambda_0<\cdots<\lambda_{h-1}\leq z-1$. 
Let $\delta_0=\lambda_0$, $\delta_{h}=z-1-\lambda_{h-1}$ and $\delta_{r}=\lambda_{r}-\lambda_{r-1}$ for all $r\in [1,h-1]$
be the distances between the consecutive positions.
Define the quantifier-free formula
\begin{align}
    \chi(y_{\min},x_0,\ldots,x_{h-1},y_{\max})
    &=  y_{\min} + \delta_0 \leq x_0
      \wedge x_{h-1}+ \delta_{h} \leq y_{\max}\\
    &\quad\wedge
    \bigwedge_{\substack{0\leq r<h-1\\ i_r=i_{r+1}}}
    x_r+ \delta_{r+1}=x_{r+1} \wedge
    \bigwedge_{\substack{0\leq r<h-1\\ i_r<i_{r+1}}}
    x_r +\delta_{r+1} \leq  x_{r+1}.
\end{align}

For each $r\in[0,h-1]$, define
\begin{align}
    \phi_r(x) = P_{a_r}(x)\wedge x\equiv\lambda_r\bmod k.
\end{align}

Let $Q$ be the $(h+2)$-ary, $(h+1)$-bounded numerical quantifier defined by $\neg\chi$. Then $Q\in\mathcal B_{[<,+\omega]}$. Define
\begin{align}
    \varphi&=\varphi_{\cont} \wedge
    \neg Q\,y_{\min}x_0\cdots x_{h-1}y_{\max} [
        y_{\min}=\min,
        \phi_0(x_0),\ldots,\phi_{h-1}(x_{h-1}),
        y_{\max}=\max].
\end{align}
We claim that $L(\varphi)=L$. Let $w$ satisfy $\varphi_{\cont}$. For each $r$, the argument $\phi_r$ defines a set of positions
of cardinality $|\Delta_{j_r\bmod k,a_r}|\leq h$
since $\lambda_r\equiv j_r\bmod k$.
Both endpoint arguments define singleton sets.
Consequently, all argument sets have cardinality less than $h+1$. By the definition of $Q$, its negation holds precisely when there exist positions $t_0,\ldots,t_{h-1}$ satisfying
$\phi_r(t_r)$ for every $r$ and $\chi(0,t_0,\ldots,t_{h-1},|w|-1)$.
These positions are strictly increasing. For each residue--letter
pair $(j,a)$, exactly $|\Delta_{j,a}|$ of the selected positions
have that pair. The exact counts and the content restriction in
$\varphi_{\mathrm{cont}}$ therefore imply that the selected
positions enumerate all positions whose letters lie outside the sets $\Gamma_{j\bmod k}$.

Suppose first that $w\in L$. Then there is a factorization $w=b_0u_0b_1\cdots u_{n-1}b_n$, where
$b_i\in B^*$ for $i \in [0,n]$ and  $u_i\in U_i$ for $i \in [0,n-1]$.
Then $w$ satisfies $\varphi_{\mathrm{cont}}$.
Let $t_r$ be the position in $w$ corresponding to the coordinate $(i_r,j_r)$.
Since every $|b_i|$ is a multiple of $k$, each $t_r$ satisfies
$\phi_r$. Moreover,
\[
    t_r=\lambda_r+\sum_{d=0}^{i_r}|b_d|.
\]
Hence $t_0\geq\delta_0$ and
$|w|-1-t_{h-1}\geq\delta_h$.
For $0\leq r<h-1$, the difference $t_{r+1}-t_r$ equals
$\delta_{r+1}$ when $i_r=i_{r+1}$ and is at least
$\delta_{r+1}$ when $i_r<i_{r+1}$.
Thus $\chi(0,t_0,\ldots,t_{h-1},|w|-1)$ holds, and
$w\models\varphi$.

Conversely, suppose that $w\models\varphi$, and choose positions
$t_0<\cdots<t_{h-1}$ as above.
To distinguish the displacements from the fixed distances
$\delta_r$, let $\eta_r=t_r-\lambda_r$ for all $r\in[0,h-1]$. The residue conditions give $\eta_r\equiv0\bmod k$.
The boundary conditions in $\chi$ give
$\eta_0\geq0$ and $\eta_{h-1}\leq|w|-z$.
Furthermore, for $0\leq r<h-1$, 
\begin{align}
\eta_{r+1}-\eta_r  = (t_{r+1}-t_r)-\delta_{r+1}.
\end{align}
This difference is zero if $i_r=i_{r+1}$ and nonnegative
otherwise. Consequently, we have
$0\leq\eta_0\leq\cdots\leq\eta_{h-1}\leq|w|-z$,  
$\eta_r\equiv0\bmod k$, and the displacements agree for all exceptional coordinates
belonging to the same block.

For each block $U_i$ containing an exceptional coordinate, let
$d_i$ be the common value of the $\eta_r$ with $i_r=i$.
For a block containing no exceptional coordinate, let $d_i$
be the value assigned to the nearest preceding block containing
an exceptional coordinate, or $0$ if there is no such block.
Then $0\leq d_0\leq\cdots\leq d_{n-1}\leq|w|-z$
 and   $d_i\equiv0\bmod k$. For $i\in[0,n-1]$, let
 \begin{align}
    s_i=k\sum_{d=0}^{i-1}q_d+d_i,
\end{align}
and let $u_i$ be the factor of $w$ starting at $s_i$ and having
length $q_i k$. These factors are well-defined and do not
overlap, since
\begin{align}
    s_{i+1}-(s_i+q_i k)=d_{i+1}-d_i\geq0
    \quad
\end{align}
for all $i \in [0,n-2]$ and $s_{n-1}+q_{n-1}k=z+d_{n-1}\leq|w|$. Each exceptional coordinate $(i_r,j_r)$ is placed at
\begin{align}
    s_{i_r}+j_r
    =\lambda_r+d_{i_r}
    =\lambda_r+\eta_r
    =t_r
\end{align}
and therefore carries the required letter $a_r$.
All other coordinates of these factors carry letters from the
corresponding sets $\Gamma_j$: every exceptional position is
among the $t_r$, and every $s_i$ is a multiple of $k$.
Thus $u_i\in U_i$ for every $i$.

Let $b_0,\ldots,b_n$ be the intervening factors, so that $w=b_0u_0b_1\cdots u_{n-1}b_n$. Their lengths are $|b_0|=d_0$, $|b_i|=d_i-d_{i-1}$ for all $i \in [1,n-1]$ and $|b_n|=|w|-z-d_{n-1}$.
All these lengths are nonnegative multiples of $k$, and each
$b_i$ starts at a position divisible by $k$.
Since no exceptional position lies in any $b_i$, the content
restriction in $\varphi_{\mathrm{cont}}$ gives $b_i\in B^*$.
Hence $w\in L$.
\end{proof}

\subsection{The Sublinear-Size Characterization Theorem}\label{S log main}

\begin{definition}[Modular Threshold Languages]\label{D Thr}
	For each $k,t\in \bbN_{>0}$, $i\in [0,k-1]$ and $a \in \Sigma$ 
	the language $\mathrm{Thr}_{i,k}(a,t)$ is defined to be
    \begin{align}
        \mathrm{Thr}_{i,k}(a, t) &= 
	    \{ w \in \Sigma^+ : \left|\pi_{i,k}(w)\right|_a \geq t \}.
    \end{align}
\end{definition}

\begin{samepage}
		\begin{makethm}{mainthm}{T log main}[Sublinear-Size Characterization]
			\unexpanded{Let $L\subseteq\Sigma^+$ be a regular language.
	Then the following statements are equivalent:
	\begin{enumerate}
		\item $L\in\SIZEDEPTH(O(\log^2 n),O(1))$.
		\item $L\in\SIZEDEPTH(O(\log n\log\log n),O((\log\log n)^2))$.
		\item $L\in\SIZE(o(n))$, or equivalently $c(L) \in o(n)$.
		\item The syntactic morphism of $L$
			is in $\bfX$.
		\item $L$ is a union of $\simeq_k$-classes
	for some $k\in\bbN_{>0}$.
\item $L$ is a finite positive Boolean combination of finite languages and languages of the form $uKv$, where $u,v\in\Sigma^+$ and $K$ is either a modular disjoint monomial, or a modular threshold language.
		\item $L$ is a finite Boolean combination of finite languages and
			languages of the form $uKv$, where $u,v\in\Sigma^+$ and 
			$K$ is a modular disjoint monomial.

        \item $L$ is definable in $\calL^{1}[\calB_{[<,+\omega]}, \Sigma, <, +\omega, \mathrm{mod}, \min, \max]$.
        \item $L$ is definable in  $\calL^{1}[\calB_{\mathrm{num}[O(n^2),O(1)]}, \Sigma, \mathrm{arb}]$.
        \item $L$ is definable in  $\calL^{1}[\calB_{\mathrm{num}[O(n\log n),O(\log^2 n)]}, \Sigma, \mathrm{arb}]$.
	\end{enumerate}
}
	\end{makethm}
\end{samepage}
\begin{proof}
We show the following implications:
\[
\begin{tikzcd}[row sep=small, column sep=small]
	& & & & &  & (9) \arrow[r, Rightarrow] & (1) \arrow[dr, Rightarrow] & \\
(3) \arrow[r, Rightarrow]
  & (4) \arrow[r, Rightarrow]
  & (5) \arrow[r, Rightarrow]
  & (6) \arrow[r, Rightarrow]
  & (7) \arrow[r, Rightarrow]
  & (8) \arrow[ur, Rightarrow] \arrow[dr, Rightarrow]
  &  
  & & (3) \\
& & & & &  & (10) \arrow[r, Rightarrow] & (2) \arrow[ur, Rightarrow] &
\end{tikzcd}
\]
	\begin{samepage}
\medskip

\noindent 
		$(1) \Longrightarrow  (3)$:\quad Trivial. 
\end{samepage}

	\begin{samepage}
\medskip

\noindent 
		$(2) \Longrightarrow  (3)$:\quad Trivial. 
\end{samepage}

	\begin{samepage}
\medskip

\noindent 
	$(3) \Longrightarrow  (4)$:\quad
		By Theorem~\ref{T log lower} which is proven in \S\,\ref{S lower}.
\end{samepage}

	\begin{samepage}
\medskip

\noindent 
	$(4) \Longrightarrow  (5)$:\quad
		By Proposition~\ref{P log tech} which is proven in \S\,\ref{S AlgCong}.
\end{samepage}

\medskip

\noindent 
	$(5) \Longrightarrow(6)$:\quad Observe that the congruence $\simeq_k$ is of finite index for any $k \in \bbN_{>0}$. Therefore, to prove the direction, it suffices to show that each equivalence class of $\simeq_k$ is either a finite language, or a positive Boolean combination of languages of the form $uKv$ where $K$ is either a modular disjoint monomial, or a modular threshold language. Since $\simeq_{2k}$ refines $\simeq_{k}$, without loss of generality we may  assume that $k\geq 2$.
    
    Take an equivalence class of $\simeq_k$. If it is a singleton then the direction follows. Otherwise, it is a set of the form $uKv$ for some words $u,v \in \Sigma^k$ and an equivalence class $K$ of $\congC_k^{k,k}$. If $K$ is a finite language, then again the claim follows. Otherwise, $K$ is an infinite set and it suffices to show that $K$ can be written as a positive Boolean combination of languages of the form $u'K'v'$ where $u',v' \in \Sigma^*$ and $K'$ is either a modular disjoint monomial, or a modular threshold language. 
    
     Let $w=w_0\ldots w_{n-1} \in K$ where $n\in \bbN_{>0}$. Let $n=qk+r$, where $0\leq r<k$. 
     Let $\Gamma_i = \cont_{i,k}^{\geq k}(w)$ for each $i \in [0,k-1]$. We have two cases: when 
     the set of rare positions $R_k^k(w)$ of $w$ is empty and when it is not.
     
Assume that the set of rare positions $R_k^k(w)$ of $w$ is empty. Define
\begin{align}
K'&= (\Gamma_0\ldots\Gamma_{k-1})^* \Gamma_0 \ldots \Gamma_{r-1} \cap \bigcap_{i\in[0,k-1]} \bigcap_{a\in\Gamma_i} \mathrm{Thr}_{i,k}(a,k).
\end{align}

We claim that $K=K'$. Note that the claim implies that $K$ can be expressed as a positive Boolean combination of languages of the form $u'K'v'$ where $u',v' \in \Sigma^*$ and $K'$ is either a modular disjoint monomial, or a modular threshold language, as required. To prove the claim, it suffices to show that
$w'\mathrel{\calC_k^{k,k}}w$ if and only if $w'\in K'$
for every word $w'\in\Sigma^+$. Clearly, $w'\mathrel{\calC_k^{k,k}}w$ if and only if
\begin{align}
    |w'|\equiv_k |w|,\qquad
R_k^k(w')=\emptyset,\qquad
\cont_{i,k}^{\geq k}(w')=\cont_{i,k}^{\geq k}(w)=\Gamma_i
\end{align}
for every $i\in[0,k-1]$. By the definition of $K'$,  these conditions hold if, and only if, 
$w'\in K'$.
     
Next assume that the set of rare positions  $R_k^k(w)$ of $w$ is nonempty. Let $R_k^k(w)=\{x_0,\ldots, x_{\ell-1}\} \subseteq [0,n-1]$ for some $\ell \geq 1$ such that $x_0<\cdots < x_{\ell-1}$. We use the following notation. 
For $x\in \bbN$ let $\Gamma_{x} = \Gamma_{x \bmod k}$.

We define the set $E$ to be
\begin{align}
        E &=  E_0 w_{x_0} E_1 w_{x_1} \ldots E_{\ell-1}  w_{x_{\ell-1}} E_\ell
    \end{align}
    where $E_0,\ldots, E_\ell$ are defined as follows.

\begin{align}
E_0&=
\begin{cases}
\varepsilon
    & \text{if } x_0=0,\\
\Gamma_0\Gamma_1\cdots\Gamma_{x_0-1}
    & \text{if } 1\le x_0<k,\\
(\Gamma_0\cdots\Gamma_{k-1})^+\Gamma_0\cdots\Gamma_{(x_0 \bmod k)-1} & \text{if } x_0\ge k.
\end{cases}
\end{align}

For each $j\in[0,\ell-2]$, let
$d_j=(x_{j+1}-x_j)\bmod k$. Define
\[
E_{j+1}=
\begin{cases}
\varepsilon
& \text{if } x_{j+1}-x_j=1,\\[1mm]
\Gamma_{x_j+1}\Gamma_{x_j+2}\cdots\Gamma_{x_j+d_j-1}
& \text{if } 1<x_{j+1}-x_j<k,\\[1mm]
(\Gamma_{x_j+1}\cdots\Gamma_{x_j+k})^*
\Gamma_{x_j+1}\cdots\Gamma_{x_j+k+d_j-1}
& \text{if } x_{j+1}-x_j\ge k.
\end{cases}
\]

Finally, let $d_{\ell-1}=(n-1-x_{\ell-1})\bmod k$.
Define
\[
E_\ell=
\begin{cases}
\varepsilon
& \text{if } n-1-x_{\ell-1}=0,\\[1mm]
\Gamma_{x_{\ell-1}+1}\cdots
\Gamma_{x_{\ell-1}+d_{\ell-1}}
& \text{if } 1\le n-1-x_{\ell-1}<k,\\[1mm]
(\Gamma_{x_{\ell-1}+1}\cdots\Gamma_{x_{\ell-1}+k})^+
\Gamma_{x_{\ell-1}+1}\cdots
\Gamma_{x_{\ell-1}+d_{\ell-1}}
& \text{if } n-1-x_{\ell-1}\ge k.
\end{cases}
\]
We define
\begin{align}
    K' &= E \cap \bigcap_{i\in[0,k-1]} \bigcap_{a\in\Gamma_i} \mathrm{Thr}_{i,k}(a,k).
    \label{eq:reg-thr}
\end{align}

We claim that $K=K'$. Let us observe that the claim implies that $K$ can be expressed as a positive Boolean combination of languages of the form $uMv$ where $u,v \in \Sigma^*$ and $M$ is either a modular disjoint monomial, or a modular threshold language, as required. Since $K$ is  infinite, the set $\{j \in [0,\ell] : |E_j|=\infty \}$ is nonempty. Therefore $E$ can be written as a finite  union of languages of the form $uMv$ where $u,v \in \Sigma^*$ and $M$ is a modular disjoint monomial, and from (\ref{eq:reg-thr}) the observation follows. 

Next we prove the claim. 
Let $w'\in K$. Since $w'$ belongs to the same
$\congC_k^{k,k}$-equivalence class as $w$, it has the same
rare-position pattern, including the labels of the rare
positions and the distances between them modulo $k$. Therefore,
$w'\in E$. Moreover, for every $i\in[0,k-1]$ and
$a\in\Gamma_i$, the letter $a$ occurs at least $k$ times
in modulo class $i$ of $w'$. Thus $w'\in K'$ and  $K\subseteq  K'$.

Conversely, let $w'\in  K'$. Since $w'\in K'$, every
position $i\in[0,|w'|-1]$ other than the rare positions contains
a letter from the corresponding set $\Gamma_i$. Furthermore,
the rare positions of $w'$ have the same labels and the same
residues modulo $k$ as those of $w$. In particular, any letter
$a\notin\Gamma_i$ occurring in modulo class $i$ of $w'$
can occur only at these rare positions. The number of such
occurrences is therefore the same as in $w$, and is strictly
less than $k$. On the other hand, since $w'$ satisfies the threshold condition in (\ref{eq:reg-thr}), every letter $a\in\Gamma_i$ occurs at least $k$ times in modulo class $i$ of $w'$. It follows that $\cont_{i,k}^{\geq k}(w')=\Gamma_i$
for every $i\in[0,k-1]$.
Thus $w'$ has precisely the same common letters in each
modulo class as $w$, as well as the same rare-position pattern.
By the characterization of $\congC_k^{k,k}$, we conclude
that $w'\in K$. Hence $K'\subseteq K$, proving the claim.

\medskip
\noindent 
	$(6) \Longrightarrow  (7)$: It suffices to show that, for all $u,v\in\Sigma^+$, every
language $u\operatorname{Thr}_{i,k}(a,t)v$ is a finite Boolean
combination of languages of the form $u'Mv'$, where
$u',v'\in\Sigma^+$ and $M$ is a modular disjoint monomial.
Let 
    $k\in \bbN_{>0}$, $a \in \Sigma$, and $i \in [0,k-1]$. We define the following sets.
    \begin{align}
    B &= \Sigma^{i} a \Sigma^{k-i-1}\\
    D &= \Sigma^k \setminus B = \Sigma^{i} (\Sigma \setminus a) \Sigma^{k-i-1}
    \end{align} 
    Let $\mathrm{EQ}_{i,k}(a,0)=D^*$ and for $t \geq 1$ let 
 $\mathrm{EQ}_{i,k}(a,t) =(D^*B)^t D^*$. Define,
\begin{align}
\mathrm{EQ}'_{i,k}(a,0) &= \bigcup_{x \in \Sigma^{<k},\atop \pi_{i,k}(x)\neq a} \mathrm{EQ}_{i,k}(a,0)x
\end{align}
and for $t\geq 1$,
\begin{align}
\mathrm{EQ}'_{i,k}(a,t) &= \bigcup_{x \in \Sigma^{<k}} \mathrm{EQ}_{i,k}\left (a, t-\left|\pi_{i,k}(x)\right|_a\right) x.
\end{align}

Then for all $t\in \bbN_{>0}$,
    \begin{align}
    \mathrm{Thr}_{i,k}(a,t) &= \Sigma^+ \setminus \bigcup_{0 \leq t' < t}  \mathrm{EQ}'_{i,k}(a,t').
    \end{align}
    For every $t\in\mathbb N$, the language $\mathrm{EQ}_{i,k}(a,t)$ is a $k$-modular disjoint monomial. 
Therefore the language $u\mathrm{EQ}'_{i,k}(a,t)v$ is a finite union
of languages of the form $uMxv$, where $M$ is a modular
disjoint monomial and $x\in \Sigma^{<k}$.
These languages have the required form, since $xv\in\Sigma^+$.

\medskip
\noindent 
	$(7) \Longrightarrow  (8)$: Every finite language is definable in
$\rmFO^1[\Sigma,\min,\max,+\omega]$ by
Example \ref{example:logic-const}. Hence, it suffices to show that every
language of the form $u(K\setminus \varepsilon)v$, where $u,v\in\Sigma^+$ and $K$ is a
modular disjoint monomial, is definable in
$\calL^1[\calB_{[<,+\omega]},\Sigma,<,+\omega,\mathrm{mod},\min,\max]$.
The desired direction then follows from closure under Boolean
operations.

Let $u=a_0\cdots a_{m-1}$ and $ v=b_0\cdots b_{n-1}$
where $m,n\in \bbN_{>0}$, and let $K$ be a $k$-modular disjoint monomial
for some $k\in\bbN_{>0}$. By
Proposition~\ref{lem:mod-dis-mono-logic}, there exists a formula
$ \psi\in\calL^1[\calB_{[<,+\omega]},\Sigma, <,+\omega,\mathrm{mod},\min,\max]$
such that $L(\psi)=K\setminus\varepsilon$. 

Define the formula
\begin{align}
 \iota(x) &=  \bigwedge_{i\in[0,m-1]} \min+i\neq x \wedge \bigwedge_{i\in[0,n-1]} x+i\neq\max.
\end{align}
Thus, $\iota(x)$ defines the interval of positions strictly between
the prefix $u$ and suffix $v$.

We define the relativization $\psi^{\iota}$ of $\psi$ to the interval defined by
$\iota(x)$. The translation of sentences commutes with Boolean operations. We first give the atomic sentence cases. We omit formulas that are identically true or false (such as $\max <\min$, $\min \equiv 1 \bmod q$ etc.) 
\begin{align}
    P_a(\min)^\iota &\defeq \exists x (\min+m=x\wedge P_a(x))\\
    P_a(\max)^\iota &\defeq \exists x (x+n=\max\wedge P_a(x))\\
        (\min < \max)^\iota &\defeq \min+m+n\neq\max,\\
    (\min = \max)^\iota &\defeq \min+(m+n)=\max,\\
    (\min + j = \max)^\iota &\defeq\min+(m+n+j)=\max,\\
    (\max \equiv r \bmod q)^\iota &\defeq \max \equiv (r+m+n) \bmod q.
\end{align}
 Next, we define
\begin{align}
    Q^{\iota}x_0\cdots x_{k-1} [\phi_0^{\iota}(x_0),\ldots,
   \phi_{k-1}^{\iota}(x_{k-1})]
\end{align}
for each subformula of the form $Qx_0\cdots x_{k-1} [\phi_0(x_0),\ldots,\phi_{k-1}(x_{k-1})]$ for $k\in \bbN_{>0}$. Let $i\in [0,k-1]$. 
The formula 
$\phi_i(x_i)$ uses atomic formulas of the form $P_a(t)$, $t < t'$, $t'=t$,  $t+j=t'$, $t\equiv r \bmod q$, where $a\in \Sigma$, $t,t' \in \{\min,\max,x_i\}$, $r\in \bbN$ and $j,q\in \bbN_{>0}$. By Remark \ref{remark:sentence-free}, we may assume that each $\phi_i(x_i)$ is a quantifier-free formula and every subformula of $\phi_i(x_i)$ contains the variable $x_i$.
Let $\phi_i(x_i)^\iota = \iota(x_i) \wedge T_{m,n}(\phi_i(x_i))$ where the translation $T_{m,n}(\phi_i(x_i))$ commutes with the Boolean operations and is defined inductively from the base cases: 
\begin{align}
    T_{m,n}(P_a(x_i)) &\defeq P_a(x_i), \\
    T_{m,n}(\min < x_i) &\defeq \min+m \neq x_i, \\
    T_{m,n}(x_i < \max) &\defeq  x_i +n \neq \max, \\
     T_{m,n}(\min+j = x_i) &\defeq \min + (j+m) = x_i, \\
    T_{m,n}(x_i+ j = \max) &\defeq x_i + (j+n)=\max, \\
    T_{m,n}(x_i=\min) &\defeq \min+m=x_i,\\
    T_{m,n}(x_i=\max) &\defeq x_i+n=\max,\\
    T_{m,n}(x_i\equiv r \bmod q) &\defeq   x_i\equiv (r+m) \bmod q.
\end{align}

The predicates $<$ and $+j$ are invariant under a common translation of all coordinates. Since this translation also preserves cardinalities, it preserves the defining condition of $Q$; hence we may take $Q^\iota = Q$.

To see the correctness of this relativization, let $w$ have length
$N=m+n+s$, where $s\in \bbN_{>0}$, and let $z$ be the factor of $w$ induced by the interval
$[m,m+s-1]$. The map $\rho:x \mapsto x+m$ identifies the positions of $z$ with
the positions of this interval. By construction, the relation
defined by the relativized subformula is the image under
$\rho$ of the relation defined by the original subformula
on $z$. Consequently,
\begin{align}
w\models Q^\iota x_0\cdots x_{k-1} [\phi_0^\iota(x_0),\ldots,\phi_{k-1}^\iota(x_{k-1})] & \Longleftrightarrow
z\models Qx_0\cdots x_{k-1}
[\phi_0(x_0),\ldots,\phi_{k-1}(x_{k-1})].
\end{align}
Hence, $\psi^{\iota}$ expresses that the factor between the prefix $u$ and suffix $v$ belongs to $K$.

We now define
\begin{align}
  \varphi_L  &=
  \bigwedge_{i\in[0,m-1]}
  \exists x\,\left(\min+i=x\wedge P_{a_i}(x)\right)
  \label{E formula 1}\\
  &\wedge
  \bigwedge_{i\in[0,n-1]}  \exists x\,\left(x+i=\max\wedge P_{b_{n-1-i}}(x)\right)\nonumber\\
  &\wedge \exists x\,\iota(x)  \wedge \psi^{\iota}.
  \label{E formula 2}
\end{align}
Clearly, $ \varphi_L\in \calL^1[\calB_{[<,+\omega]},\Sigma, <, +\omega, \mathrm{mod}, \min,\max]$.

The first two conjuncts of $\varphi_L$ express that the word begins
with $u$ and ends with $v$, respectively. The third conjunct ensures
that the factor between the prescribed prefix and suffix is nonempty.
Finally, by the correctness of relativization, the last conjunct
holds precisely when this factor belongs to $K$. Consequently,
$L(\varphi_L)=u(K\setminus\varepsilon)v$.
    
\medskip
\noindent 
	$(8) \Longrightarrow  (9)$:\quad Since $\{<,+\omega\} \subseteq \mathrm{num}[O(n),O(1)]$ and $\{<,+\omega,\mathrm{mod},\min,\max \}\subseteq \mathrm{arb}$.
    
\medskip
\noindent 
	$(8) \Longrightarrow  (10)$:\quad Since $\{<,+\omega\} \subseteq \mathrm{num}[O(n),O(1)]$ and $\{<,+\omega,\mathrm{mod},\min,\max \}\subseteq \mathrm{arb}$.

\medskip
\noindent 
	$(9) \Longrightarrow  (1)$:\quad 
It suffices to show that every language, not necessarily regular,  defined by an  $\calL^1[\,\calB_{\mathrm{num}[O(m^2),O(1)]}, \Sigma,\mathrm{arb}]$ sentence is computed by a constant-depth circuit family of size $O(\log^2 n)$. Observe that constant-depth circuit families of size $O(\log^2 n)$ are closed under finite Boolean combinations.

Let $\varphi$ be an $\calL^1[\,\calB_{\mathrm{num}[O(m^2),O(1)]}, \Sigma,\mathrm{arb}]$ sentence defining the language $L$. We may assume by Remark \ref{remark:sentence-free} that $\varphi$ is a Boolean combination of atomic sentences and formulas of the form $\psi = Q x_0\ldots x_{k-1}\, [\phi_0(x_0),\ldots, \phi_{k-1}(x_{k-1})]$, where $Q\in \calB_{\mathrm{num}[O(m^2),O(1)]}$ and each $\phi_i(x_i)$ is a quantifier-free formula over the variable $x_i$. Atomic sentences have constant-size circuits: numerical sentences are hardwired for each input length, and letter predicates on constants are input gates. Thus, to prove the claim it suffices to construct a constant-depth circuit family of size $O(\log^2 n)$ for each formula of the form $\psi$. 

 For a  subset $\Lambda \subseteq \Sigma$ we denote by $\Lambda(x)$ the formula $\vee_{a\in \Lambda} P_a(x)$ with the convention $\emptyset(x) \equiv \bot$. Note that the negative literal $\neg P_a(x)$ is equivalent to the formula $(\Sigma\setminus a)(x)$. Also, for all  subsets $\Lambda_1, \Lambda_2 \subseteq \Sigma$ we have $\Lambda_1(x) \wedge \Lambda_2(x) = (\Lambda_1 \cap \Lambda_2) (x)$ and $\Lambda_1(x) \vee \Lambda_2(x) = (\Lambda_1 \cup \Lambda_2) (x)$. 
Let $C=\{c_0,\ldots, c_{\ell-1}\}$, where $\ell\in \bbN$, be the set of constants used in the formula $\psi$. Let 
\begin{align}
    \Delta(C) &= \left \{ \bigwedge_{i=0}^{\ell-1} P_{a_i}(c_i) \mid (a_0,\ldots, a_{\ell-1})\in \Sigma^\ell \right\}
\end{align} 
\noindent be the set of unary letter types over the constants $C$. Note that the formulas in $\Delta(C)$ are mutually exclusive and exhaustive, i.e., on any given word precisely one of them is true.

Let $i \in [0,k-1]$ and consider the formula $\phi_i(x)$ that is quantifier-free.
We can rewrite $\phi_i(x)$ equivalently as a conjunction  
\begin{align}
    \phi_i'(x) &= \bigwedge_{\delta \in \Delta(C)} (\delta \rightarrow \phi_{i,\delta}'(x))
\end{align}
\noindent where $\delta \in \Delta(C)$,  and $\phi_{i,\delta}'(x)$ is a Boolean combination of numerical predicates using the variable $x$ and constants in $C$, and letter predicates using the variable $x$.
It is easy to see that 
\begin{align}
\psi &\equiv
    \bigwedge_{\delta \in \Delta(C)} \left (\delta \rightarrow Q x_0\ldots x_{k-1}\, [\phi_{0,\delta}'(x_0),\ldots, \phi_{k-1,\delta}'(x_{k-1})] \right).
\end{align}

Thus it suffices to construct a circuit family of the desired size and depth for each formula of the form 
$Q x_0\ldots x_{k-1}\, [\phi_{0}(x_0),\ldots, \phi_{k-1}(x_{k-1})]$ where each $\phi_i(x)$ is a Boolean combination of numerical predicates using the variable $x$ and constants in $C$, and letter predicates using the variable $x$.

Further, each $\phi_{i}(x)$ is equivalent to a formula 
\begin{align}
   \phi_{i}(x) \equiv \bigwedge_{j \in J} (o_{i,j}(x,C) \rightarrow \Gamma_{i,j}(x)).
\end{align}
 where $o_{i, j}(x,C)$ is a Boolean combination of numerical predicates using the variable $x$ and constants in $C$, and $\Gamma_{i,j} \subseteq \Sigma$ such that the formulas $\{o_{i,j}(x,C) \mid j \in J\}$ are mutually exclusive and exhaustive, i.e., precisely one of them is true at any given position. 
For each fixed input length, the numerical conditions $o_{i,j}$ determine a unique $j$ at each position $p$. For a word $w$ of this length, the characteristic word of $\phi_i^w$ is therefore obtained directly from the
subset-encoding input gates $(p,\Gamma_{i,j})$, without additional gates. Let $w$ be a word of length $n$. It is easy to see that 
\begin{align}
    \phi_i^w = \bigcup_{j\in J} \{p \in o_{i,j}^w \mid  w_p \in \Gamma_{i,j}\}.\label{Eq:phi_i^w}
\end{align}

Assume that the quantifier $Q$ is $t$-bounded, where $t\in \bbN_{>0}$, and it is defined by the  quantifier-free formula $\chi(x_0,\ldots, x_{k-1}
)$ using the predicates in $\mathrm{num}[O(m^2),O(1)]$.
 The formula $\psi$ is true on the word $w$ if precisely one of the following conditions holds:
\begin{enumerate}
    \item $\phi_i^w = \emptyset$ for some $i\in [0,k-1]$.    
    \item  $\phi_i^w \neq \emptyset$ for all $i\in [0,k-1]$,
    \begin{enumerate}[label=\Alph*.]
    \item $1 \leq |\phi_i^w| < t$ for each $i\in [0,k-1]$, and 
    \item $\barv \models \chi $ for each $\barv \in \prod_{i \in [0,k-1]} \phi_i^w $.
    \end{enumerate}  
\end{enumerate}

It suffices to show that for each length $n\in \bbN_{>0}$ there exist constant depth circuits of size $O(\log^2 n)$ for checking each condition. Using (\ref{Eq:phi_i^w}) Condition (1) can be checked by a circuit of size $O(k)$ and depth two. Let $r\in \bbN$. By Corollary \ref{cor:cir-thr} for each $i\in [0,k-1]$ and $j\in J$ there is a circuit $C_{i,j}^r$ of size $O(\log n)$ and depth $O(1)$ that computes whether 
\begin{align}
    \sum_{a\in \Gamma_{i,j}}\left |\{ p \in o_{i,j}^w : w_p = a \} \right | = r.
\end{align} 
Since the sets of positions $\{o_{i,j}^w \mid j \in J\}$ are mutually exclusive and exhaustive, the circuit $C_{i}^r$ defined below computes whether $|\phi_i^w| = r$: without loss of generality, let $J=[0,s-1]$ for some $s \in \bbN_{>0}$, and
\begin{align}
C_{i}^r=\bigvee_{\left (r_0,\ldots, r_{s-1}\right)\in \bbN^{s} \atop
r_0 + \cdots + r_{s-1} = r} \bigwedge_{j \in J} C_{i,j}^{r_j}
\end{align}
 The circuit $C_{i}^r$ has size $O(\log n)$ and depth $O(1)$.

Next we detail how to verify Condition (2).B. 
Given that $|\phi_i^w|=r$ (verified by the circuit $C_i^r$), by Theorem \ref{thm:occ-log2n}, there is a circuit of size $O(\log^2n)$ and depth $O(1)$ that computes the set of positions $\phi_i^w$. 
Each occurrence circuit is used only when its promise holds. Since $\chi(x_0,\ldots, x_{k-1})$ is a formula over the predicates in $\mathrm{num}[O(m^2),O(1)]$, there is a circuit of size $O(\log^2 n)$ and depth $O(1)$ that computes $\chi$. We verify $\chi$ on each tuple in $\prod_{i \in [0,k-1]} \phi_i^w$. The number of tested tuples depends only on the fixed formula. Thus Condition (2).B is verified in constant-depth and size $O(\log^2 n)$.
\medskip

\noindent 
	$(10) \Longrightarrow  (2)$:\quad As in the previous direction, it suffices to show that every language, not necessarily regular,  defined by an  $\calL^1[\,\calB_{\mathrm{num}[O(m\log m),O(\log^2 m)]}, \Sigma,\mathrm{arb}]$ sentence is computed by a circuit family of size $O(\log n\log\log n)$ and depth $O((\log\log n)^2)$.  

    Let $\varphi$ be an $\calL^1[\,\calB_{\mathrm{num}[O(m\log m),O(\log^2 m)]}, \Sigma,\mathrm{arb}]$ sentence defining the language $L$. 
We may assume that $\varphi$ is a Boolean combination of atomic sentences and formulas of the form $\psi=Qx_0\ldots x_{k-1}[\phi_0(x_0),\ldots,\phi_{k-1}(x_{k-1})]$. Atomic sentences are handled as in the preceding direction.
Here each $\phi_i(x_i)$ is a quantifier-free formula
over the variable $x_i$.
   Thus, to prove the claim it suffices to construct a circuit family of size $O(\log n\log\log n)$ and depth $O((\log \log n)^2)$ for each formula of the form $\psi$. 

    We repeat the proof of the previous direction. The only difference is in checking Condition (2).B. 
We use Theorem~\ref{T upper loglog} to compute the set of positions $\phi_i^w$ using a circuit family of size  $O(\log n\log\log n)$ and depth $O((\log\log n)^2)$. Since $\chi(x_0,\ldots, x_{k-1})$ is a formula over the predicates in $\mathrm{num}[O(m\log m),O(\log^2 m)]$, there is a circuit of size $O(\log n\log \log n)$ and depth $O((\log\log n)^2)$ that computes $\chi$. By verifying $\chi$ on each tuple in $\prod_{i \in [0,k-1]} \phi_i^w$ we verify Condition (2).B in size $O(\log n\log\log n)$ and depth $O((\log \log n)^2)$. 
\end{proof}
\subsubsection{Neutral Letter Languages of Sublinear Circuit Complexity}

\label{neutral letter}
A language $L\subseteq \Sigma^+$ has a \emph{neutral letter} if there exists a letter $c\in \Sigma$ such that $uv\in L$ if, and only if, $ucv\in L$ for all 
$u,v\in \Sigma^*$ such that $uv\in\Sigma^+$. 
A language having a neutral letter is called a \emph{neutral letter language}.
The syntactic semigroup of a neutral letter language $L$ with the neutral letter $c$ is a monoid with the identity $\eta_L(c)$. Moreover, the stable semigroup of its syntactic morphism is the entire syntactic semigroup.

Let $\bfM\mathbf{Nil}$ denote the pseudovariety of finite aperiodic monoids $M$ satisfying the identity $ex=xe$ for all elements $x\in M$ and $e\in E(M)$ \cite{mnil}.

\begin{lemma}
    Let $L \subseteq \Sigma^+$ be a language whose syntactic stamp is in $\bfX$. If $L$ has a neutral letter then the syntactic semigroup of $L$ is in $\bfM\mathbf{Nil}$. 
    \label{lem:X-MNil}
\end{lemma}
\begin{proof}
Let $\eta_L : \Sigma^+ \to S(L)$ be the syntactic stamp of $L$. If $L$ has a neutral letter then $ S(L)$ is a monoid and moreover $\stab(\eta_L)=S(L)$. Let $1$ denote the identity of $S(L)$. Let $x\in S(L)$ and $e \in E(S(L))$. Assume that the stability index of $\eta_L$ is $s$ and  let $u,v,w\in \Sigma^s$ be words with the images $e,x,1$ under $\eta_L$. Let $n \in \bbN_{>0}$ be an exponent of $S(L)$. 
Then, we observe that
\begin{align}
    v^n &=_{\eta_L} v^n\underbracket[.5pt]ww^n\\
    &=_{\eta_L} v^nvw^n\\
    &=_{\eta_L} v^{n+1}
\end{align}
Therefore we conclude that $S(L)$ obeys the identity $x^\omega=x^{\omega+1}$. Thus $S(L)$ is aperiodic. Further, we observe that
\begin{align}
    uv &=_{\eta_L} (u\underbracket[.5pt]u)^nv\underbracket[.5pt]w(w\underbracket[.5pt]w)^n \\
    &=_{\eta_L} (uu)^nvu(ww)^n && (\text{Applying $(x_1x_2)^\omega p_1y_2(y_1y_2)^\omega = (x_1x_2)^\omega p_1x_2(y_1y_2)^\omega $})\\
       &=_{\eta_L} uvu\\
       &=_{\eta_L} (\underbracket[.5pt]ww)^n \underbracket[.5pt]uv (\underbracket[.5pt]uw)^n\\
       &=_{\eta_L} (ww)^n wv (uw)^n && (\text{Applying $(x_1x_2)^\omega y_1p_2(y_1y_2)^\omega = (x_1x_2)^\omega x_1p_2(y_1y_2)^\omega $})\\
       &=_{\eta_L} vu.
\end{align}
Therefore we conclude that $S(L)$ obeys the identity $ex=xe$.
\end{proof}

\begin{samepage}
\begingroup
	\renewcommand{\themainthm}{\getrefnumber{intro:neutral}}
\begin{makethm}{mainthm}{T Neutral Letter}
\unexpanded{
    Let $L\subseteq \Sigma^+$ be a regular neutral letter language. Then, 
	the following statements are equivalent:
    \begin{enumerate}
    \item $L\in\SIZEDEPTH(O(\log^2 n),O(1))$.
		\item $L\in\SIZEDEPTH(O(\log n\log\log n),O((\log\log n)^2))$.
		\item $L\in\SIZE(o(n))$.
	   
        \item  The syntactic semigroup of $L$ is in $\bfM\mathbf{Nil}$.
	   \item $L$ is a disjoint union of $\congC_1^{t,0}$-classes for some $t\in \bbN_{>0}$.
      \item $L$ is a finite Boolean combination of disjoint monomials.
      
\item $L$ is definable in  $\calL^1[\calB_{[<]}, \Sigma]$.  
\end{enumerate}
	}
\end{makethm}
\endgroup
\end{samepage}
\begin{proof}
We show the following implications:
\[
\begin{tikzcd}[row sep=small, column sep=small]
	& & (5) & & (1) \arrow[dr, Rightarrow] & \\
(3) \arrow[r, Rightarrow]
  & (4) \arrow[ur, Leftrightarrow]  \arrow[r, Leftrightarrow]
  & (6) \arrow[r, Rightarrow]
  & (7) \arrow[ur, Rightarrow] \arrow[dr, Rightarrow]
  & & (3) \\
& &   & & (2) \arrow[ur, Rightarrow] &
\end{tikzcd}
\]

$(1) \Longrightarrow (3)$ and $(2) \Longrightarrow (3)$ are trivial and
$(7) \Longrightarrow (1)$ and $(7) \Longrightarrow (2)$ follow from Theorem~\ref{T log main}. The equivalence
$(4)\Longleftrightarrow (6)$ is proved in  \cite{mnil}. Thus it suffices to show 
$(3)\Longrightarrow (4)$, $(4)\Longleftrightarrow (5)$ and $(6)\Longrightarrow (7)$.

\medskip
\noindent
	$(3)\Longrightarrow(4)$:\quad 
By Theorem~\ref{T log main}, if a regular language $L$ is in $\SIZE(o(n))$ then $\eta_L \in \bfX$. Further, if $L$ has a neutral letter then $S(L) \in \bfM\mathbf{Nil}$ by Lemma~\ref{lem:X-MNil}.

\medskip
\noindent
$(4)\Longleftrightarrow(5)$:\quad For $t\in\bbN_{\geq0}$, let $\sim^t$ be the congruence on
$\Sigma^+$ defined by
\begin{align}
    u\sim^t v &\Longleftrightarrow
\Gamma_t(u)=\Gamma_t(v) \wedge 
u_{< t}=v_{< t},
\end{align}
where $\Gamma_t(w)=\{a\in\Sigma:|w|_a\geq t\}$ and
$w_{< t}$ is obtained from $w$ by erasing all occurrences of letters in $\Gamma_t(w)$. In $\cite{mnil}$ it is shown that $S(L) \in \bfM\mathbf{Nil}$
if, and only if, there exists $t \in \bbN_{>0}$ such that $L$ is a union of equivalence classes of $\sim^t$, i.e., $u \sim^t v \implies \eta_L(u)=\eta_L(v)$.

Assume that $L$ has a neutral letter $c$.
To prove the equivalence, it suffices to show that $S(L) \in \bfM\mathbf{Nil}$
if, and only if, there exists $t \in \bbN_{>0}$ such that $L$ is a union of equivalence classes of $\calC_1^{t,0}$.
Clearly, $u \mathbin{\calC_1^{t,0}} v \implies u \sim^t v$. Thus, $\congC_1^{t,0}$ refines $\sim^t$, and hence
the implication $(4)\Longrightarrow(5)$ follows.

For the converse, suppose that $L$ is a union of
$\congC_1^{t,0}$-classes for some $t\in\bbN_{>0}$.
We observe that $u\sim^t v
\implies c^tuc^t\mathbin{\congC_1^{t,0}}c^tvc^t$. Since $L$ is a union of
$\congC_1^{t,0}$-classes and $\congC_1^{t,0}$ is
a congruence, we have
$u\mathbin{\congC_1^{t,0}}v
\implies \eta_L(u)=\eta_L(v)$.
Since $c$ is a neutral letter,
$\eta_L(c)=1$. Consequently, for any $u,v\in\Sigma^+$
such that $u\sim^t v$, we obtain
\begin{align}
    \eta_L(u)
    &=\eta_L(c^tuc^t)\\
    &=\eta_L(c^tvc^t) &&(\text{Since $u\sim^t v
\implies c^tuc^t\mathbin{\congC_1^{t,0}}c^tvc^t$})\\
    &=\eta_L(v).
\end{align}
Thus, $L$ is a union of equivalence classes of $\sim^t$.
Hence, it follows that $S(L)\in\bfM\mathbf{Nil}$.

\medskip
\noindent
	$(6)\Longrightarrow(7)$:\quad 
Since $\calL^1[\calB_{[<]}, \Sigma]$ is closed under Boolean operations, it suffices to observe that each disjoint monomial $L=\Gamma^*a_0\dots\Gamma^*a_{n-1}\Gamma^*$, where $a_0,\dots, a_{n-1}\in\Sigma\setminus\Gamma$ and $\Gamma\subseteq\Sigma$, is expressed by a formula in $\calL^1[\calB_{[<]}, \Sigma]$. 
This is given by Proposition~\ref{lem:dis-mono-logic}.

\end{proof}

\subsubsection{Corollaries Relating $\bfX$ with Other Complexity Classes}\label{S Corollaries}

The following corollary states that the problem $\Occ_t$ can be seen
as a complexity bottleneck when improving the circuit complexity
of regular languages associated with $\bfX$.
\begin{corollary}
	Assume that for every fixed $t \in \bbN_{>0}$, the problem $\Occ_t$ (Problem~\ref{Pro Occ}) 
	were computable by circuit families of size $O(\log n)$ and
	depth $O(1)$. Then every regular language associated
	with $\bfX$ is in $\SIZEDEPTH(O(\log n),O(1))$.
	In particular the regular languages in $\SIZE(o(n))$
	and in $\SIZEDEPTH(O(\log n),O(1))$ coincide.
\end{corollary}
\begin{proof}

 By Item (8) of Theorem (\ref{T log main}), every regular language in $\SIZE(o(n))$ is definable by a formula in  $\calL^{1}[\calB_{[<,+\omega]}, \Sigma, <, +\omega, \mathrm{mod}, \min, \max]$. 
We repeat the proof of the direction $(9) \implies (1)$ of Theorem (\ref{T log main}) and it suffices to consider bounded quantifiers defined by fixed quantifier-free formulas over $[<, +\omega]$. Since the operation of evaluating the order relation and adding a fixed constant is computed by constant depth circuits of logarithmic size, the complexity of evaluating $\chi_Q$ on a fixed number of tuples is bounded by the complexity of computing a fixed number of indices. Hence if for every fixed $t \in \bbN_{>0}$, the problem $\Occ_t$ were computable by circuit families of size $O(\log n)$ and depth $O(1)$ then $\chi_Q$ can be evaluated in logarithmic size and constant depth. Hence every regular language with sublinear circuit complexity, i.e., those with syntactic stamp in $\bfX$, would be computable in constant depth and logarithmic size.
\end{proof}

The pseudovarieties of finite semigroups $\bfD\bfA$ and $\bfL\bfD\bfA$ are defined respectively by the identities
$(xy)^\omega x(xy)^\omega=
(xy)^\omega$ and 
$(z^\omega x z^\omega y z^\omega)^\omega
  \,x\,
  (z^\omega x z^\omega y z^\omega)^\omega
  =
  (z^\omega x z^\omega y z^\omega)^\omega
$.
We remark that $\bfD\bfA$ is strictly contained in $\bfL\bfD\bfA$.

\begin{samepage}
	\begin{proposition}
    \label{E prop QDA}
	Let $\{x_i,y_i\mid i\in\bbN\} \subseteq V$ be a set of variables.
		Let $\varphi$ be a stamp that 
		satisfies the identities
\begin{align}
&(x_0 \cdots x_{m-1})^\omega
  p_0 \cdots p_{i-1} x_i p_{i+1} \cdots p_{m-1}
  (y_0 \cdots y_{m-1})^\omega \notag\\
={}&(x_0 \cdots x_{m-1})^\omega
  p_0 \cdots p_{i-1} y_i p_{i+1} \cdots p_{m-1}
  (y_0 \cdots y_{m-1})^\omega
	\label{eq QDA}
\end{align}
		for all $m\in \bbN_{>0}$, $i\in [0,m-1]$ and 
		$p_j\in\{x_j,y_j\}$ for all $j\in[0,m-1]\setminus\{i\}$.
Then $\varphi\in\bfQ\bfD\bfA$.
\end{proposition}
\end{samepage}
	\begin{proof}
		Let $\varphi:\Sigma^+\to S$ be a stamp with stability index $s\in\bbN_{>0}$
		that satisfies the identities stated in the proposition.
		We show that $\stab(\varphi)\in\bfD\bfA$.
		Let $x,y\in\stab(\varphi)$ and  $e=(xy)^\omega\in\stab(\varphi)$.
		Fix words $u,v,w\in\Sigma^s$ such that
		\begin{align}
			\varphi(u)=x,\quad \varphi(v)=y,\quad \text{ and }\quad
			\varphi(w)=e\quad.
		\end{align}
		By assumption, the following identity holds, where
$m=2$ and $i=0$:
		\begin{align}
			(x_0x_1)^\omega x_0y_1(y_0y_1)^\omega=
			(x_0x_1)^\omega y_0y_1(y_0y_1)^\omega
		\end{align}
 When applying this for the substitution
		$\sigma:\{x_0,x_1,y_0,y_1\}\to\Sigma^s$, where $\sigma(x_0)=u$,
		$\sigma(x_1)=v$ and $\sigma(y_0)=\sigma(y_1)=w$
		we obtain
		\begin{align}
			(xy)^\omega xe(xy)^\omega &= (xy)^\omega ee(ee)^\omega.
		\end{align}
		As $(xy)^\omega=e$ and $e$ is an idempotent the
		latter simplifies to 
			$exe=e$,
		thus
			$(xy)^\omega x(xy)^\omega = (xy)^\omega$.
	\end{proof}

Since every identity of Proposition~\ref{E prop QDA} is also an 
identity of $\bfX$ the following can be concluded.

\begingroup
\renewcommand{\themainlemma}{\getrefnumber{intro:QDA}}
\begin{makethm}{mainlemma}{L X in QDA}
\unexpanded{
We have $\bfX\subsetneq\QDA$.
For the language $L_2= (a+b)^*c a c (a+b)^*$ from Example~\ref{E not in X} we have 
		$\eta_{L_2}\in\QDA\setminus\bfX$.
		}
\end{makethm}
\endgroup
	\begin{proof}
		In Example~\ref{E not in X} we show that $\eta_{L_2}\not\in\bfX$.
		Since the regular languages whose syntactic stamp is
		in $\bfQ\bfD\bfA$ coincide with
the ones definable in $\mathrm{FO}^2[\Sigma,<,\bmod]$ by 
		Theorem 6 in~\cite{Papermanfo2}
		it suffices to define $L_2$ in 
$\mathrm{FO}^2[\Sigma,<,\bmod]$.
The formula
\[
B(y) =
\bigl(\exists x\,(P_c(x)\land x<y)\bigr)
\land
\bigl(\exists x\,(P_c(x)\land y<x)\bigr)
\]
expresses that $y$ lies strictly between two occurrences of $c$.
The sentence 
\[
	\varphi_{L_2} =
\exists x\Bigl(
    P_a(x)\land
    \forall y\bigl(B(y)\leftrightarrow y=x\bigr)
\Bigr)
\]
defining $L_2$
expresses that there is exactly one position between occurrences of
$c$ and that position carries an $a$.
	\end{proof}

\begin{samepage}
\begingroup
\renewcommand{\themaincorollary}{\getrefnumber{intro:Sigma}}
\begin{makethm}{maincorollary}{C polylog QDA}
\unexpanded{
    Let $L$ be a regular language in $\SIZE(o(n))$. Then its syntactic morphism belongs to $\mathbf{QDA}$, and consequently, $L$ is definable in $\rmFO^2[\Sigma,<,\bmod]$, as well as in both the $\Sigma_2$ and $\Pi_2$ fragments of $\rmFO[\Sigma,<,\bmod]$. 
	}
\end{makethm}
\endgroup

\end{samepage}
\begin{proof}
	Let $L$ be a regular language in $\SIZE(o(n))$.
	Then $\eta_L$ is in $\bfX$ by Points (3) and (4) of
	Theorem~\ref{T log main}.
	By Lemma~\ref{L X in QDA} we have that $\eta_L$ is in $\QDA$.
	By Theorem 6 in~\cite{Papermanfo2}, $L$ is definable
	in $\rmFO^2[\Sigma, <,\bmod]$.
	Since $\rmFO^2[\Sigma, <,\bmod]$ coincides with the intersection 
	of the 
$\Sigma_2$ fragment and the $\Pi_2$ fragment
	of $\rmFO[\Sigma, <,\bmod]$
	by Corollary 5.10 in \cite{KW15}, the corollary follows.
\end{proof}

\begin{samepage}
\begingroup
\renewcommand{\themaincorollary}{\getrefnumber{intro:WLAC}}
\begin{makethm}{maincorollary}{C polylog WLAC0}
\unexpanded{
	Every regular language in $\SIZE(o(n))$ belongs to $\WLACO$.
}
\end{makethm}
\endgroup
\end{samepage}
\begin{proof}
	Let $L$ be a regular language in $\SIZE(o(n))$. Then
	$\eta_L\in\bfX\subseteq \bfQ\bfD\bfA$ by Points (3) and (4) of 
	Theorem~\ref{T log main} and Lemma~\ref{L X in QDA}.
	The corollary follows since $\bfQ\bfD\bfA$ is contained in $\bfQ\bfL\bfD\bfA$
	and the regular languages associated with 
$\bfQ\bfL\bfD\bfA$ coincide with the ones in $\WLACO$
	by Theorem 11 in~\cite{CadilhacP22}.
\end{proof}

%!TEX root = main.tex

\section{From Algebra to Circuit Complexity Lower Bounds}\label{S lower}

This section is devoted to proving the following theorem.

\begin{samepage}
\begin{theorem}\label{T log lower}
Let $L$ be a regular language. If $\eta_L$ is not in $\bfX$,
	then $\opc(L)\in\OmegaHL(n)$.
\end{theorem}
\end{samepage}

In \S\,\ref{S Witness} we prove a non-membership witness lemma (Lemma~\ref{L s 1 dist}).
In \S\,\ref{S Lower n} we give the actual $\OmegaHL(n)$ circuit complexity
lower bound proof for any regular language not in $\bfX$.

\subsection{The $\bfX$ Non-Membership Witness Lemma}\label{S Witness}

The following lemma summarizes a necessary substitution property
obeyed by languages whose syntactic stamps are in $\bfX$.
	\begin{lemma}\label{L s 1 dist}
		If a stamp $\varphi:\Sigma^+\to S$ is not in $\bfX$, then 
		$uv[i\leftarrow (u)_i]w\not=_{\varphi}uv[i\leftarrow (w)_i]w$
		for some $u,v,w\in\Sigma^k$, where $k$ is a multiple
		of the stability index of $\varphi$,
		$i\in[0,k-1]$ and $u$ and $w$ are idempotents.
	\end{lemma}

	This section is devoted to proving Lemma~\ref{L s 1 dist}.

\begin{definition}\label{D kl dist}
	Let $V=\{x_i,y_i,p_i\mid i\in\N\}$ denote the set of variables occurring
	in the identities for $\bfX$.
	We say a stamp $\varphi:\Sigma^+\to S$ 
	{\em is $(k,\ell)$-distinguishable} if
	there exists an identity $t=t'$ in $\bfX_k$ and some 
	$\sigma:V\to\Sigma^\ell$ such that
	$\varphi(\sigma(t))\not=\varphi(\sigma(t'))$.
\end{definition}

\begin{definition}
Let $x=x_0\ldots x_{\ell-1},y=y_0\ldots y_{\ell-1} \in \Sigma^+$ be two words 
	of length $\ell$. The \emph{$i$-splice} of the pair $(x,y)$ for $i\in [0,\ell]$ is the word 
    \begin{align}
        \alpha_i(x)\omega_{\ell-i}(y)=x_0\ldots x_{i-1} y_{i}\ldots y_{\ell-1}.
    \end{align}  
 Let $z_i$ denote the $i$-splice of $(x,y)$. Clearly, $z_0=y$ and $z_\ell=x$. 
	Moreover,
 $z_{i+1}=z_i[i\leftarrow x_{i}]$.
\end{definition}
\begin{lemma}
\label{lem:splice}
	Let $\varphi:\Sigma^+ \to S$ be a stamp. Let $\ell\in\bbN_{>0}$ 
	and $x,y \in \Sigma^\ell$ be two words of length $\ell$. 
	For all $u,v\in \Sigma^*$, if $uxv\neq_{\varphi} uyv$, then 
	$uz_iv\neq_\varphi uz_{i+1}v$ for some $i\in[0,\ell-1]$, 
	where $z_0,\ldots, z_\ell$ represent the splices of $(x,y)$.
\end{lemma}
\begin{proof}
Assume that $u,v\in \Sigma^*$.
    Clearly if $uz_iv=_\varphi uz_{i+1}v$ for all $i\in[0,\ell-1]$, then 
	$uyv=_\varphi uz_0v=_\varphi uz_1v=_\varphi\ldots =_\varphi uz_\ell v
	=_\varphi uxv$. A contradiction. Therefore, there is some 
	$i\in[0,\ell-1]$ such that $uz_iv\neq_\varphi uz_{i+1}v$.
\end{proof}

\begin{lemma}\label{L k 1 dist}
If a stamp $\varphi$ is not in $\bfX$, then $\varphi$ is $(k,1)$-distinguishable 
	for some $k\in\bbN_{>0}$.
	\end{lemma}
\begin{proof}
    Let $\varphi: \Sigma^+ \to S$ be a stamp that is not in $\bfX$.
	Hence 
	there exists some $r\in\bbN_{>0}$ such that $\varphi$ is not in $\bfX_r$,
	where $r\in\bbN_{>0}$ is chosen minimal with this property.
	Thus, $\varphi$ is $(r,\ell)$-distinguishable for some $\ell\in\bbN_{>0}$.
	In case $\ell=1$ we are done, so let us assume $\ell>1$.
		By minimality of $r$ there exists $\sigma:V\to\Sigma^\ell$
		such that 
		$\varphi(\sigma(t_1))\neq\varphi(\sigma(t_1'))$ 
		for some $i_0\in[0,r-1]$, where 
	\begin{align}
		t_1&=(x_0\dots x_{r-1})^\omega 
		p_0\dots p_{i_0-1}x_{i_0} p_{i_0+1}\dots p_{r-1}	
		(y_0\dots y_{r-1})^\omega,\\
		t_1'&=(x_0\dots x_{r-1})^\omega 
	p_0\dots p_{i_0-1}y_{i_0} p_{i_0+1}\dots p_{r-1}	
		(y_0\dots y_{r-1})^\omega.
		\label{eq unsat0}
	\end{align}
For each $i \in [0,r-1]$ let $u_i,q_i,v_i \in \Sigma^\ell$ be the words
	such that $u_i = \sigma(x_i)$, $q_i = \sigma(p_i)$, and
	$v_i = \sigma(y_i)$. Let 
	$u=u_0\cdots u_{r-1}$ and $ v=v_0\cdots v_{r-1}$. 
Let $q'=q_0\cdots q_{i_0-1}$ and $q''=q_{i_0+1}\cdots q_{r-1}$ and 
	let $n\in\bbN_{>0}$ be an exponent of $S$. We have
\begin{align}
    u^nq'u_{i_0}q''v^n \neq_\varphi u^nq'v_{i_0}q''v^n.
\end{align}
By \Cref{lem:splice}, there exists some $j\in [0,\ell-1]$ such that 
\begin{align}
    u^nq'z_jq''v^n \neq_\varphi u^nq'z_{j+1}q''v^n,
\end{align}
where $z_j$ denotes the $j$-splice of $(u_{i_0},v_{i_0})$. 
As a consequence, notice for the identity $t_2=t_2'$, where
\begin{align}
	t_2&=(x_0\dots x_{r\ell-1})^\omega  p_0\dots p_{i-1}x_ip_{i+1}\dots
	p_{r\ell-1}(y_{0}\dots y_{r\ell-1}) ^\omega, \\
	t_2'&= (x_0\dots x_{r\ell-1})^\omega 
p_0\dots p_{i-1}y_ip_{i+1}\dots p_{r\ell-1}
	(y_0\dots y_{r\ell-1})^\omega,
\end{align}
	we have $\varphi(\sigma'(t_2))\not=\varphi(\sigma'(t_2'))$
	for all substitutions $\sigma':V \to \Sigma$ satisfying
\begin{align}
	\sigma'(x_0\ldots x_{r\ell-1})&=u\\
	\sigma'(p_0\ldots p_{r\ell-1})&=q'z_jq''\\
	\sigma'(y_0\ldots y_{r\ell-1})&=v
\end{align}
at the position $i=i_0\ell+j$.
\end{proof}

We are now ready to prove Lemma~\ref{L s 1 dist}.
Let $\varphi:\Sigma^+\to S$ be a stamp  that is not in $\bfX$.
Let $s$ be the stability index of $\varphi$. 
	Then $\varphi$ is $(k,1)$-distinguishable for some 
	$k\in\bbN_{>0}$ by Lemma~\ref{L k 1 dist}.
	Hence there exist $\sigma:V\to\Sigma$
				such that 
		$\varphi(\sigma(t_1))\not=
		\varphi(\sigma(t_1'))$,
		where for some $i_0\in[0,k-1]$ 
		we have
	\begin{align}
		t_1&=(x_0\dots x_{k-1})^\omega 
		p_{0}\dots p_{i_0-1}
		x_{i_0}p_{i_0+1}\dots p_{k-1} 
		(y_0\dots y_{k-1})^\omega\\
		t_1'&=(x_0\dots x_{k-1})^\omega 
		p_{0}\dots p_{i_0-1}
		y_{i_0}p_{i_0+1}\dots p_{k-1} 
		 (y_0\dots y_{k-1})^\omega.
		\label{eq unsat1}
	\end{align}
Let $x,y,z\in \Sigma^k$ be the words such that 
\begin{align}
	x&=\sigma(x_0\dots x_{k-1})\\
	y&=\sigma(p_0\dots p_{k-1})\\
	z&=\sigma(y_0\dots y_{k-1})
\end{align}
Let $n$ be an exponent of $S$.
Then we have 
\begin{align}
	x^ny[i_0\leftarrow (x)_{i_0}]z^n \neq_\varphi  x^ny[i_0\leftarrow (z)_{i_0}]z^n. \label{E xy'z}
\end{align}
Let $m$ be a positive common multiple of $n$ and $s$. 
We define the following words:
\begin{align}
	u&=x^m \label{E u}\\
	v&=yz^{m-1} \label{E v}\\
	w&=z^m\label{E w}
\end{align}
Note that $|u|=|v|=|w|=mk$ and $u$ and $w$ are idempotents since $n$ divides $m$.
To finish the proof it suffices to show
\begin{align}
	uv[i_0\leftarrow (u)_{i_0}]w\neq_\varphi uv[i_0\leftarrow (w)_{i_0}]w.
\end{align}
Indeed, if equality were to hold, then, by appending $z$ 
we obtain from (\ref{E u})-(\ref{E w})
\begin{align}
	x^my[i_0\leftarrow (x)_{i_0}]z^{2m}=_\varphi 
	x^my[i_0\leftarrow (z)_{i_0}]z^{2m}
\end{align}
clearly contradicting (\ref{E xy'z}) since $x^n=_\varphi x^m$
and $z^n=_\varphi z^{m}=_\varphi z^{2m}$.
	This concludes the proof of Lemma~\ref{L s 1 dist}.
Observing that $mk$ is a multiple of the stability index $s$  
we obtain the following corollary.
	\begin{corollary}\label{C k1 dist}
		If a stamp $\varphi$ is not in $\bfX$, then
$\varphi$ is $(k,1)$-distinguishable
		for some multiple $k$ of the stability index of $\varphi$.
	\end{corollary}

\subsection{The Linear Lower Bound}\label{S Lower n}

We will be working with the following central notion,
where all circuits are assumed to have precisely one output gate.
\begin{definition}
	An $n$-circuit $C$ over the alphabet $\Sigma$ 
	{\em separates a pair of words $(x,y)\in\Sigma^n\times\Sigma^n$} 
	if $x\in L(C)$ and $y\not\in L(C)$.
	Analogously, a gate $g$ of $C$ separates $(x,y)$ if $\eval(x,g)=1$
	and $\eval(y,g)=0$.
\end{definition}

For the rest of this section fix a regular language $L$ such that
$\eta_L$ is not in $\bfX$.
By Lemma~\ref{L s 1 dist} there exist words $u=u_0\dots u_{k-1}$, 
$v=v_0\dots v_{k-1}$, and 
$w=w_0\dots w_{k-1}$, where $k\in\bbN_{>0}$ is a multiple of
the stability index of $\eta_L$, $j_0\in[0,k-1]$,
$u$ and $w$ are idempotents and moreover
\begin{align}
	uv[j_0\leftarrow u_{j_0}]w\ \not=_{\eta_L} uv[j_0\leftarrow w_{j_0}]w,
\end{align}
so in particular $u_{j_0}\not=w_{j_0}$.
In the following let $v'=v[j_0\leftarrow u_{j_0}]$ and 
$v''=v[j_0\leftarrow w_{j_0}]$,
thus 
\begin{align}
	uv'w\ \not=_{\eta_L} uv''w,\ \quad
	(v')_{j_0}=u_{j_0},\quad (v'')_{j_0}=w_{j_0}\quad.
	\label{E uv'w}
\end{align}
In the following let $K=\eta_L^{-1}(\eta_L(uv'w))$.
By Lemma~\ref{L reduce} we have $K\reduce L$,
so $\opc(K)\in\OmegaHL(n)$
implies $\opc(L)\in\OmegaHL(n)$ by Theorem~\ref{T asymptotics}.
Thus it will be sufficient to prove that $\opc(K)\in\OmegaHL(n)$.
To this end, for each $n\geq 2$ and each $i\in[1,n-1]$ we define the following 
	words from $\Sigma^{k(n+1)}$, where the claimed (non-)membership in $K$
	follows from (\ref{E uv'w}), $i,n-i\geq 1$ and the fact that 
	$u,w$ are idempotents:
	\begin{align}
		x_{n,i}&=u^iv'w^{n-i}\in K\\
		y_{n,i}&=u^iv''w^{n-i}\not\in K
	\end{align}
For all $i\in[1,n-1]$ let us define
\begin{align}
	p_i=ik+j_0.
\end{align}
\begin{samepage}
\begin{remark}\label{R sep xy}
	The words $x_{n,i}$ and $y_{n,i}$ only differ in position
	$p_i$, namely $(x_{n,i})_{p_i}=(u)_{j_0}$ and 
	$(y_{n,i})_{p_i}=(w)_{j_0}$.
\end{remark}
\end{samepage}

The following proposition implies $\opc(K)\in\OmegaHL(n)$.
\begin{proposition}\label{P lower}
For each $n\geq 2$ each $k(n+1)$-circuit $C$ that 
	separates all of the pairs
	$(x_{n,1},y_{n,1}),\dots,(x_{n,n-1},y_{n,n-1})$,
	satisfies $|C|\geq n-1$.
\end{proposition}
Indeed, since any $k(n+1)$-circuit computing $K\cap\Sigma^{k(n+1)}$ must clearly 
separate 
all of the pairs $(x_{n,1},y_{n,1}),\dots,(x_{n,n-1},y_{n,n-1})$ by 
Proposition~\ref{P lower} we have:
\begin{align}
	\forall n\geq 2:\opc(K)(k(n+1))\geq n-1 &\quad\Longrightarrow\quad
	\exists^\infty n\in\bbN_{>0}:\opc(K)(n)\geq n/k-2
	& \\
&\quad\Longrightarrow\quad
	\exists^\infty n\in\bbN_{>0}:\opc(K)(n)\geq n/(k+1)
	& \\
	&\quad\Longrightarrow\quad \opc(K)\in\OmegaHL(n)
\end{align}

We devote the rest of this section to the proof of Proposition~\ref{P lower}.
For this let us fix some $n\geq 2$ and some $k(n+1)$-circuit 
$C=(G,\prec,\lambda,g_{out})$
that separates all of the pairs $(x_{n,1},y_{n,1}),\dots,(x_{n,n-1},y_{n,n-1})$.

\begin{samepage}
\begin{definition}
For each set of gates $H\subseteq G$ we denote by 
	$\pre(H)=\{g\in G\mid \exists h\in H, g\prec h\}$ 
the set of (immediate) {\em predecessors} of some gate in $H$.
When $H$ is a singleton we just write $\pre(h)$.
\end{definition}
\end{samepage}
\begin{definition}
	In the following for any $H\subseteq G$ let $H_1$ be an abbreviation
	for $H\cap G_1$ and $H_0$ be an abbreviation for $H\cap G_0$.
	Let $I(g)=\{i\in[1,n-1]\mid g \text{ separates }(x_{n,i},y_{n,i})\}$
	for each $g\in G$ and $I(H)=\bigcup_{h\in H}I(h)$ for each $H\subseteq G$.
	We define
	\begin{align}
		J(g)=I(g)\setminus I(\pre(g)_1).
	\end{align}
\end{definition}

\begin{remark}\label{R output gate}
	Note that if $g$ is the unique output gate of $C$, then
$I(g)=\{1,\dots,n-1\}$.
\end{remark}

	\begin{samepage}
		\begin{lemma}\label{L Lemma A}
Let $g\in G_1$ be a non-input gate.
	For each $i\in J(g)$
	there exists $g'\in\pre(g)_0$ with
	$\lambda(g')=(p_i,\Gamma)$
	for some $\Gamma\subseteq\Sigma$
	such that $i\in I(g')$.
\end{lemma}
	\end{samepage}
\begin{proof}
	Let $g\in G_1$ be a non-input gate and let 
	$i\in J(g)$.

		Let us first assume that $g$ is an $\vee$-gate.
		Let $g'\in\pre(g)_1$ be arbitrary. We must have
		$\eval(x_{n,i},g')=0$ or $\eval(y_{n,i},g')=1$
		since $i\not\in I(g')$ by assumption.
		Now the case $\eval(y_{n,i},g')=1$
			 can be ruled out since that would imply
			 $\eval(y_{n,i},g)=1$ as 
			 $g$ is an $\vee$-gate,
			contradicting $i\in I(g)$.
			Hence we must have $\eval(x_{n,i},g')=0$
			for all $g'\in\pre(g)_1$.
		As $g$ is an $\vee$-gate
			satisfying $\eval(x_{n,i},g)=1$ and
			$\eval(y_{n,i},g)=0$, and moreover
			$x_{n,i}$ and $y_{n,i}$ only differ in position
			$p_i$ by Remark~\ref{R sep xy},
	there must exist an input gate $g_0\in\pre(g)_0$
	such that $\lambda(g_0)=(p_i,\Gamma)$ for some $\Gamma\subseteq\Sigma$
			satisfying $\eval(x_{n,i},g_0)=1$ and 
			$\eval(y_{n,i},g_0)=0$.
			Hence $i\in I(g_0)$ as required.

		Let us now assume that $g$ is an $\wedge$-gate.
		Let $g'\in\pre(g)_1$ be arbitrary. Again we have
		$\eval(x_{n,i},g')=0$ or $\eval(y_{n,i},g')=1$
		since $i\not\in I(g')$ by assumption.
		Now the case
			$\eval(x_{n,i},g')=0$ can be ruled out
			since that would imply
			$\eval(x_{n,i},g)=0$ as $g$ is an $\wedge$-gate,
			contradicting $i\in I(g)$.
			Hence we must have $\eval(y_{n,i},g')=1$
			for all $g'\in\pre(g)_1$.
		As $g$ is an $\wedge$-gate
			satisfying both $\eval(x_{n,i},g)=1$ and
			$\eval(y_{n,i},g)=0$, and moreover
			$x_{n,i}$ and $y_{n,i}$ only differ in position
			$ki+j_0$ by Remark~\ref{R sep xy},
	there must exist an input gate $g_0\in\pre(g)_0$
	such that $\lambda(g_0)=(ik+j_0,\Gamma)$ for some $\Gamma\subseteq\Sigma$
			satisfying $\eval(x_{n,i},g_0)=1$ and 
			$\eval(y_{n,i},g_0)=0$.
			Hence $i\in I(g_0)$ as required.
		\end{proof}

		The next lemma upper bounds the number of pairs
		$(x_{n,i},y_{n,i})$ 
		a non-input gate can separate that are not already
		separated by its non-input predecessors.
\begin{samepage}
	\begin{lemma}\label{L Lemma B}
		$|J(g)|\leq 1$ for all $g\in G_1$.
	\end{lemma}
\end{samepage}
		\begin{proof}
			For the sake of contradiction assume $|J(g)|\geq 2$
			so assume some $i_1,i_2\in J(g)$ with $i_1<i_2$.
			By Lemma~\ref{L Lemma A} there exist 
			$g_{1}\in\pre(g)_0$ and $g_2\in\pre(g)_0$
			with $\lambda(g_1)=(p_{i_1},\Gamma_1)$
			and $\lambda(g_2)=(p_{i_2},\Gamma_2)$
			for some $\Gamma_1,\Gamma_2\subseteq\Sigma$
			and moreover $i_1\in I(g_1)$ and
			$i_2\in I(g_2)$.

			Let us treat the case when $g$ is an $\vee$-gate.
			Since $i_1\in I(g_1)$ we have $\eval(x_{n,i_1},g_1)=1$.
			Due to $i_1<i_2$ note that $y_{n,i_2}$ has letter 
			$(u)_{j_0}$ at position $p_{i_1}$,
			and so does $x_{n,i_1}$ by Remark~\ref{R sep xy}.
			Hence, $\eval(y_{n,i_2},g_1)=1$.
			Since $g$ is an $\vee$-gate and $g_1\in\pre(g)_0$
			we have $\eval(y_{n,i_2},g)=1$, so
			$g$ does not separate $i_2$, contradicting
			our assumption that $i_2\in J(g)$.

			Let us now treat the case when $g$ is an $\wedge$-gate.
			Since $i_1<i_2$ note that
			$x_{n,i_1}$ has letter $(w)_{j_0}$ at position
			$p_{i_2}$, and so does $y_{n,i_2}$ by Remark~\ref{R sep xy}.
			Hence, $\eval(x_{n,i_1},g_2)=0$.
			Since $g$ is an $\wedge$-gate and $g_2\in\pre(g)_0$
			we have $\eval(x_{n,i_1},g)=0$,
			so $g$ does not separate $i_1$, contradicting
			our assumption that $i_1\in J(g)$.
			\end{proof}

	We are now ready to prove Proposition~\ref{P lower}.
	Let $g_1,\dots,g_s$ be an enumeration of the non-input gates $G_1$
	of $C$
	in topological order, i.e. $g_i\prec^+ g_j$ implies
	$i<j$ for all $i,j\in[1,s]$. Note that in particular $|C|=s$.

	For each $r\in[0,s]$ let
	\begin{align}
		A_r=\bigcup_{\ell=1}^r I(g_\ell),
	\end{align}
	so $A_0=\emptyset$.
	Thus $A_r$ contains precisely those indices
	that are separated by one of the gates $g_1,\dots,g_r$.
	Note that if $i\in A_r\setminus A_{r-1}$,
	then $i\in I(g_r)\setminus I(\pre(g_r)_1)$,
	hence $i\in J(g_r)$.
	Thus, $A_r\setminus A_{r-1}\subseteq J(g_r)$ for
	each $r\in[1,s]$. Consequently, by Lemma~\ref{L Lemma B}
	we have
	\begin{align}
		|A_r|\leq |A_{r-1}|+1.
	\end{align}

	By induction one proves $|A_r|\leq r$.

	Remark~\ref{R output gate} implies $|A_s|=n-1$,
	thus we obtain
	\begin{align}
		n-1=|A_s|\leq s =|C|,
	\end{align}
	which concludes the proof of Proposition~\ref{P lower}.

%!TEX root = main.tex

\section{From Algebra to Congruence}\label{S AlgCong}

The section is devoted to the proof of the following proposition. 

\begin{proposition}\label{P log tech}
Let $\varphi$ be a stamp in $\bfX$ with stability index $s$. 
	Let $N=|\stab(\varphi)|$ and $k=4s (N+1)!$. 
	For each $u,v \in \Sigma^{+}$ we have that 
	if $u \simeq_{k} v$, then $\varphi(u)=\varphi(v)$. 
\end{proposition}

We immediately deduce the following corollary and the direction 
$(4) \Longrightarrow (5)$ of the Sublinear-Size Characterization 
Theorem~(Theorem~\ref{T log main}).

\begin{corollary}\label{C log main}
If the syntactic morphism $\eta_L$ of a regular language
$L\subseteq\Sigma^+$ is in $\bfX$, then 
	there exists a $k\in\bbN_{>0}$ such that $L$ is a union of 
$\simeq_k$-classes.
\end{corollary}
\begin{proof}
Assume the syntactic morphism 
$\eta_L:\Sigma^+\rightarrow S(L)$ of $L$ is in $\bfX$ with stability index $s$. Let $N=|\stab(\eta_L)|$. By Proposition~\ref{P log tech} 
	for $k=4s(N+1)!$ for all $u,v\in\Sigma^+$ we have that
	if $u\simeq_k v$, then $\eta_L(u)=\eta_L(v)$. 
		Thus $\eta_L^{-1}(x)$ is a union of $\simeq_k$-classes 
		for each $x\in S(L)$. 
		Hence $L=\eta_L^{-1}(\eta_L(L))$ itself is a 
		union of $\simeq_k$-classes.
\end{proof}

The rest of this section is devoted to proving  \Cref{P log tech}.

We have the following corollary of the exchange principle that gives a sufficient condition for two words to have the same morphic image in an idempotent context.  

\begin{lemmarep}
\label{lem:cor-exchange-principle}
Let $\varphi: \Sigma^+ \to S$ be a morphism in $\bfX$ with stability index $s$. 
	Let $u\in \Sigma^{s\bbN_{>0}}$ be an idempotent. 
	Assume that $v,v' \in \Sigma^{s\bbN_{>0}}$ satisfy 
	$|v|=|v'|$ and 
	$\{(v)_i,(v')_i\} \subseteq \cont_{i,s}(u) $ for each 
	$i\in H(v,v') \subseteq [0,|v|-1]$.
Then, we have
\begin{align}
	uvu =_{\varphi} uv'u.
\end{align}
\end{lemmarep}
\begin{appendixproof}
	Let $u \in \Sigma^{s\bbN_{>0}}$ be an idempotent and assume that 
	$v,v'\in \Sigma^{s\bbN_{>0}}$ are as in the premise. 
    We prove by induction on the size of the set $H(v,v')$ that
     $uvu =_{\varphi} uv'u$. 
	Clearly, if $|H(v,v')|=0$ then $v=v'$ and the claim is trivial. For the induction step, assume that $|H(v,v')| =  n+1>0$. 
	Let $i\in H(v,v')$. By assumption, there exist positions $j,k\in [0,|u|-1]$ such that $j\equiv_s k \equiv_s i$, $(v)_i = (u)_j$ and $(v')_i = (u)_k$. Therefore,
     \begin{align}
	     uvu &= uv[i \leftarrow (u)_j]u&&(\text{Since $(v)_i=(u)_j$})\\
         &=_\varphi uv[i\leftarrow(u)_k] u&&(\text{By exchange principle})\\
         &=_\varphi uv'u.&&(\text{Since $|H(v[i\leftarrow(u)_k], v')|=n$ and by IH})
     \end{align}
\end{appendixproof}

The following observation follows from the pigeonhole principle.

\begin{lemmarep}
\label{lem:prefidem1}
Let $S$ be a finite semigroup and let $N = |S|$. 
	For every sequence $s_0, \dots, s_{N}$ of $N+1$ elements of $S$, 
	there exist
	indices $i,j \in [0,N]$, $i<j$  such that for all $\ell\in\bbN$
	we have
	\begin{align}
    s_0\ldots s_i &= s_0\ldots s_i \cdot (s_{i+1}\cdots s_j)^\ell.
\end{align}
\end{lemmarep}
\begin{appendixproof}
Let  $s_0, \dots, s_{N}$ be $N+1$ elements of $S$. 
	Applying the pigeonhole principle to the $N+1$ prefix
	products $\{s_0, s_0s_1, \ldots, s_0\cdots s_N\} \subseteq S$, there exist $i,j \in [0,N]$, $i<j$ such that 
$s_0\ldots s_i = s_0\ldots s_j$. Therefore, for all $\ell\in \bbN$,
\begin{align}
    s_0\ldots s_i &= s_0\ldots s_i \cdot (s_{i+1}\cdots s_j) \\
    &=s_0\ldots s_i \cdot (s_{i+1}\cdots s_j)^\ell.
\end{align}
\end{appendixproof}

As an immediate consequence we obtain the following lemma which we use later.

\begin{samepage}
\begin{lemma}[Proposition~6.34 in \cite{Pin2025Mathematical}]
\label{lem:prefidem}
Let $S$ be a finite semigroup and let $n = |S|$. 
	For every sequence $s_0, \dots, s_{n-1}$ of $n$ elements of $S$, 
	there exists
	an index $i \in [0,n-1]$ and an idempotent $e \in E(S)$ such that 
	$s_0 \cdots s_i \, e \;=\; s_0 \cdots s_i$
	and dually, there exists an index $j \in[0,n-1]$ 
	and an idempotent $f \in E(S)$ such that
	$f \, s_j s_{j+1} \cdots s_{n-1} \;=\; s_j s_{j+1} \cdots s_{n-1}$.
\end{lemma}
\end{samepage}

Using Lemma~\ref{lem:prefidem1} we derive the following lemma.

\begin{samepage}
\begin{lemma}
\label{lem:pump0}
Let $\varphi:\Sigma^+\to S$ be a stamp with stability index $s$. 
Let $N=|\stab(\varphi)|$ and $k=s(N+1)!$. 
	Assume that $u\in \Sigma^+$ is a word of length $\geq s(N+1)$. Then there is 
	a factorization $u=u'u''u'''$, where 
	$u',u''\in\Sigma^{s\bbN_{>0}}$ and $u'''\in\Sigma^*$ 
	such that for each 
	$m\in\bbN_{>0}$ there is a word $v \in u'(u'')^+u'''$ that 
	satisfies the following
	\begin{enumerate}
		\item $u=_\varphi v$,
		\item $|v|=|u|+mk$, and
		\item $|\pi_{i,s}(v)|_a\geq |\pi_{i,s}(u)|_a$ for all $i\in[0,s-1]$
			and all $a\in\Sigma$.
	\end{enumerate}
\end{lemma}
\end{samepage}
\begin{proof}
	Let $u\in \Sigma^+$ be such that $|u|\geq s(N+1)$.
Let $u=u_0u_1\ldots u_{N}u_{N+1}$  be the factorization of $u$ such that
$|u_j|=s$ for each $j\in [0,N]$ and $u_{N+1}\in \Sigma^*$. 
	By applying Lemma~\ref{lem:prefidem1} to $\stab(\varphi)$
	there exist indices $i< j\in[0,N]$ such that for all
	$\ell\in\bbN$ we have
\begin{align}
u_0\ldots u_{j} &=_\varphi  u_0\ldots u_{i}(u_{i+1}\ldots u_{j})^{\ell}.
\end{align}

We define $u'=u_0\ldots u_{i}$,  $u''=u_{i+1}\ldots u_{j}$ and 
	$u'''=u_{j+1}\ldots u_{N}u_{N+1}$.
Clearly, $|u'|\equiv_s |u''|\equiv_s 0$. 
Moreover, 
	\begin{align}
		|u''|=|u_{i+1}\ldots u_{j}|=(j-i)s\label{E u''}.
	\end{align}

	Let $m\in\bbN_{>0}$. We define $r=m(N+1)!/(j-i)$ 
	which is a whole number since $(j-i)|(N+1)!$.
We define the word $v = u'(u'')^{r+1}u'''\in \Sigma^+$ as follows:
\begin{align}
v&=u_0\ldots u_{i}(u_{i+1}\ldots u_{j})^{r+1} u_{j+1}\ldots u_{N}u_{N+1}
\end{align}
	Clearly, $v=_\varphi u$, so Point~(1) follows.
	Moreover we have
\begin{align}
    |v|&=|u|+ r \cdot |u''|\\
	&= |u| + \frac{m(N+1)!}{(j-i)}\cdot (j-i)s&\text{(By (\ref{E u''}))}\\
    &= |u|+mk,
\end{align}
	so Point~(2) holds.
	Since $|u'|,|u''|\in s\bbN_{>0}$
	we have $|\pi_{i,s}(v)|_a=
	|\pi_{i,s}(u)|_a+r|\pi_{i,s}(u'')|_a\geq |\pi_{i,s}(u)|_a$
	for all $i\in[0,s-1]$ and all $a\in\Sigma$.
	Thus Point~(3) holds.
\end{proof}

The previous lemma can be applied to two words $u$ and $v$ simultaneously to obtain equivalent words of identical length, yielding the following corollary.

\begin{samepage}
\begin{lemma}
\label{lem:pump}
Let $\varphi:\Sigma^+\to S$ be a stamp with stability index $s$. 
Let $N=|\stab(\varphi)|$ and $k=s(N+1)!$. Assume that $u,v\in \Sigma^+$ are words such that $|u|\equiv_k^k |v|$. Then, there exist $u',v'\in\Sigma^+$ such that
	\begin{enumerate}
		\item $x=_\varphi x'$ for all $x\in\{u,v\}$,
		\item $|u|\equiv_s|u'|=|v'|\equiv_s|v|$, and
		\item $|\pi_{i,s}(x')|_a\geq |\pi_{i,s}(x)|_a$
		for all $i\in[0,s-1]$, $x\in\{u,v\}$ and $a\in\Sigma$.
	\end{enumerate}
\end{lemma}
\end{samepage}
\begin{proof}
  Let $u,v\in  \Sigma^+$ be words such that $|u|\equiv_k^k |v|$. 

	If $|u|=|v|$ we are done since we can take 
	$u'=u$ and $v'=v$ and the lemma is immediate. 

	Hence, let us assume that 
	$|u|,|v|\geq k$, $|u|\equiv_k |v|$ and $|u|\not=|v|$.
Further assume without loss of generality that $|u|>|v|\geq k$, 
	the other case being symmetric. 
	Since $|u|\equiv_k |v|$,  $|u|-|v|\equiv_k 0$. Let $q\in\bbN_{>0}$ 
	be such that $|u|-|v|=qk$.  
	Since $|v|\geq k=s(N+1)!\geq s(N+1)$ we can apply
	Lemma~\ref{lem:pump0}. It yields a factorization 
	$v=v_1v_2v_3 \in \Sigma^+$
	with $v_1,v_2\in\Sigma^{s\bbN_{>0}}$ and a word 
	$v' \in v_1(v_2)^+v_3$ such that 
	$v'=_\varphi v$, $|v'|=|v|+qk$
	and $|\pi_{i,s}(v')|_a\geq|\pi_{i,s}(v)|_a$ for all $i\in[0,s-1]$
	and all $a\in\Sigma$.
Since $|u|-|v'|=|u|-(|v|+qk)=|u|-|v|-qk=qk-qk=0$, the words $u'=u$ and $v'$ 
	satisfy the claim.
\end{proof}

 With the added assumption that $u \congC_K^{K,K} v$, where $K=2k$, the corollary above can be strengthened to yield two equivalent words of identical length that moreover have their difference set drawn from their frequent letters in the respective modular classes. 

 \begin{samepage}
\begin{lemma}\label{L redux substitution}
Let $\varphi:\Sigma^+\to S$ be a stamp with stability index $s$. 
Let $N=|\stab(\varphi)|$, $k=s(N+1)!$, and $K=2k$.
	Assume that $u,v \in \Sigma^{s\bbN_{>0}}$ are words
	such that $u \congC_K^{K,K} v$. 
	Then, there exist $u',v' \in \Sigma^{s\bbN_{>0}}$ such that the following holds.
\begin{enumerate}
\item $u=_\varphi u'$ and $v=_\varphi v'$,
\item $|u'|=|v'|$, 
\item $u'\congC_s^{k,k} v'$, and
\item $\rmR_{s}^{k}(u')=\rmR_{s}^{k}(v')$.
\end{enumerate}
\end{lemma}
 \end{samepage}
\begin{proof}
Let $u,v \in \Sigma^{s\bbN_{>0}}$ be such that $u \congC_K^{K,K} v$. 
	Hence $u\congC_k^{k,K} v$ by Points (2) and (3) of Lemma~\ref{L congC}.
	In particular we have $|\rmR_k^k(u)|=|\rmR_k^k(v)|$.

	Let us first consider the case when
	$\rmR_k^k(u)=\rmR_k^k(v)=\emptyset$. We apply the construction
	from the proof of Lemma~\ref{lem:pump} with parameter $s(N+1)!$.
	We obtain words $u',v'\in\Sigma^{s\bbN_{>0}}$ such that
$u=_\varphi u'$,$v=_\varphi v'$, and $|u'|=|v'|$,
which establishes Points~(1) and~(2).
	Since $s|k$ we have that $\rmR_k^k(u)=\rmR_k^k(v)=\emptyset$
	implies $\rmR_s^k(u)=\rmR_s^k(v)=\emptyset$.
Moreover, $|\pi_{j,s}(u')|_a\geq|\pi_{j,s}(u)|_a$ and
	$|\pi_{j,s}(v')|_a\geq|\pi_{j,s}(v)|_a$ for all $j\in[0,s-1]$
	and all $a\in\Sigma$
	by Point~(3) of Lemma~\ref{lem:pump}.
	By construction we have $\cont_{j,s}(x')=\cont_{j,s}(x)$ for all
	$x\in\{u,v\}$ and $j\in[0,s-1]$.
	We obtain $\rmR_s^k(u')=\rmR_s^k(u)=\rmR_s^k(v)=\rmR_s^k(v')=\emptyset$,
	which proves Point~(4).
	Thus, $u'\congC_s^{k,k}u$ and $v'\congC_s^{k,k}v$.
	By refinement of our assumption we have $u\congC_s^{k,k}v$,
	so we obtain $u'\congC_s^{k,k}v'$ by transitivity. This shows Point~(3).

	Let us therefore assume $\rmR_{k}^k(u),\rmR_{k}^k(v)\not=\emptyset$.
 Then there exist factorizations 
\begin{align}
	u&=u_0a_0u_1\dots a_{\ell-1} u_\ell,\label{E mu k u2}\\
	v&=v_0a_0v_1\dots a_{\ell-1} v_\ell,
\end{align}
	where $u_i,v_i\in \Sigma^*$ and moreover the letters $a_0,\ldots, a_{\ell-1}$ are precisely the rare letters, i.e., those letters
	appearing $<k$ times in their respective residue class 
	\textrm{mod} $k$.

	More precisely, for each $i\in[0,\ell-1]$ define
	\begin{align}
		p_i = i+\sum_{j=0}^{i}|u_j|\quad\text{ and }\quad
q_i = i+\sum_{j=0}^{i}|v_j|.
	\end{align}
Then
	\begin{align}
\rmR_k^k(u)=\{p_i\mid 0\le i<\ell\},
\quad\text{ and }\quad
\rmR_k^k(v)=\{q_i\mid 0\le i<\ell\},
	\end{align}
and thus $(u)_{p_i}=(v)_{q_i}=a_i$ for each $i\in[0,\ell-1]$.
	The order-preserving bijection $\beta$ that
	witnesses $u\congC_{k}^{k,K}v$ from 
	$\rmR_k^k(u)\cup\{0,|u|-1\}$
	to 
	$\rmR_k^k(v)\cup\{0,|v|-1\}$
	satisfies $\beta(0)=0$, $\beta(p_i)=q_i$ and $\beta(|u|-1)=|v|-1$.
	Note that $p_i-p_{i-1}=|u_i|+1$ and $q_i-q_{i-1}=|v_i|+1$
	for each $i\in[1,\ell-1]$.
	That is, $|u_i|+1\equiv_k^K |v_i|+1$ for each $i\in[1,\ell-1]$.
	Hence $|u_i|\equiv_k^{K-1}|v_i|$ and, due to $K-1=2k-1\geq k$,
	we obtain $|u_i|\equiv_k^k|v_i|$ for each $i\in[1,\ell-1]$.

	Moreover $|u_0|\equiv_k^K|v_0|$ 
	and $|u_\ell|\equiv_k^K|v_\ell|$,
	so $|u_0|\equiv_k^k|v_0|$ and
	so $|u_\ell|\equiv_k^k|v_\ell|$.

	We have thus shown $|u_i|\equiv_k^k|v_i|$ for all $i\in[0,\ell]$.

For each $i\in[0,\ell]$ there exists a pair of words $(u_i',v_i')$ 
such that either $u_i'=v_i'=\varepsilon$ in case $u_i=v_i=\varepsilon$,
or both $u_i'$ and $v_i'$ are non-empty and satisfy the
following conditions provided by Lemma~\ref{lem:pump}:
	\begin{enumerate}
		\item $u_i=_\varphi u_i'$,
			$v_i=_\varphi v_i'$.
		\item $|u_i|\equiv_s|u_i'|=|v_i'|\equiv_s|v_i|$, and
		\item $|\pi_{j,s}(u_i')|_a\geq|\pi_{j,s}(u_i)|_a$
			and 
			$|\pi_{j,s}(v_i')|_a\geq|\pi_{j,s}(v_i)|_a$
			 for all $j\in[0,s-1]$ and all $a\in\Sigma$.
	\end{enumerate}
	We define 
	\begin{align}
		u'&= u_0'a_0u_1'\dots a_{\ell-1}u_{\ell}'\\
		v'&= v_0'a_0v_1'\dots a_{\ell-1}v_{\ell}'.
	\end{align}
	Point (1) immediately implies Point (1) of the lemma
	and so does Point (2) imply Point (2) of the lemma.
	By construction we have $\cont_{j,s}(u_i')=\cont_{j,s}(u_i)$
	and $\cont_{j,s}(v_i')=\cont_{j,s}(v_i)$ for all $j\in[0,s-1]$,
	all $i\in[0,\ell]$ and all $a\in\Sigma$.
	Moreover, $|\pi_{j,s}(x')|_a\geq|\pi_{j,s}(x)|_a$
	for all $x\in\{u,v\}$ and $j\in[0,s-1]$.
	Every occurrence inside a block is frequent modulo $s$. 
	By construction, pumping adds occurrences 
	only to these frequent letter–residue pairs, 
	so counts below $k$ remain unchanged.
	Thus, $|\pi_{j,s}(u')|_a\equiv^k|\pi_{j,s}(u)|_a\equiv^k
	|\pi_{j,s}(v)|_a\equiv^k|\pi_{j,s}(v')|_a$ for all $j\in[0,s-1]$
	and all $a\in\Sigma$.
	Since $|u_i'|=|v_i'|$ for all $i\in[0,\ell]$ by construction, we
	obtain $\rmR_s^k(u')=\rmR_s^k(v')$.
	This proves Point~(4) of the lemma.

	Let us show Point~(3) of the lemma.
	The preceding threshold-count equivalences and $|u'|=|v'|$ imply
$u'\congC_s^k v'$. Moreover, the rare positions coincide and carry
the same separator letters. Therefore, the identity map on
\begin{align}
\rmR_s^k(u')\cup\{0,|u'|-1\}
\end{align}
witnesses the ordered condition. Hence $u'\congC_s^{k,k}v'$,
proving Point~(3).
\end{proof}

\begin{samepage}
	\begin{lemma}\label{L X in Y}
		Let $\varphi$ be a stamp in $\bfX$. Then $\stab(\varphi)$
		satisfies the following identities,
		where $e$ and $f$ denote idempotents:
		\begin{align}
exf=efxf=exef
		\end{align}
	\end{lemma}
\end{samepage}
	\begin{proof}
		Let $\varphi:\Sigma^+\to S$ be a stamp in $\bfX$
		and let $s$ be the stability index of $\varphi$.
		Let $x\in\stab(\varphi)$, $e,f\in E(\stab(\varphi))$
		and let $u,v,w\in\Sigma^s$ be words such that
		$\varphi(u)=e$, $\varphi(v)=x$, 
		and $\varphi(w)=f$. Let $n$ be an exponent of $S$. 
		We observe the following.
		\begin{align}
			uvw&=_\varphi 
			(u\underbracket[.5pt]u)^nv\underbracket[.5pt]w
		(w\underbracket[.5pt]w)^n\\
&=_\varphi  (u\underbracket[.5pt]u)^n v\underbracket[.5pt]u
		(w\underbracket[.5pt]w)^n && (\text{Since  $(x_1x_2)^\omega px_2 (y_1y_2)^\omega = (x_1x_2)^\omega py_2 (y_1y_2)^\omega$})\\
			    &=_\varphi  uvuw.
		\end{align}
        Thus $exf=exef$. Analogously one proves $exf=efxf$.
	\end{proof}

\begin{samepage}
\begin{lemma}
\label{lem:freq-idempotent}
Let $\varphi: \Sigma^+ \to S$ be a morphism with stability index $s$ and let
	$N=|\stab(\varphi)|$. 
	Let $i\in[0,s-1]$. Assume that $u\in \Sigma^{s\bbN_{>0}}$ and 
	$a\in \cont_{i,s}^{>N}(u)$. Then, there exists an 
	idempotent $x\in \Sigma^{s\bbN_{>0}}$ and a factorization 
	$u=u'u''$ such that 
\begin{itemize}
    \item $u=_\varphi u'xu''$,
    \item $(x)_0=a$, and
    \item $|u'|\equiv_s i$.
\end{itemize} 
\end{lemma}
\end{samepage}
\begin{proof}
Let $i\in[0,s-1]$ and let $u\in \Sigma^{s\bbN_{>0}}$ be such that 
	$a\in \cont_{i,s}^{>N}(u)$. 
	Hence, there exists a factorization
\begin{align}
	u&=u_{-1} au_0au_1\ldots u_{N-1}au_N
\end{align}
	such that $|u_{-1}|\equiv_s i$
and $|au_j|\equiv_s 0$ for each $j\in [0,N-1]$.
Let $v_j=au_j$ for each $j\in[0,N-1]$ and let $v_{N}=v_0$.

Applying \Cref{lem:prefidem1} to the sequence 
	$\varphi(v_0),\dots, \varphi(v_{N})$ there exist indices
	$0\leq p<q\leq N$ such that 
\begin{align}
    v_0\dots v_p
	&=_\varphi v_0\dots v_p(v_{p+1}\dots v_{q})^\ell
\end{align}
	for all $\ell\in\bbN$. 
	Let $n$ be an exponent of $S$, 
	let $x=(v_{p+1}\dots v_q)^n$, $u'=u_{-1}v_0\dots v_p$, 
	and $u''=v_{p+1}\ldots v_{N-1}au_N$.
	Clearly $u=_\varphi u'xu''$, $(x)_0=a$, and 
	$|u'|\equiv_s|u_{-1}|\equiv_s i$.
\end{proof}

Let $u=u_0\ldots u_{n-1}\in \Sigma^+$ and let $i\in[0,n-1]$. 
For a word $v\in \Sigma^*$ we define the  word 
$u[i\downarrow v]$ to be
\begin{align}
    u[i\downarrow v]=  u_0\ldots u_{i-1} v u_{i}\ldots u_{n-1}.
\end{align}

\begin{lemma}
\label{lem:pup-sub}
Let $\varphi: \Sigma^+ \to S$ be a stamp in $\bfX$ with stability index $s$. 
	Let $p\in \Sigma^s$ be an idempotent.
Let $i\in [0,s-1]$ and let $u=u'xu''\in \Sigma^{s\bbN_{>0}}$ such that
	$|u'|\equiv_s i$ and 
	$x \in \Sigma^{s\bbN_{>0}}$ is an idempotent. Then we have 
\begin{align}
	pup &=_\varphi pp[i\downarrow x] (pup)\\
	&=_\varphi (pup) p[i\downarrow x]p.
\end{align}
\end{lemma}
\begin{proof}
	Let $p=p_0\ldots p_{s-1}\in\Sigma^s$ be an idempotent and let $i\in[0,s-1]$.
We define the following words:
    \begin{align}
r_i&=p_{i}\ldots p_{s-1}\in\Sigma^+\\
\ell_i&=p_0\ldots p_{i-1}=p(r_i)^{-1}\in\Sigma^*\\
    e_i&=(r_i\ell_i)^2
\end{align}
Clearly, $\ell_ir_i=p$, 
	$|r_i|= s-i$, and $|\ell_i|= i$.
	Moreover, $p^3=\ell_ir_i\ell_ir_i\ell_ir_i=\ell_ie_ir_i$ and $e_{i}$ is an 
	idempotent as the below computation shows:
\begin{align}
e_i^2 &=_\varphi(r_i\ell_i)^2(r_i\ell_i)^2\\
&=_\varphi r_i (\ell_ir_i)^3\ell_i\\
&=_\varphi r_ip^3\ell_i\\
&=_\varphi r_ip\ell_i\\&=_\varphi e_i
\end{align}
Let $u=u'xu''\in \Sigma^{s\bbN_{>0}}$ be such that $|u'|\equiv_s  i$
	and $x\in\Sigma^{s\bbN_{>0}}$ is an idempotent.
	We show that $pup =_\varphi pp[i\downarrow x] (pup)$. Indeed:
\begin{align}
    pup&=_\varphi p^3pup &&(\text{Since $p$ is idempotent})\\
    &=_\varphi \ell_{i}\underbracket[.5pt]{e_{i}}\underbracket[.5pt]{r_{i} pu'}\underbracket[.5pt]{x}u''p&& (\text{Since $p^3=\ell_ie_ir_i$ and $u=u'xu''$})\\
	&=_\varphi \ell_{i}\underline{e_{i}}xr_{i}pu'xu''p&& (\text{Since $|r_ipu'|\equiv_s 0$ \& applying $exf=efxf$ by Lemma~\ref{L X in Y}})\\
   &=_\varphi \underline{\ell_{i}r_i}p\underline{\ell_ixr_{i}}pup&& (\text{Substituting $e_i=r_ip\ell_i$ \& $u=u'xu''$})\\
	&=_\varphi p^2 p[i\downarrow x] (pup) && (\text{Since $\ell_ir_i=p$ \& $\ell_ixr_{i}=p[i\downarrow x]$})\\
	&=_\varphi pp[i\downarrow x] (pup)&& (\text{Since $p$ is idempotent})
\end{align}

	By a symmetric argument one can prove $pup=_\varphi (pup) 
	p[i\downarrow x]p$.
\end{proof}

From \Cref{lem:freq-idempotent} and \Cref{lem:pup-sub}, we get
\begin{samepage}
\begin{corollary}
\label{cor:log-idem-ins0}
Let $\varphi: \Sigma^+ \to S$ be a stamp in $\bfX$ with stability index $s$ 
	and let $N=|\stab(\varphi)|$. 
	Let $p\in\Sigma^s$ be an idempotent and let $i\in[0,s-1]$. 
	Assume that $u\in \Sigma^{s\bbN_{>0}}$ and 
	$a\in \cont_{i,s}^{>N}(u)$. Then there exists an idempotent 
	$x\in \Sigma^{s\bbN_{>0}}$ such that $(x)_0=a$ and
\begin{align}
	pup &=_\varphi pp[i\downarrow x](pup)\\
	&=_\varphi (pup) p[i\downarrow x]p.
\end{align}
\end{corollary}
\end{samepage}

Let $\sigma=w_0,\ldots, w_{n-1}$ be a sequence of words,
let $u\in\Sigma^+$ and $i\in[0,|u|-1]$.
Then $u[i\downarrow \sigma]$ is the word
\begin{align}
	u[i\downarrow \sigma]=u[i\downarrow w_0]\cdot \prod_{j=1}^{n-1} 
	\left(u\cdot u[i\downarrow w_j]\right)
\end{align} 
in case $\sigma$ is non-empty.
If $\sigma$ is empty we define $u[i\downarrow\sigma]=\varepsilon$.

    When $\sigma$ is non-empty, for each $i\in\bbN$ with
    $i<\min\{|w_j|: j\in[0,n-1]\}$
    we define $(\sigma)_i=\{(w_j)_i \mid j \in [0,n-1]\}$
    and $(\sigma)_i=\emptyset$ otherwise.
    In case $\sigma$ is empty we define $(\sigma)_i=\emptyset$ for
    all $i\in\bbN$.

\begin{lemma}
\label{lem:log-idem-ins}
    Let $\varphi: \Sigma^+ \to S$ be a stamp in $\bfX$ with stability index $s$
	and let $N=|\stab(\varphi)|$. Let $p \in \Sigma^s$ be an idempotent and let
	$i\in[0,s-1]$.
	For each $u\in \Sigma^{s\bbN_{>0}}$ there exists a sequence of 
	idempotents $\sigma\in (\Sigma^{s\bbN_{>0}})^*$ such that
\begin{enumerate}
    \item $(\sigma)_0=\cont_{i,s}^{>N}(u)$ and
    \item $pup =_\varphi p p[i\downarrow\sigma](pup)=_\varphi
(pup)p[i\downarrow\sigma] p=_\varphi
	    p p[i\downarrow\sigma](pup)p[i\downarrow\sigma] p $.
\end{enumerate} 
\end{lemma}
\begin{proof}
Assume that $u\in \Sigma^{s\bbN_{>0}}$. If 
$\cont_{i,s}^{>N}(u) = \emptyset$ then the lemma follows trivially. Otherwise, we have $|\cont_{i,s}^{>N}(u)| = n \in \bbN_{>0}$. 
	By \Cref{cor:log-idem-ins0} there exists a sequence of 
	idempotents $\sigma=w_0,w_1,\ldots, w_{n-1} \in (\Sigma^{s\bbN_{>0}})^+$ 
	such that 
	$(\sigma)_0=\{(w_j)_0 \mid j\in [0,n-1]\} = \cont_{i,s}^{>N}(u)$ and for each $j\in [0,n-1]$ we have
\begin{align}
	pup &=_\varphi pp[i\downarrow w_j]  (pup)\label{E pup left0}\\
	&=_\varphi (pup)  p[i\downarrow w_j]p.  \label{eq:barsigma}
\end{align}
	By repeatedly applying (\ref{E pup left0}) one proves
	by induction on $m\in[0,n-1]$:
\begin{align}
pup &=_\varphi\left [ \prod_{j=0}^{m} p p[i\downarrow w_j] \right ] (pup)
\end{align}
In particular
\begin{align}
pup &=_\varphi\left [ \prod_{j=0}^{n-1} p p[i\downarrow w_j] \right ] (pup)=
	pp[i\downarrow\sigma](pup)\label{E pup left}
\end{align}
	Analogously, making use of (\ref{eq:barsigma})
	one proves 
	\begin{align}
		pup=_\varphi (pup) p[i\downarrow\sigma]p.\label{E pup right}
	\end{align}
	By combining (\ref{E pup left}) and (\ref{E pup right}) we obtain
in total:
	\begin{align}
		pup&=_\varphi (pup) p[i\downarrow\sigma]p&\text{(By (\ref{E pup right}))} \\
		&=_\varphi 
		pp[i\downarrow\sigma](pup) p[i\downarrow\sigma]p&\text{(By (\ref{E pup left}))}
	\end{align}
\end{proof}

For each word $u\in\Sigma^+$ and each sequence
$\overline{\sigma}=(\sigma_i)_{i\in [0,n-1]}$, where $1\leq n\leq |u|$
and each $\sigma_i$ is a sequence of words
we define $u[ \overline{\sigma}]$ to be
\begin{align}
    u[\overline{\sigma}]=
u[0\downarrow \sigma_0] \cdot\prod_{j=1}^{n-1} \left(u u[j\downarrow \sigma_j]\right).
\end{align} 
When $\overline{\sigma}$ is empty we define $u[\overline{\sigma}]=\varepsilon$.

\begin{samepage}
	\begin{corollary}\label{C idem vec}
    Let $\varphi: \Sigma^+ \to S$ be a stamp in $\bfX$ with stability index $s$
	and let $N=|\stab(\varphi)|$. Let $p \in \Sigma^s$ be an idempotent.
	For each $u\in \Sigma^{s\bbN_{>0}}$ there exists 
	 a vector $\overline{\sigma}=(\sigma_i)_{i\in[0,s-1]}$ such that
	 \begin{enumerate}
		 \item $\sigma_i\in(\Sigma^{s\bbN_{>0}})^*$ is a sequence of idempotents 
for each $i\in[0,s-1]$,
		 \item $(\sigma_i)_0=\cont_{i,s}^{>N}(u)$ for each $i\in[0,s-1]$, 
			 and
		 \item $pup=_\varphi (pp[\overline{\sigma}])^\ell(pup) 
			 (p[\overline{\sigma}]p)^\ell$ for each $\ell\in\bbN$.
	 \end{enumerate}
  \end{corollary}
  \end{samepage}

\begin{proof}
	Let $p\in\Sigma^s$ be an idempotent and let $u\in\Sigma^{s\bbN_{>0}}$.
	By Lemma~\ref{lem:log-idem-ins} for each $i\in[0,s-1]$ there
	exists a sequence of idempotents $\sigma_i\in(\Sigma^{s\bbN_{>0}})^*$
	such that $(\sigma_i)_0=\cont_{i,s}^{>N}(u)$.
	Thus Points (1) and (2) follow.
	It remains to prove Point (3).
	Let $\overline{\sigma}=(\sigma_i)_{i\in[0,s-1]}$.
	By repeatedly applying Point (2) of Lemma~\ref{lem:log-idem-ins}
	we obtain:
\begin{align}
pup &=_\varphi pp[0\downarrow \sigma_0] \cdots  pp[s-1\downarrow \sigma_{s-1}] pup\\
	&=  pp[\overline {\sigma}]  (pup)\label{E pup sleft}
\end{align} 
Analogously one proves
\begin{align}
	pup &=_\varphi  (pup) p[\overline {\sigma}]p.\label{E pup sright}
\end{align}
	By combining (\ref{E pup sleft}) and (\ref{E pup sright}) we obtain
	\begin{align}
		pup &=_\varphi  pp[\overline {\sigma}] (pup) 
		p[\overline {\sigma}] p\\
		&=_\varphi  (pp[\overline {\sigma}])^\ell (pup) 
		(p[\overline {\sigma}]p)^\ell
	\end{align}
	for each $\ell\in\bbN$.
\end{proof}

Finally, we combine Corollary~\ref{C idem vec} and the exchange principle.

\begin{lemma}
Let $\varphi : \Sigma^+ \to S$ be a stamp in $\bfX$ with stability index 
	$s\in\bbN_{>0}$ and let $N=|\stab(\varphi)|$. 
	Let $k=s(N+1)!$, $K=2k$ and let 
	$u,v \in \Sigma^{s\bbN_{>0}}$ be  words such that 
	$u\congC_K^{K,K} v$.
Then, $pup=_\varphi pvp$ for all idempotents $p\in \Sigma^s$.
\label{lemma:log-pup=pvp}
\end{lemma}
\begin{proof}
Let $u,v \in \Sigma^{s\bbN_{>0}}$ be  words such that $u\congC_K^{K,K} v$. 
	By \Cref{L redux substitution}, there exist
	$u',v' \in \Sigma^{s\bbN_{>0}}$ such that 
	\begin{enumerate}
		\item $u=_\varphi u'$ and $v=_\varphi v'$,
		\item $|u'|=|v'|$,
		\item $u'\congC_s^{k,k} v'$, and
		\item $\rmR_s^k(u')=\rmR_s^k(v')$.
	\end{enumerate}
	Points (2), (3) and (4) together imply
	\begin{align}
		\forall j\in H(u',v'):\quad \{(u')_j,(v')_j\}\subseteq
		\cont_{j,s}^{\geq k}(u')=
		\cont_{j ,s}^{\geq k}(v')\ \label{E jH}
	\end{align}
	and moreover, since $k> N$, we have:
	\begin{align}
		\forall i\in[0,s-1]:\ \cont_{i,s}^{>N}(u')=\cont_{i,s}^{>N}(v')
		\label{E cont>N}
	\end{align}

	By Corollary~\ref{C idem vec} there exist vectors
of sequences of idempotents
	$\overline{\sigma}=(\sigma_i)_{i\in [0,s-1]}$ and
	 $\overline{\tau}=(\tau_i)_{i\in [0,s-1]}$ such that 
	\begin{enumerate}
		\item[(5)] $\sigma_i\in(\Sigma^{s\bbN_{>0}})^*$ and
			$(\sigma_i)_0=\cont_{i,s}^{> N}(u')$ for each $i\in [0,s-1]$,
		\item[(6)] $\tau_i\in(\Sigma^{s\bbN_{>0}})^*$ and $(\tau_i)_0=\cont_{i,s}^{> N}(v')$ for each $i\in [0,s-1]$, 
		\item[(7)] $pu'p =_\varphi (p p[\overline{\sigma}])^\ell (pu'p) 
			(p[\overline{\sigma}]p)^\ell$ for each $\ell\in\bbN$, and
		\item[(8)]  $pv'p =_\varphi (pp[\overline{\tau}])^\ell (pv'p)
			(p[\overline{\tau}]p)^\ell$ for each $\ell\in\bbN$.
	\end{enumerate}
Let $n>N$ be an exponent of $S$. We define the following idempotents:
\begin{align}
	z_1&=(pp[\overline{\sigma}])^np\label{Def z}\\
    z_2&=p(p[\overline{\tau}]p)^n\label{Def z'}
\end{align}
To see that they are idempotents, observe that $p$ is idempotent and 
	that $n$ is an exponent of $S$, so we have
$z_1^2 =_\varphi (pp[\overline{\sigma}])^{2n}p
      =_\varphi z_1$
	and $
z_2^2 =_\varphi p(p[\overline{\tau}]p)^{2n}
      =_\varphi z_2$.

Observe that for all $j\in H(u',v')$ we have:
	\begin{align}
		\{(u')_j,(v')_j\}&\subseteq\cont_{j,s}^{\geq k}(u')& 
		\text{(By (\ref{E jH}))}\label{E Hu'v'1}\\
		&=\cont_{j,s}^{\geq k}(v')& 
		\text{(By (\ref{E jH}))}\\
		&\subseteq \cont_{j,s}^{>N}(u')&\text{(Since $k>N$)}\\
		&= \cont_{j,s}^{>N}(v')&\text{(By (\ref{E cont>N}))}\\
		&\subseteq \cont_{j,s}^{>N}(z_1)\cap\cont_{j,s}^{>N}(z_2)
		&\text{(By (5),(6), (\ref{Def z}) and (\ref{Def z'}))}\\
		&\subseteq \cont_{j,s}(z_1)\cap\cont_{j,s}(z_2) &
		\label{E Hu'v'2}
	\end{align}
We observe that 
\begin{align}
	pu'p &=_\varphi (pu'p)p&&(\text{Since $p$ is an idempotent})\\
	&=_\varphi (pp[\overline{\sigma}])^n (pu'p)(p[\overline{\sigma}]p)^n  p && \text{(By (7))}\label{eq:pupeq1}\\
	&=_\varphi (pp[\overline{\sigma}])^n p(pu'p)(p[\overline{\sigma}]p)^n  p && (\text{Since $p$ is an idempotent})\\
	&=_\varphi z_1(pu'p)  z_1. && (\text{By (\ref{Def z})
	and idempotence of $p$})\label{eq:pupeq2}
\end{align}
Analogously one proves
\begin{align}
	pv'p &=_\varphi z_2(pv'p)z_2.\label{eq:pupeq3}
\end{align}

	Since $u=_\varphi u'$ and $v=_\varphi v'$ by Point (1), 
	to prove the lemma it suffices to show that $pu'p =_\varphi pv'p$. 
Indeed, we have
\begin{align}
	pu'p &=_\varphi p(pu'p)p && (\text{Since $p$ is an idempotent})\\
	&=_\varphi pz_1  (pu'p) z_1p && (\text{By (\ref{eq:pupeq2})})\\
	&=_\varphi pz_1 (pv'p)  z_1p && (\text{By (\ref{E Hu'v'1},\ref{E Hu'v'2}) and \Cref{lem:exchange-principle}})\\
	&=_\varphi \underbracket[.5pt]{p}\underbracket[.5pt]{z_1}
	\underbracket[.5pt]{z_2} (pv'p) z_2
	z_1p  &&(\text{By (\ref{eq:pupeq3})})\\
	&=_\varphi pz_2z_1z_2(pv'p)\underbracket[.5pt]{z_2}
	\underbracket[.5pt]{z_1}\underbracket[.5pt]{p} &&
	(\text{Since $\stab(\varphi)$ satisfies $exf=efxf$ by
	Lemma~\ref{L X in Y}})\\
	&=_\varphi pz_2z_1z_2(pv'p) z_2z_1z_2p &&(\text{Since $\stab(\varphi)$ satisfies
	$exf=exef$ by Lemma~\ref{L X in Y}})\\
	&=_\varphi pz_2z_1 (pv'p) z_1z_2p  &&(\text{By \ref{eq:pupeq3}})\\
	&=_\varphi pz_2z_1(pu'p)z_1z_2p  && (\text{By (\ref{E Hu'v'1},\ref{E Hu'v'2}) and \Cref{lem:exchange-principle}})\\
	&=_\varphi pz_2(pu'p) z_2p && (\text{By \ref{eq:pupeq2}})\\
	&=_\varphi pz_2(pv'p) z_2p &&  (\text{By (\ref{E Hu'v'1},\ref{E Hu'v'2}) and \Cref{lem:exchange-principle}})\\
	&=_\varphi p(pv'p)p && (\text{By \ref{eq:pupeq3}})\\
	&=_\varphi pv'p. &&(\text{Since $p$ is an idempotent})
\end{align}
\end{proof}

\begin{samepage}
	\begin{lemma}[Lemma 4.30 in \cite{GM26arxiva}]\label{L pq idem}
    Let $\varphi : \Sigma^+ \to S$ be a stamp with stability index 
	$s\in\bbN_{>0}$ and let $N=|\stab(\varphi)|$ be the size of its stable semigroup.
	Then for all words $u\in\Sigma^{s\N_{>0}}$ with $|u|\geq 2sN$ there exist idempotents
	$p,q\in\Sigma^s$ and a factorization $u=xu'y$
	such that
	$x=_\varphi xp$, $qy=_\varphi y$, $|x|=(i+1)\cdot s$, 
	and $|y|=(j+1)\cdot s$ for some $i,j\in[0,N-1]$.
\end{lemma}
\end{samepage}

The following lemma can be proved analogously to Lemma 4.31 in \cite{GM26arxiva}.

\begin{lemma}
Let $\varphi:\Sigma^+\to S$ be a stamp in $\bfX$ with stability index $s$. Let $N=|\stab(\varphi)|$ and $k=2s(N+1)!$. 
For each $u,v \in \Sigma^{+}$ we have that 
	if $u \simeq_{k} v$ and $|u|\equiv_k |v| \equiv_k 0 $, 
	then $\varphi(u)=\varphi(v)$. 
\label{lemma:log-alg-to-cong}
\end{lemma}

\begin{proof} 
	Let $u,v\in \Sigma^+$ be such that
	$u \simeq_k v$ and $|u|\equiv_k |v|\equiv_k 0 $.
	If $\min(|u|,|v|)\leq 2k$ then $u=v$ and therefore 
	$u=_\varphi v$ and the lemma is proved. 
	Hence, let us assume that $|u|,|v|> 2k$.

	Since $\alpha_k(u)=\alpha_k(v)$, $\omega_k(u)=\omega_k(v)$
	and $|\alpha_k(u)|=|\omega_k(u)|=k\geq 2sN$ 
	by Lemma~\ref{L pq idem} there exist idempotents $p,q\in \Sigma^s$ and 
	factorizations $u=xu'y$, $v=xv'y$ 
	such that $|x| = (i+1)\cdot s$, $|y| = (j+1)\cdot s$ 
	for some $i,j\in[0,N-1]$,
$u=_\varphi xpu'qy$ and $v=_\varphi xpv'qy$. 
	Note that $|p|,|u'|, |v'|, |q|\in s\bbN_{>0}$.
	Thus, $\varphi(p),\varphi(u'),
	\varphi(v'),\varphi(q)\in\stab(\varphi)$.
	Since moreover $\stab(\varphi)$ satisfies the identity $exf=exef$,
we have
	\begin{align}
		pu'q=_\varphi pu'pq\qquad{\text{and}}\qquad
pv'q=_\varphi pv'pq.
	\end{align}
	Consequently, $u=_\varphi xpu'qy=_\varphi xpu'pqy$ 
	and $v=_\varphi xpv'qy=_\varphi xpv'pqy$.
Thus, to prove the lemma, it suffices to show that 
	$pu'p=_\varphi pv'p$. 

Note that $u'=x^{-1}uy^{-1}=x'\kappa_k(u)y'$ and $v'=x^{-1}vy^{-1}=x'\kappa_k(v)y'$, where 
	$x'=x^{-1}\alpha_{k}(u)=x^{-1}\alpha_{k}(v)$ and
$y'=\omega_{k}(u)y^{-1}=\omega_{k}(v)y^{-1}$.
	Since $u\simeq_k v$ implies $\kappa_k(u)\congC_k^{k,k} \kappa_k(v)$ by 
	definition and since 
	$\congC_k^{k,k}$ is a congruence 
	by Point (1) of Lemma~\ref{L congC}
	we have $u'=x'\kappa_k(u)y' \congC_k^{k,k} x'\kappa_k(v)y' = v'$. 
Recalling that the lengths of $u'$ and $v'$ are both divisible by
$s$ we therefore have
	$pu'p=_\varphi pv'p$ by Lemma~\ref{lemma:log-pup=pvp}.
\end{proof}

\subsection{Proof of Proposition~\ref{P log tech}}
\begin{proof}
	The proof is analogous to the proof of Proposition 4.24 in~\cite{GM26arxiva}.
	We repeat the proof here for completeness. 
	Let $k=2s (N+1)!$. By \Cref{lemma:log-alg-to-cong},
	for all $u,v\in\Sigma^+$ we have that
	if $u\simeq_{k} v$ 
	and $|u|\equiv_{k} |v| \equiv_{k} 0$, then $\varphi(u)=\varphi(v)$. 
	To prove the proposition we 
	show that 
	for all $u,v \in \Sigma^+$ such that $u \simeq_{2k} v$ we
	have $\varphi(u)=\varphi(v)$.

Assume that $u,v\in \Sigma^+$ are words such that 
	$u\simeq_{2k}v$. If $|u|$ or $|v|$ is $\leq 4k$,
	then $u=v$ and $\varphi(u)=\varphi(v)$ and the proof is completed. 
Next assume that $|u|,|v|>4k$. Since $u\simeq_{2k}v$, 
	clearly $|u|\equiv_{2k} |v|$. 
	Let $u',v'\in \Sigma^+$ and $u'',v''\in\Sigma^*$ be such that
	$u=u'u''$, $v=v'v''$, $|u'|\equiv_{k} |v'|\equiv_{k} 0 $ 
	and $|u''|=|v''|<k$. Since $u$ and $v$ agree on their $2k$-length suffix, clearly $u''=v''$. Hence to show that $\varphi(u)=\varphi(v)$ it suffices to show that $\varphi(u')=\varphi(v')$.
	To show $\varphi(u')=\varphi(v')$ in turn, we will prove 
	$u'\simeq_{k} v'$, which is sufficient by \Cref{lemma:log-alg-to-cong}.
 
	So let us prove $u'\simeq_{k} v'$. 
	Since $u\simeq_{2k} v$, we have $\alpha_{2k}(u)=\alpha_{2k}(v)$ 
	and $\omega_{2k}(u)=\omega_{2k}(v)$. 
	Therefore, $\alpha_{k}(u')=\alpha_{k}(u)=\alpha_{k}(v)=
	\alpha_{k}(v')$.  Also,
	$\omega_{k}(u')=\omega_{k}(\omega_{2k}(u)(u'')^{-1})=\omega_{k}(\omega_{2k}(v)(v'')^{-1})
	=\omega_{k}(v')$. 
	It remains to show that $\kappa_{k}(u')\congC_{k}^{k,k} \kappa_{k}(v')$.
We observe that 
\begin{align}
	\kappa_{k}(u')&=(\omega_k\circ \alpha_{2k}(u))\cdot \kappa_{2k}(u) \cdot 
	(\alpha_{k-|u''|}\circ \omega_{2k}(u))\label{E ao1}\\
	\kappa_{k}(v')&=(\omega_k\circ \alpha_{2k}(v))\cdot \kappa_{2k}(v) \cdot 
	(\alpha_{k-|v''|}\circ \omega_{2k}(v))\label{E ao2}\\
	&=(\omega_k\circ \alpha_{2k}(u))\cdot \kappa_{2k}(v) \cdot 
	(\alpha_{k-|u''|}\circ \omega_{2k}(u))\label{E ao3}
\end{align}
    
	Since $u\simeq_{2k}v$ we have $\kappa_{2k}(u)\congC_{2k}^{2k,2k} \kappa_{2k}(v)$ by definition.
	By Points (2) and (3) of Lemma~\ref{L congC} 
	we have $\kappa_{2k}(u)\congC_{k}^{k,k} \kappa_{2k}(v)$. 
	Since $\congC_{k}^{k,k}$ is a congruence
	by Point (1) of Lemma~\ref{L congC}, we have
	$\kappa_{k}(u')\congC_{k}^{k,k} \kappa_{k}(v')$ by (\ref{E ao1}) and (\ref{E ao3}), 
	as required.
\end{proof}

%!TEX root = main.tex

\section{Circuit Complexity Upper Bounds}\label{S upper}

\newcommand{\bin}{\mathrm{bin}}

This section is devoted to establishing circuit-size upper bounds.

In \S\,\ref{S Threshold} we recall the folklore result that threshold functions with constant thresholds (and hence counting the occurrences of a letter up to a fixed threshold) can be computed by circuit families of size $O(\log n)$ and depth $O(1)$.

In \S\,\ref{S Rare Comp} we establish two circuit complexity upper bounds for computing the binary representations of all positions at which a fixed letter occurs, under the promise that the letter occurs only a constant number of times in the input word. The first construction yields circuits of size $O(\log^2 n)$ and depth $O(1)$, whereas the second achieves size $O(\log n\log\log n)$ and depth $O((\log\log n)^2)$.

\subsection{Computing the Constant Threshold Function}\label{S Threshold}

In this section we concern ourselves with the computation
of the threshold function for fixed threshold via circuit
families of logarithmic size and constant depth.
For this we introduce the following notion due to Friedman.

\begin{samepage}
	\begin{definition}\cite[p.~642]{Friedman86}\label{D Friedman}
	A tuple $(A_0,\dots,A_{t-1})$ of subsets of a finite set $S$ is called a 
	subpartition of $S$
	if the subsets $A_0,\dots,A_{t-1}\subseteq S$ are pairwise disjoint,
	but it is not required that $S=\bigcup_{i=0}^{t-1}A_i$.
	A family of subpartitions $(A_{0}^{j},\dots,A_{t-1}^j)_{j\in J}$
	is an {\em $(S,t)$-scheme} of size $|J|$ if 
	for every $S'\subseteq S$ with $|S'|=t$ 
	there exists $j\in J$ such that each set $A_{0}^j,\dots, A_{t-1}^j$ 
	contains exactly one element of $S'$. 
\end{definition}
\end{samepage}

\begin{theorem}\cite[Theorem 1.2]{Friedman86}\label{T Friedman}
	Let $t\in\bbN_{>0}$ be fixed. Then for every finite set $S$ 
	one can construct an 
	$(S,t)$-scheme of size $O(\log |S|)$ in time polynomial in $|S|$.
\end{theorem}

The previous theorem allows us to state the following proposition.
We will only apply the existence of $(S,t)$-schemes of size $O(\log |S|)$ in 
what follows.

\begin{samepage}
\begin{proposition}
Let $t\in\bbN_{>0}$ be a constant.
The following problem is decidable by a circuit family of size
	$O(\log n)$ and depth $O(1)$:

	INPUT: $w\in\{0,1\}^n$.

	QUESTION: $|w|_1\geq t$?
\end{proposition}
\end{samepage}
\begin{proof}
Let $S=[0,n-1]$.
	By Theorem~\ref{T Friedman} there exists an $(S,t)$-scheme
$(A_0^j,\dots,A_{t-1}^j)_{j\in J}$ of
size $|J|\in O(\log n)$.
The following predicate $P_t$ can be realized by an $n$-circuit of 
	size $O(\log n)$ and depth $O(1)$:
	\begin{align}
		P_t:\Longleftrightarrow\quad\bigvee_{j\in J} 
		\bigwedge_{m=0}^{t-1}\bigvee_{p\in A_m^j} (w)_p=1
	\end{align}
	If $|w|_1\geq t$ there exists a $t$-element subset
	$S'\subseteq[0,n-1]$ of positions at which
	$w$ has a $1$. Then by Definition~\ref{D Friedman}
	there exists $j\in J$ such that 
	each set $A_0^j,\dots,A_{t-1}^j$ contains at least one
	position $p\in S'$ with $(w)_p=1$. Hence the predicate
	$P_t$ holds.
	Conversely, if the predicate $P_t$ holds, then for the index
	$j$ that witnesses the predicate to hold, 
	due to pairwise disjointness of $A_0^j,\dots,A_{t-1}^j$,
	we are guaranteed that $|w|_1\geq t$ indeed holds.
\end{proof}

\begin{corollary}
\label{cor:cir-thr}
	Let $\Sigma$ be a fixed finite alphabet, let $a\in\Sigma$
	and let $k,t\in\bbN_{>0}, i\in[0,k-1]$  be fixed. 
	Then the following is decidable by a circuit family of
	size $O(\log n)$ and depth $O(1)$:

	INPUT: $w\in\Sigma^n$.

	QUESTION: $|\pi_{i,k}(w)|_a>t$?
\end{corollary}

\subsection{Computing the Positions of Words Containing Rare Letters}
\label{S Rare Comp}

For the rest of this section, whenever we refer to the {\em lexicographic order} 
we mean the {\em right-to-left lexicographic order}.
Moreover $\log$ denotes the base two logarithm.

In this section, binary representations are written
least significant bit first. For each non-negative integer $n\in\N$ and 
$\ell\geq\max\{1,\lceil\log(n+1)\rceil\}$, 
let $\bin_{\ell}(n)=b_0\dots b_{\ell-1}\in\{0,1\}^+$
denote its {\em binary representation with $\ell$ bits}, i.e.
$n=\sum_{i=0}^{\ell-1} b_i2^i$. 
We also write $\bin_{\ell,i}(n)$ to
denote $b_i$, i.e. the $i^{\text{th}}$ (least significant) bit.
Observe that for each $n,n'\in[0,2^\ell-1]$, we have $n<n'$ if, and only if,
$\bin_\ell(n)$ is (right-to-left) lexicographically smaller
than $\bin_\ell(n')$.
			In this section we concern ourselves with the
			following promise problem for each fixed
			$t\in\bbN_{>0}$.
\begin{samepage}
	\Needspace{8\baselineskip}
	\begin{problem}[\textsl{Occ$_t$}]\label{Pro Occ}
\ \\
\noindent
				INPUT: A word $w\in\{0,1\}^n$, where 
				$n\in\bbN_{>0}$ and $|w|_1=t$.

				\medskip
		\noindent
				OUTPUT: Bit strings
				$\bin_\ell(p_0),\dots,\bin_\ell(p_{t-1})$
			where $\ell=\lceil\log(n+1)\rceil$ and
		$p_j$ is the position of the $j^{\text{th}}$ occurrence of $1$ 
		in $w$, with indexing starting at $0$.
	\end{problem}
\end{samepage}

\subsubsection{A Size $O(\log^2 n)$ and Depth $O(1)$ Upper Bound}

\begin{samepage}
The following standard lemma states that one can lexicographically
sort constantly many $n$-bit strings on a circuit of linear size
and constant depth.
	\begin{lemma}\label{L sort}
		Let $t\in\bbN_{>0}$ be fixed.
		The following is computable by a circuit family of size
		$O(n)$ and depth $O(1)$, where $<_{\textrm{lex}}$ denotes 
		the strict lexicographic order on $\{0,1\}^+$.
\ \\
\noindent
				INPUT: $t$ pairwise distinct $n$-bit strings 
				$w_0,\dots,w_{t-1}\in\{0,1\}^n$.

				\medskip
		\noindent
				OUTPUT: $w_{\beta(0)},\dots,w_{\beta(t-1)}$
				for the unique permutation 
				$\beta:[0,t-1]\to[0,t-1]$
				satisfying 
				$i< j\Rightarrow w_\beta(i)<_{\textrm{lex}}w_{\beta(j)}$ for all $i,j\in[0,t-1]$.
\end{lemma}
\end{samepage}
\begin{proof}
	For each $i,j\in[0,t-1]$ define the predicate $P_{i,j}$ expressing that $w_i<_{\textrm{lex}}w_j$
	as
	\begin{align}
		P_{i,j}:\Longleftrightarrow\quad \bigvee_{k=0}^{n-1}\left(
		(w_{i})_k=0\wedge (w_{j})_k=1\wedge 
		\left(\bigwedge_{\ell\in[k+1,n-1]}
		(w_{i})_\ell=(w_{j})_\ell\right)\right),
	\end{align}
	which is computable by a circuit of size $O(n)$ and depth $O(1)$.
	One can compute the bit $v_{i,k}=(w_{\beta(i)})_k$ for 
	each $i\in[0,t-1]$ and each $k\in[0,n-1]$ as 
	\begin{align}
		v_{i,k}=\quad
		\bigvee_{\rho:[0,t-1]\to[0,t-1]\atop
		\text{$\rho$ is permutation}}
		\left[\left(\bigwedge_{0\leq a<b\leq t-1} P_{\rho(a),\rho(b)}\right)
		\wedge (w_{\rho(i)})_k=1\right],
	\end{align}
	which is achievable by $O(1)$ additional gates per output bit.
	The overall circuit size is $O(n)$ and depth $O(1)$, as required.
\end{proof}

\begin{samepage}
\begin{theorem}
{\textsl{Occ$_t$}} is computable by a circuit family
	of size $O(\log^2 n)$ and depth $O(1)$.
\label{thm:occ-log2n}
\end{theorem}
\end{samepage}
\begin{proof}
	Let $S=[0,n-1]$ and let $w\in\{0,1\}^n$ with $|w|_1=t$.
	By Theorem~\ref{T Friedman} there exists an $(S,t)$-scheme
	$(A_0^j,\dots,A_{t-1}^j)_{j\in J}$ of size $O(\log n)$.
	Let $P=\{p_0,\dots,p_{t-1}\}=\{p\in[0,n-1]\mid (w)_p=1\}$
	where $p_0<p_1<\dots< p_{t-1}$.
	An index $j\in J$ is {\em good} if 
	each of the sets $A_0^j,\dots,A_{t-1}^j$ contains precisely
	one element of $P$. In this case there exists
	a bijection $\beta_j:[0,t-1]\to[0,t-1]$ such that 
	$p_{i}\in A_{\beta_j(i)}^j$
	for each $i\in[0,t-1]$.
	The following predicate
\begin{align}
	P_j:\Longleftrightarrow\quad\bigwedge_{m=0}^{t-1}\bigvee_{p\in A_m^j} (w)_p=1
\end{align}
evaluates to true if, and only if, $j$ is good --- moreover it can be realized 
	 by a circuit of size $O(1)$.
However, there could be more than one $j\in J$ that is good.
To compute a unique element of $J$ that witnesses this consider the
	following predicates $U_j$ for each $j\in J$, where
$<$ denotes some total ordering of $J$:
\begin{align}
	U_j:\Longleftrightarrow\quad P_j\wedge\bigwedge_{j'<j} \neg P_{j'}
\end{align}
For fixed $t$, each flag $P_j$ and its complement can be
computed using $O(1)$ gates and constant depth, by De Morgan's
laws. Once these values are computed and shared, each selector
$U_j$ requires only one additional unbounded fan-in AND gate.
Since $|J|=O(\log n)$, all flags, their complements, and all
selectors together require $O(\log n)$ gates and constant depth.
	Observe that $U_j$ holds for precisely one $j$, let us call it $j_1\in J$.
Our next goal is to compute
	\begin{align}
		\prod_{i=0}^{t-1}\textrm{bin}_\ell(p_{\beta^{-1}(i)}),
	\end{align}
	where $\beta=\beta_{j_1}$.
	 Let $v_{i,k}$ denote $\textrm{bin}_{\ell,k}(p_{\beta^{-1}(i)})$, 
	 that is,
	the $k^{\text{th}}$ bit of $\text{bin}_\ell(p_{\beta^{-1}(i)})$.
	It can be realized as
\begin{align}
	v_{i,k}=\quad
\bigvee_{j\in J}\left( U_j\wedge
\bigvee_{p\in A_i^j:\atop
	\bin_{\ell,k}(p)=1} w_p=1
	\right),
\end{align}
	and hence by an $n$-circuit of size $O(\log n)$ and depth $O(1)$.
	Since there are $t\in O(1)$ values for $i$ and $\ell\in O(\log n)$ 
	values for $k$ 
	our total circuit size so far sums up to $O(\log^2 n)$.
	The complements of the extracted bits are obtained by the same De 
	Morgan construction, with only constant-factor size overhead and 
	unchanged depth.
	It remains to lexicographically sort the $\ell$-bit strings
	$\textrm{bin}_\ell(p_{\beta^{-1}(0)}),\dots,
	\textrm{bin}_\ell(p_{\beta^{-1}(t-1)})$,
	which is possible by a circuit family of size $O(\log n)$
	and depth $O(1)$ by Lemma~\ref{L sort}, and we are done.
\end{proof}

\subsubsection{A Size $O(\log n\log\log n)$ and Depth $O((\log\log n)^2)$ 
Upper Bound}\label{S loglogn}

Before we state and prove the main result of this section we require
Spielman's error-correcting encoders computable by circuits of linear
size and logarithmic depth.
The first result is a consequence of a celebrated result of Spielman.
\begin{samepage}
	\begin{theorem}[Consequence of Theorem 19 in \cite{Spielman96}]\label{T Spielman}
	\Needspace{10\baselineskip}
	There is a constant $0<\delta<1$ such that
	for all $b\in\bbN_{>0}$ there is a mapping
	\begin{align}
		S_b:\{0,1\}^b\to\{0,1\}^{M_b},
	\end{align}
	where 
	\begin{enumerate}
		\item $M_b\in O(b)$,
		\item $\left|H(S_b(u),S_b(v))\right|\geq \delta M_b$
			for all $u,v\in\{0,1\}^b$ with $u\not=v$, and
		\item there is a $b$-circuit of size $O(b)$, depth
			$O(\log (b+1))$ and fan-in two computing $S_b$.
	\end{enumerate}
\end{theorem}
\end{samepage}

The following lemma will be derived from the previous theorem by
a classical sampling argument.

\begin{lemma}\label{L Spielman}
	For each fixed $t\in\bbN_{>0}$ there exists a finite alphabet
	$\Sigma_t$ such that for all $b\in\bbN_{>0}$ there
	is a mapping 
	\begin{align}
		E_b:\{0,1\}^b\to\Sigma_t^{N_b},
	\end{align}
	where 
	\begin{enumerate}
		\item $\Sigma_t=\{0,1\}^k$ for some constant 
			$k\in\bbN_{>0}$ that does not depend on $b$,
		\item $N_b=tb$,
		\item $\left|H(E_b(u),E_b(v))\right|> N_b(1-1/t)$
			for all $u,v\in\{0,1\}^b$ with $u\not=v$, and
		\item there is a $b$-circuit  
			of size
			$O(b)$, depth $O(\log (b+1))$ and fan-in two
			computing $E_b$ by using $kN_b$ designated 
			output gates.
	\end{enumerate}
\end{lemma}
\begin{proof}
	Let $0<\delta<1$ be the constant of Theorem~\ref{T Spielman}.
	Fix any $t\in\bbN_{>0}$. We choose constants $k=k(t)$ and $p>0$ such that 
	\begin{align}
		p=(1-\delta)^k\leq \frac{1}{16t}.\label{P ineq}
	\end{align}
	Moreover, for each $b\in\bbN_{>0}$ let
	\begin{align}
		\Sigma_t=\{0,1\}^k\qquad\text{ and }\qquad N_b=tb\quad,
	\end{align}
	thus Points (1) and (2) follow.
	For each $b\in\bbN_{>0}$ let $M_b\in O(b)$ be as 
	given by Theorem~\ref{T Spielman}.
	For each $r\in[0,N_b-1]$ choose indices $i_{r,0},\dots,i_{r,k-1}$
	uniformly from $[0,M_b-1]$ over all choices with replacement.
	Let  
	\begin{align}
		(E_b(u))_r=((S_b(u))_{i_{r,0}},\dots, (S_b(u))_{i_{r,k-1}}).
	\end{align}
	For each fixed $u,v\in\{0,1\}^b$ with $u\not= v$ and
	each $r\in[0,N_b-1]$ the probability
	that $(E_b(u))_r=(E_b(v))_r$ holds is at most $p$.
	These events are independent across $r\in[0,N_b-1]$.
	By taking a union bound over subsets of positions of
	size $b$ of the $N_b=tb$ output positions we obtain:
	\begin{align}
		\Pr\left\lbrack|\{r\mid 
		(E_b(u))_r=(E_b(v))_r\}|\geq b\right\rbrack&\leq
		{tb\choose b}p^b\\
		&\leq (et)^bp^b&
		\left(\text{Since ${n\choose s}\leq \left(\frac{en}{s}\right)^s$}\right)\\
		&\leq (et)^b(16t)^{-b}&
		(\text{By (\ref{P ineq})})
		\\
		&=(e/16)^b
	\end{align}
	There are strictly fewer than $(2^b)^2=4^b$ pairs of distinct messages.
	Therefore the probability that there are two 
	distinct words from $\{0,1\}^b$ whose image under
	$E_b$ agrees on at least $b$ positions can be upper bounded
	as follows:
	\begin{align}
		\Pr\left\lbrack \exists u,v\in\{0,1\}^b: u\not= v\wedge
		\left|\{r\mid (E_b(u))_r=(E_b(v))_r\}\right|\geq b
		\right\rbrack &<4^b(e/16)^b\\
		&=(e/4)^b\\
		&<1
	\end{align}
	Thus, there exists a fixed choice of all indices
$i_{r,s}$  with $r\in[0,N_b-1]$ and $s\in[0,k-1]$
such that the associated map $E_b$ satisfies
	\begin{align}
		\forall u,v\in\{0,1\}^b:\quad u\not= v\ \Longrightarrow\
		\left|\{r\in[0,N_b-1]\mid (E_b(u))_r=(E_b(v))_r\}\right|<
		b=N_b/t,
	\end{align}
	or equivalently
	\begin{align}
\forall u,v\in\{0,1\}^b:\quad u\not= v\ \Longrightarrow\ 
		\left|H(E_b(u),E_b(v))\right|> N_b(1-1/t),
	\end{align}
 so Point (3) holds.

 By Theorem~\ref{T Spielman} there is a circuit $C_b$ of size $O(b)$,
depth $O(\log(b+1))$, and fan-in at most two computing $S_b$.
Compute $S_b(u)$ once. For each $r\in[0,N_b-1]$ and
$s\in[0,k-1]$, add a fresh one-input $\vee$-gate connected
to the gate computing $(S_b(u))_{i_{r,s}}$.
These $kN_b$ new gates, ordered first by $r$ and then by $s$,
are the designated output gates computing $E_b(u)$.
Remove any unused non-input gates.

Since $k$ and $t$ are fixed and $N_b=tb$, this adds
$kN_b=O(b)$ gates and one layer. The resulting circuit
therefore has size $O(b)$, depth $O(\log(b+1))$, and
fan-in at most two, proving Point~(4).
\end{proof}

The rest of this section is devoted to proving the following theorem.

\begin{samepage}
	\begin{theorem}\label{T upper loglog}
{\textsl{Occ$_t$}} is computable by a circuit family
		of size $O(\log n\log\log n)$ and depth $O((\log\log n)^2)$.
\end{theorem}
\end{samepage}

	Let $n\in\bbN_{>0}$ and let $w\in\{0,1\}^n$ be a word
	with $|w|_1=t$
	that we fix for the rest of this section.
	Let $\ell=\lceil \log (n+1)\rceil$,
 so $\ell$ bits suffice to write down each 
	position in $0,\dots,n-1$ in binary.
	Let $L\in 2^\N$ be the least power of two 
	such that $L\geq\ell$.

	Moreover let
	\begin{align}
		P(w)=\{p\in[0,n-1]\mid (w)_p=1\}
	\end{align}
	denote the positions of $w$ containing the bit $1$.
	Our overall goal is to compute with a circuit of size 
	$O(\log n\log\log n)$ and depth $O((\log \log n)^2)$ the binary 
	representation of all numbers
	in $P(w)$ in increasing order, or equivalently the set
	\begin{align}
		\bin_L(P(w))=\{\bin_L(p)\mid p\in P(w)\}\label{E bin ell}
	\end{align}
	in lexicographically increasing order.
	For each word $x=x_0\dots x_{L-1}\in\{0,1\}^L$ and each 
	subinterval $B\subseteq[0,L-1]$ 
	let $x_B=\prod_{i\in B}x_i$ denote the {\em factor of $x$} induced by $B$,
	where $\prod$ denotes concatenation.
	We generalize (\ref{E bin ell}) to each subinterval 
	$B\subseteq[0,L-1]$ 
	and introduce the following notation for each $p\in[0,n-1]$
	and each $X\subseteq[0,n-1]$:
	\begin{align}
		\bin_B(p)&=(\bin_L(p))_B\\
		\bin_B(X)&=\{\bin_B(p)\mid p\in X\}
	\end{align}

	Let $L=2^k$, thus $k\in O(\log\log n)$.
	\begin{samepage}
		\begin{lemma}\label{L partition}
	For each $i\in[0,k]$ the subintervals of length
	$2^i$ that partition $[0,L-1]$ are
		\begin{align}B_{i,j}=[j2^i,(j+1)2^i-1]\text{, where $0\leq j<\frac{L}{2^i}$.}\label{E partition}
		\end{align}
In case $i>0$ we have 
		$B_{i,j}=B_{i-1,2j}\uplus B_{i-1,2j+1}$
		for each $0\leq j<\frac{L}{2^i}$.
	\end{lemma}
	\end{samepage}
	\begin{proof}
		Fix any $i\in[0,k]$.
		Clearly, each $B_{i,j}$ has size $2^i$.
		Moreover, the intervals are pairwise disjoint since
		$\max (B_{i,j})=(j+1)2^i-1<(j+1)2^i=\min(B_{i,j+1})$.
		The first interval starts at
		$\min(B_{i,0})=0$ and the last at 
		\begin{align}
			\max\left(B_{i,\frac{L}{2^i}-1}\right)=\left(\frac{L}{2^i}-1+1\right)\cdot 2^i-1=2^k-1=L-1,
		\end{align}
		which shows (\ref{E partition}).
		\noindent
	Finally, for each $i>0$ we have
		\begin{align}
			B_{i,j}&=[j2^i,(j+1)2^i-1]\\
			&=[j2^i,j2^i+2^{i-1}-1]\uplus[j2^i+2^{i-1},(j+1)2^i-1]\\
			&=[2j\cdot 2^{i-1},(2j+1)2^{i-1}-1]\uplus
			[(2j+1)2^{i-1},(2j+2)2^{i-1}-1]\\
			&= B_{i-1,2j}\uplus B_{i-1,2j+1}.
		\end{align}
	\end{proof}

\begin{samepage}
	\begin{lemma}\label{L ijb}
	Let $i\in[0,k]$, $0\leq j<\frac{L}{2^i}$ and $b=|B_{i,j}|$. 
		There is an $(n+b)$-circuit of 
		size $O(b)$ and depth $O(\log (b+1))$
	deciding the following problem:
	\smallskip

	INPUT: Our original input $w\in\{0,1\}^n$ with $|w|_1=t$ 
		and $u\in\{0,1\}^b$.

		QUESTION: $u\in\bin_{B_{i,j}}(P(w))$?
	\end{lemma}
\end{samepage}
	\begin{proof}
By Lemma~\ref{L Spielman} there exists a constant $\kappa\in\bbN_{>0}$
		such that for each $b\in\bbN_{>0}$ there exists a mapping
$E_b:\{0,1\}^b\to\Sigma_t^{N_b}$
	with $\Sigma_t=\{0,1\}^\kappa$ and $N_b=tb$
	such that for all $u,v\in\{0,1\}^b$ with $u\not=v$ we have
	\begin{align}
		\left|\big.\{r\in[0,N_b-1]:(E_b(u))_r=(E_b(v))_r\}\right|&=
		N_b-|H(E_b(u),E_b(v))|\label{E_b 1}\\
		&< N_b-N_b(1-1/t)\\
		&=N_b/t=b.\label{E_b 2}
	\end{align}
	Moreover $E_b$ is computable by 
		a $b$-circuit of size $O(b)$, depth 
		$O(\log (b+1))$ and fan-in two
using $\kappa N_b$ designated output gates.

		In the following let $B=B_{i,j}$.
		For each $r\in[0,N_b-1]$ and each $\sigma\in\Sigma_t$ define
		the predicate
		\begin{align}
			Z_{B,r,\sigma}(w):\Longleftrightarrow\quad\bigvee_{p\in[0,n-1]\atop
			(E_b(\bin_{B}(p)))_r=\sigma} (w)_p=1
		\end{align}
		which can be realized with a single $\vee$-gate
		accessing hard-wired positions of $w$. 
		With respect to our fixed $B$ there are 
		$N_b|\Sigma_t|=N_b2^\kappa\in O(b)$ many such gates.
	Now consider the predicate
	\begin{align}
		\textrm{Mem}_B(w,u):\Longleftrightarrow\quad\bigwedge_{r\in[0,N_b-1]\atop
		\sigma\in\Sigma_t}
		E_b(u)_r=\sigma\rightarrow Z_{B,r,\sigma}(w)
	\end{align}
	which can be realized by a circuit of size $O(b)$
	that first computes $E_b(u)$ and then accesses its $O(b)$ output
	gates.
		The total circuit size is $O(b)$ and depth $O(\log (b+1))$.
	To finish the proof of the lemma it suffices to prove
		that 
		$u\in\bin_B(P(w))$ if, and only if, $\textrm{Mem}_B(w,u)$ holds.

		Let us first assume $u\in\bin_B(P(w))$. 
		Hence there exists some $p_0\in P(w)$ such that $\bin_B(p_0)=u$.
	Let $E_b(u)=\sigma_0\dots\sigma_{N_b-1}\in\Sigma_t^{N_b}$,
	that is $(E_b(u))_r=(E_b(\bin_B(p_0)))_r=\sigma_r\in\Sigma_t$ 
	for each $r\in[0,N_b-1]$.
Hence each predicate $Z_{B,r,\sigma_r}(w)$ is witnessed
	by $p_0$, so the predicate $\textrm{Mem}_B(w,u)$ holds.

		Conversely, assume $u\not\in\bin_B(P(w))$. 
		Thus, $u\not=\bin_B(p)$ for all
		$p\in P(w)$.
	By (\ref{E_b 2})
	we have that 
	\begin{align}
		\left|\bigl.\{r\in[0,N_b-1]:(E_b(u))_r=(E_b(\bin_B(p)))_r\}\right|<N_b/t
	\end{align}
		for all $p\in P(w)$. Recalling that $|P(w)|=t$ it follows
	\begin{align}
		\sum_{p\in P(w)}\left|\bigl.\{r\in[0,N_b-1]:(E_b(u))_r=
		(E_b(\bin_B(p)))_r\}\right|<N_b
	\end{align}
		Hence there exists $r_0\in[0,N_b-1]$ such that
$(E_b(u))_{r_0}\not=(E_b(\bin_B(p)))_{r_0}$ for each $p\in P(w)$.
Thus, for $\sigma_0=(E_b(u))_{r_0}$ the predicate $Z_{B,r_0,\sigma_0}(w)$
does not hold. Hence
$\textrm{Mem}_B(w,u)$ does not hold.
	\end{proof}

Recall that $\ell=\lceil\log(n+1)\rceil$ and that $L=2^k$ is the
least power of two satisfying $L\geq\ell$. We explain how
Theorem~\ref{T upper loglog} follows from Lemma~\ref{L Bij} below.

At the root, $B_{k,0}=[0,L-1]$. Since $|P(w)|=t$ and the map
$p\mapsto\operatorname{bin}_L(p)$ is injective, we have
$|\bin_L(P(w))|=t$. Thus, writing
$P(w)=\{p_0<\cdots<p_{t-1}\}$, the root circuit outputs
$\bin_L(p_0),\ldots,\bin_L(p_{t-1})$
in increasing order, with all validity flags $\chi_0,\dots,\chi_{t-1}$ 
equal to one.

Discard the validity flags and retain only the first $\ell$
bits of each output word. These are the least significant bits,
and all discarded higher bits are zero. The retained outputs
are therefore precisely
$\operatorname{bin}_\ell(p_0),\ldots,\operatorname{bin}_\ell(p_{t-1})$,
as required by Problem~\ref{Pro Occ}.
Remove every non-input gate that
does not lie on a path to a retained output gate. This adds
neither gates nor depth.

Since $\ell\leq L<2\ell$, Lemma~\ref{L Bij}  yields size
\begin{align}
O\bigl(L\log(L+1)\bigr)
   =O(\log n\log\log n)
\end{align}
and depth
\begin{align}
O\bigl((\log(L+1))^2\bigr)
   =O\bigl((\log\log n)^2\bigr),
\end{align}
proving Theorem~\ref{T upper loglog}.

\begin{samepage}
	\begin{lemma}\label{L Bij}
Let $i\in[0,k]$, $0\leq j<\frac{L}{2^i}$ and $b_i=|B_{i,j}|$. 
	There is a circuit of size $O(b_i\log (b_i+1))$
	and depth $O((\log (b_i+1))^2)$ computing the following:
	
	INPUT: Our original input $w\in\{0,1\}^n$, where $|w|_1=t$.

	OUTPUT: $\bin_{B_{i,j}}(P(w))$ in lexicographically increasing order.
\end{lemma}
\end{samepage}
\begin{proof}
By Lemma~\ref{L ijb} there exists a constant $c>0$, independent
of $n,i,j$, and an $(n+b_i)$-circuit $C'_{i,j}$ such that
for each $w\in\{0,1\}^n$ with $|w|_1=t$ and each
$u\in\{0,1\}^{b_i}$ we have
	\begin{align}
 C'_{i,j}(w,u)=1
 \quad\Longleftrightarrow\quad
 u\in\bin_{B_{i,j}}(P(w)),
	\end{align}
and
	\begin{align}
 |C'_{i,j}|\leq cb_i,
 \qquad
 \depth(C'_{i,j})\leq c\log(b_i+1).
		\label{E Cij'}
	\end{align}
		Since $|w|_1\leq t$ we have that $|\bin_{B_{i,j}}(P(w))|\leq t$.
	For a set of binary words $B\subseteq\{0,1\}^+$ where $|B|=m$ let 
	$(B)_0,\dots,(B)_{m-1}$ denote the lexicographically increasing
	enumeration of the elements of $B$.

	We will construct a circuit $C_{i,j}$ with $t(b_i+1)$ designated output 
	gates that will evaluate to 
	a bit string $\prod_{s=0}^{t-1}u_{s}\chi_s$ where
	$u_0,\dots,u_{t-1}\in\{0,1\}^{b_i}$ 
	and $\chi_0,\dots,\chi_{t-1}\in\{0,1\}$ satisfying the following
	invariant, where $|\bin_{B_{i,j}}(P(w))|=m\leq t$:
	\begin{samepage}
	\begin{enumerate}
		\item $\chi_0=\chi_1=\dots=\chi_{m-1}=1$ and 
			$\chi_m=\dots=\chi_{t-1}=0$
		\item $u_s=(\bin_{B_{i,j}}(P(w)))_s$ for all $s\in[0,m-1]$
		\item $u_{m}=\dots=u_{t-1}=0^{b_i}$
	\end{enumerate}
	\end{samepage}
	The construction of the circuits $C_{i,j}$ is by induction on $i\in[0,k]$.

For the induction base let $i=0$, so $b_i=2^i=1$. 
	Fix any 
	$0\leq j<\frac{L}{2^i}$.
	Let $B=B_{0,j}=\{j\}$.
	Then $\bin_{B_{i,j}}(P(w))\subseteq\{0,1\}=\{0,1\}^{b_i}$.
For each $a\in\{0,1\}$, compute
\[
 z_a=
 \bigvee_{\substack{p\in[0,n-1]\\
                    (\operatorname{bin}_{L}(p))_j=a}}
       ((w)_p=1).
\]
Since $|P(w)|=t\geq1$, at least one of $z_0,z_1$ is true.
Set $u_0=\neg z_0$.
If $t\geq2$, additionally set $\chi_1=u_1=z_0\wedge z_1$.

Indeed, $u_0$ is the smallest bit present in
$\bin_B(P(w))$. A second entry is valid precisely
when both bits occur, in which case its value is $1$.
	For each $s\in[2,t-1]$ we set $\chi_s=u_s=0$.
Since $|P(w)|=t>0$ we set $\chi_0=1$.
This can all be realized by a circuit of size and
depth $O(1)$.

	For the induction step let $i\in[1,k]$.
	Fix any 
	$0\leq j<\frac{L}{2^i}$ and let $B=B_{i,j}$.
	By Lemma~\ref{L partition} we have
	$B=B'\uplus B''$, where 
	$B'=B_{i-1,2j}$, $B''=B_{i-1,2j+1}$.
	Moreover, $b_i=2^i=2b_{i-1}$.
	By induction hypothesis one can compute the lexicographically
	increasing representation of $\bin_{B'}(P(w))\subseteq\{0,1\}^{b_{i-1}}$ 
	and $\bin_{B''}(P(w))\subseteq\{0,1\}^{b_{i-1}}$
	via the outputs (each of length $t(b_{i-1}+1)$)
	\begin{align}
		\prod_{s=0}^{t-1}u_s'\chi_s'\text{ and }
		\prod_{s=0}^{t-1}u_s''\chi_s''
	\end{align}
	satisfying the invariants
	of the circuits $C_{i-1,2j}$ and $C_{i-1,2j+1}$
respectively.
Thus, $\bin_B(P(w))$ is contained in 
$\{u_q'u_r''\mid q,r\in[0,t-1]\}\subseteq\{0,1\}^{b_i}$, yet this inclusion could of course
be strict. Therefore our circuit $C_{i,j}$ 
reserves $t^2 b_i$ intermediate
gates to store (in constant depth) each word in
$\{u_q'u_r''\mid q,r\in[0,t-1]\}$.
For each $(q,r)\in[0,t-1]^2$, we retain the candidate
$u'_q u''_r$ precisely when both child entries are valid
and their concatenation belongs to $\operatorname{bin}_B(P(w))$,
using the bit
\begin{align}
	\beta_{q,r}= C_{i,j}'(w,u_q'u_r'')\wedge \chi_q'\wedge\chi_r''.
\end{align}
In total, this can be achieved by taking $t^2$ copies of $C_{i,j}'$.

Let $m'=|\bin_{B'}(P(w))|$ and $m''=|\bin_{B''}(P(w))|$.
By Points (1) and (2) of induction hypothesis 
$u_0',\dots,u_{m'-1}'$ and 
$u_0'',\dots,u_{m''-1}''$ are the lexicographically increasing
enumeration of $\bin_{B'}(P(w))$ and 
$\bin_{B''}(P(w))$, respectively.
Let $m=|\bin_{B}(P(w))|$.
Hence there are precisely $m$ pairs $(q,r)\in[0,t-1]^2$ such that 
$\beta_{q,r}=1$.
Let $(q_0,r_0),\dots,(q_{m-1},r_{m-1})$ be the lexicographically
increasing enumeration of these pairs, in particular
$(\bin_{B}(P(w)))_d=u_{q_d}'u_{r_d}''$ for each $d\in[0,m-1]$.
Let $<_{\text{lex}}$ denote the (right-to-left) lexicographic order
on $[0,t-1]^2$: formally $(q,r)<_{\text{lex}}(q',r')$ if $r<r'$
or $r=r'$ and $q<q'$.

We therefore introduce a predicate $S_{q,r}^d$ that for each 
$(q,r)\in[0,t-1]^2$ and each $d\in[0,t^2-1]$ expresses
whether $(q,r)$ is lexicographically the $d^{\text{th}}$ smallest
pair satisfying $\beta_{q,r}$, where, as usual, our indexing starts at $0$.
For this, let  
$L(q,r)=\{(q',r')\in[0,t-1]^2\mid (q',r')<_{\text{lex}}(q,r)\}$:
\begin{align}
	S^{d}_{q,r}:\Longleftrightarrow\quad\beta_{q,r}\wedge
\bigvee_{\theta:L(q,r)\to\{0,1\}\atop
	|\theta^{-1}(1)|=d}
	\left(\bigwedge_{(q',r')\in\theta^{-1}(1)}
	\beta_{q',r'}\wedge\bigwedge_{(q',r')\in\theta^{-1}(0)}
	\neg\beta_{q',r'}\right)
\end{align}
By recalling the before-mentioned remarks on the circuit size
increasing by applying De Morgan dualization and 
by making use of the circuit computing the $\beta_{q,r}$ the
latter predicates can be realized by constantly many circuits of
constant size each.

We are now ready to state the computation of the bit string
$\prod_{s=0}^{t-1} u_s\chi_s\in\{0,1\}^{t(b_i+1)}$ satisfying
the invariant (1),(2), and (3).
For each $s\in[0,t-1]$ and each $h\in[0,b_i-1]$ we define
\begin{align}
	\chi_s&=\bigvee_{(q,r)\in[0,t-1]^2} S_{q,r}^{s}\\
	(u_s)_h&=\begin{cases}
		\bigvee_{(q,r)\in[0,t-1]^2} S_{q,r}^{s} \wedge 
		(u_{q}')_h& 
		\text{ if $h\in[0,b_{i-1}-1]$}\\
\bigvee_{(q,r)\in[0,t-1]^2} S_{q,r}^{s} \wedge 
		(u_{r}'')_{h-{b_{i-1}}}& 
		\text{ if $h\in[b_{i-1},b_{i}-1]$}.
	\end{cases}
\end{align}
Points (1),(2) and (3) of the invariant are satisfied by construction.

It remains to analyze $|C_{i,j}|$ and $\depth(C_{i,j})$.

We begin with the analysis of $|C_{i,j}|$.
For each $i\in[0,k]$ and each $0\leq j<\frac{L}{2^i}$ 
we have the following recurrence, where $d_1,d_2,d_3>0$ are constants
\begin{align}
	|C_{i,j}|&\leq\begin{cases}d_1&\quad\text{if $i=0$}\\
		|C_{i-1,2j}|+|C_{i-1,2j+1}|+d_2\cdot
		(|C_{i,j}'|+b_i)+d_3&\quad\text{else,}
	\end{cases}
\end{align}
and by (\ref{E Cij'}) that involves the constant $c>0$, we obtain:
\begin{align}
	|C_{i,j}|&\leq\begin{cases}d_1&\quad\text{if $i=0$}\\
		|C_{i-1,2j}|+|C_{i-1,2j+1}|+
		(c+1)d_2\cdot b_i
		+d_3&\quad\text{else.}
	\end{cases}
\end{align}
Put $A=(c+1)d_2$ and
\begin{align}
M_i=\max_{0\leq j<L/2^i}|C_{i,j}|.
\end{align}
Since $b_i=2^i$, we have
$M_0\leq d_1$
and 
\begin{align}
M_i\leq 2M_{i-1}+A2^i+d_3 
\end{align}
if $i\geq 1$.
Unrolling this recurrence gives
\begin{align}
M_i
\leq d_1 2^i+A i2^i+d_3(2^i-1)
\in O\bigl(2^i(i+1)\bigr).
\end{align}
Consequently,
\begin{align}
|C_{i,j}|\in O\bigl(b_i\log(b_i+1)\bigr).
\end{align}

Let us now analyze $\depth(C_{i,j})$. 
We have the following recurrence,
where $d_4,d_5>0$ are constants:
\begin{align}
	\depth(C_{i,j})&\leq \begin{cases}d_4&\quad\text{if $i=0$}\\
		\max\{\depth(C_{i-1,2j}),\depth(C_{i-1,2j+1})\}+
		\depth(C_{i,j}')+d_5&\quad\text{else}
	\end{cases}
\end{align}
and by (\ref{E Cij'}) we obtain
\begin{align}
	\depth(C_{i,j})&\leq \begin{cases}d_4&\quad\text{if $i=0$}\\
		\max\{\depth(C_{i-1,2j}),\depth(C_{i-1,2j+1})\}+
		c\log (b_i+1)+d_5&\quad\text{else.}
	\end{cases}
\end{align}

We define 
	$D_i=\max_{0\leq j<L/2^i}\depth(C_{i,j})$.
Again since $b_i=2^i$ we obtain
$D_0\leq d_4$ and 
\begin{align}
D_i\leq D_{i-1}+c\log(2^i+1)+d_5
\leq D_{i-1}+c(i+1)+d_5
\end{align}
if $i\geq 1$.

Unrolling this recurrence gives
\begin{align}
D_i
\leq d_4+\sum_{r=1}^{i}\bigl(c(r+1)+d_5\bigr)
= d_4+\frac{c\,i(i+3)}{2}+d_5i
\in O\bigl((i+1)^2\bigr).
\end{align}
Since $b_i=2^i$, we conclude that
\begin{align}
\depth(C_{i,j})
\leq D_i \in O\bigl((\log(b_i+1))^2\bigr).
\end{align}
\end{proof}

%!TEX root = main.tex
\section{Decidability of Subclass Membership }\label{S Decidability}

This section is devoted to the proof of Theorem~\ref{T intro decidability}.

\begin{samepage}
\begingroup
\renewcommand{\themainthm}{\getrefnumber{intro:decidability}}
	\begin{makethm}{mainthm}{T intro decidability}
\unexpanded{
		Whenever $\calC$ is either the class of regular languages
		of almost logarithmic or linear circuit complexity
		as stated by Theorem~\ref{T Trichotomy},
		given a nondeterministic finite automaton recognizing
		a regular language $L$, deciding if $L$ is in $\calC$
		is $\mathbf{PSPACE}$-complete.
	}
	\end{makethm}
	\endgroup
\end{samepage}

Before we state the main result of this section we introduce
nondeterministic finite automata.

Recall that a {\em nondeterministic finite automaton (NFA)}
is a tuple $\calA=(Q,\Sigma,\Delta,q_0,F)$, where $Q$ is a finite set of
{\em states}, $\Sigma$ is a finite {\em alphabet}, $\Delta\subseteq Q\times\Sigma\times Q$
is a {\em transition relation}, $q_0\in Q$ is the {\em initial state} and $F\subseteq Q$
is a set of {\em final states}.
By $|\calA|=|Q|+|\Sigma|+|\Delta|$ we denote the {\em size} of $\calA$ and by
$L(\calA)\subseteq\Sigma^+$
we denote the {\em language} of $\calA$.

	For all of the three classes $\calC$ of regular
	languages of constant, almost logarithmic and linear
	circuit complexity, according to Theorem~\ref{T Trichotomy}, 
	we introduce the following problem.

\begin{problem}
	The {\em $\calC$-membership problem} asks the following:

	INPUT: An NFA $\calA$.
	
	QUESTION: Is $L(\calA)$ in $\calC$?
\end{problem}

\begin{example}\label{E almost log}
	Consider the language $L_1=(a+b)^*c(a+b)c(a+b)^*$ from 
	Example~\ref{E in X}.
	The multiplication table of $\stab(\eta_{L_1})$ is displayed in 
	Figure~\ref{fig:stable-semigroup-isomorphism}.
	We claim that $\stab(\eta_{L_1})$ does 
	not satisfy the identity $epxyqf=epyxqf$,
	where $e$ and $f$ range over idempotents, as 
	mentioned in Point~(3) of Theorem~\ref{T const main}.
	Indeed, 
take
	\begin{align}
    e=f=p=q=\eta_{L_1}(aa),\qquad
    x=\eta_{L_1}(aac),\qquad
    y=\eta_{L_1}(acaa).
	\end{align}
	These elements belong to $\stab(\eta_{L_1})$,
and $\eta_{L_1}(aa)$ is idempotent. By the multiplication
table, we obtain
	\begin{align}
    epxyqf=\eta_{L_1}(cac)
    \not= \eta_{L_1}(cc)=epyxqf,
	\end{align}
where the inequality follows from $cac\in L_1$ and
$cc\notin L_1$.
	Hence $\opc(L_1)\not\in O(1)$ by Theorem~\ref{T const main}.
	Since $\eta_{L_1}\in\bfX$ as shown in Example~\ref{E in X},
	it follows that $\opc(L_1)\in O(\log n\log\log n)$ by Theorem~\ref{T log main}.
	Hence the circuit complexity of $L_1$ is almost
	logarithmic as formulated in Theorem~\ref{T Trichotomy}.
\end{example}

The constant circuit complexity case has recently been studied.

\begin{theorem}[{\cite[Theorem 5.1]{GM26arxiva}}]\label{T constant}
For the class $\calC$ of regular languages of constant circuit
	complexity the $\calC$-membership problem is $\mathbf{PSPACE}$-complete.
\end{theorem}

\begin{samepage}
	\begin{theorem}\label{T trich decidability}
		Whenever $\calC$ is either the class of regular languages
		of almost logarithmic or linear circuit complexity
		as stated by Theorem~\ref{T Trichotomy},
		the $\calC$-membership problem 
		is $\mathbf{PSPACE}$-complete.
\end{theorem}
\end{samepage}

Hence, Theorem~\ref{T constant} and Theorem~\ref{T trich decidability} imply
Theorem~\ref{T intro decidability}.
Theorem~\ref{T trich decidability} in turn follows from 
Proposition~\ref{P sublinear PSPACE upper}
and 
Proposition~\ref{P sublinear PSPACE lower}.

\subsection{Upper Bound}

\begin{samepage}
\begin{proposition}\label{P X small}
Let $\varphi:\Sigma^+\to S$ be a stamp.  
	If $\varphi\not\in\bfX$, then $\varphi$ is
	$(k,1)$-distinguishable for some $k\in [1,3|S|^{3}]$.
\end{proposition}
\end{samepage}
\begin{proof}
Let $\varphi:\Sigma^+\to S$ be a stamp that is not in $\bfX$.
	Then $\varphi$ is $(k,1)$-distinguishable for some $k\in\bbN_{>0}$
	by Lemma~\ref{L k 1 dist}.
	Assume $k\in\bbN_{>0}$ is minimal such that $\varphi$ is $(k,1)$-distinguishable.
	It suffices to show that $k\leq 3|S|^3$.
	For the sake of contradiction, let us assume that $k>3|S|^3$.
	Since $\varphi$ is $(k,1)$-distinguishable there exist 
	$i\in[0,k-1]$ and 
	words $u,v,w\in\Sigma^k$ such that:
	\begin{align}
		\varphi(u)^\omega \varphi(v[i\leftarrow (u)_i]) \varphi(w)^\omega
		\not=
		\varphi(u)^\omega \varphi(v[i\leftarrow (w)_i]) \varphi(w)^\omega
	\end{align}
	Let $N=|S|^3$. Then 
	\begin{align}
		\max\{i,k-i-1\}\geq\frac{k-1}{2}\geq\frac{3N}{2}>N.
	\end{align}
	Without loss of generality let us assume $k-i-1>N$,
	the case when $i>N$ can be treated analogously.

	For each word $z\in\Sigma^k$ define the equivalence relation $\sim_z$
	on $[0,k-1]$ where $p\sim_z q$ if 
	$\omega_{k-p}(z)=_\varphi\omega_{k-q}(z)$.
	We claim that there exists a pair of positions $(p,q)\in[i+1,k-1]^2$ with
	$p<q$ such that $p\sim_z q$ for all $z\in\{u,v,w\}$.
	Indeed, since \begin{align}|[i+1,k-1]|=k-1-(i+1)+1=k-i-1>N,\end{align}
	by the pigeonhole principle
	there exist strictly more than $|S|^2$ positions all of which are
	$\sim_u$-equivalent. By applying the pigeonhole principle on these, 
	there are strictly more than $|S|$ positions all of which are 
	$\sim_v$-equivalent, and thus both $\sim_u$- and $\sim_v$-equivalent. 
	By applying the pigeonhole principle once more, among the latter
	positions, there exist two positions $p,q$
	with $p<q$ with $p\sim_w q$, and thus $p\sim_z q$ for all $z\in\{u,v,w\}$.
	Let us assume 
	\begin{align}
		u&=u_0\dots u_{k-1},\\
		v&=v_0\dots v_{k-1},\text{ and}\\
		w&=w_0\dots w_{k-1}.
	\end{align}

	Hence, for the words 
	\begin{align}
		u'&=\alpha_p(u)\omega_{k-q}(u)=u_0\dots u_{p-1}u_q\dots u_{k-1}\\
		v'&=\alpha_p(v)\omega_{k-q}(v)=v_0\dots v_{p-1}v_q\dots v_{k-1}\\
		w'&=\alpha_p(w)\omega_{k-q}(w)=w_0\dots w_{p-1}w_q\dots w_{k-1}
	\end{align}
	we have $u=_\varphi u'$, $v=_\varphi v'$, and $w=_\varphi w'$.
	Since $p>i$ we have 
	$u'[i\leftarrow a]=_\varphi u[i\leftarrow a]$
	and 
	$v'[i\leftarrow a]=_\varphi v[i\leftarrow a]$
	for all $a\in\Sigma$.
	Therefore we get
	\begin{align}
		\varphi(u')^\omega \varphi(v'[i\leftarrow (u')_i]) \varphi(w')^\omega
		\not=
		\varphi(u')^\omega \varphi(v'[i\leftarrow (w')_i]) \varphi(w')^\omega.
	\end{align}
	This implies that $\varphi$ is $(k-(q-p),1)$-distinguishable,
	a contradiction to minimality of $k$.
\end{proof}

The proof of Proposition~\ref{P sublinear PSPACE upper} is analogous to the proof of Proposition 5.2 in~\cite{GM26arxiva}.

\begin{samepage}
\begin{proposition}\label{P sublinear PSPACE upper}
Whenever $\calC$ is either the class of regular languages
		of almost logarithmic or linear circuit complexity
		as stated by Theorem~\ref{T Trichotomy},
		the $\calC$-membership problem 
		is in $\mathbf{PSPACE}$.
\end{proposition}
\end{samepage}
\begin{proof}
For the $\mathbf{PSPACE}$ upper bound let $L=L(\calA)$ for some given 
	NFA $\calA=(Q,\Sigma,\Delta,q_0,F)$.
To show the proposition it will be sufficient to prove
	that the problem to decide if
	\begin{align}
		\eta_L\text{ is not in }\bfX.\label{E eta}
	\end{align}
	is in $\mathbf{PSPACE}$.
	
	On the one hand, according to Theorem~\ref{T Trichotomy}, we have 
	that the circuit complexity
	of $L$ is almost logarithmic, if and only if, 
	\begin{itemize}
		\item $\opc(L)\not\in O(1)$, which is 
			decidable in $\mathbf{PSPACE}$ by 
			Theorem~\ref{T constant}, and
		\item  $\opc(L)\in o(n)$, which is equivalent to 
			$L\in\SIZE(o(n))$ and hence to
			the negation of (\ref{E eta})
	by Theorem~\ref{T log main}.
	\end{itemize}

	On the other hand, according to Theorem~\ref{T Trichotomy}, we have 
	that the circuit complexity
	of $L$ is linear, if and only if, 
	$\opc(L)\in \Omega(n)$, which is equivalent to $L\not\in\SIZE(o(n))$,
	which in turn is equivalent to (\ref{E eta}) 
	by Theorem~\ref{T log main}.

	Let $\mathbb{B}^{Q\times Q}$ denote the multiplicative 
	semigroup of Boolean matrices of dimension $Q\times Q$.
	Let $\mu:\Sigma^+\to\mathbb{B}^{Q\times Q}$
	be the unique morphism satisfying $\mu(a)=t_a$,
	where $t_a(q,q')=1\Leftrightarrow (q,a,q')\in\Delta$ for each $a\in\Sigma$.
	Let $T$ denote the subsemigroup $\mu(\Sigma^+)$ of 
	$\mathbb{B}^{Q\times Q}$.
	
	Clearly, each element of $\mathbb{B}^{Q\times Q}$ can be stored in space polynomial in $|\calA|$
	and so can (matrix) multiplication in $\mathbb{B}^{Q\times Q}$ 
	be performed in polynomial space in $|\calA|$.

	Let $\beta:T\to S(L)$ denote the unique surjective morphism that 
	factors $\eta_L$ as $\eta_L=\beta\circ\mu$.

Put $m=|Q|$ and $B=2^{m^2}$. Since $\beta$ is surjective,
	we have $|S(L)|\leq |T|\leq B$. By Proposition~\ref{P X small}
	we have that
$\eta_L\not\in\bfX$ if, and only if, $\eta_L$ is
$(k,1)$-distinguishable for some $k\in[1,3B^3]$.
We give a nondeterministic polynomial-space procedure for
this condition. First, guess $k\in[1,3B^3]$ and
$i\in[0,k-1]$, both represented in binary.

For these choices of $k$ and $i$, it remains to guess
words $u,v,w\in\Sigma^k$ satisfying
	\begin{align}
		\eta_L(u)^\omega \eta_L(v[i\leftarrow (u)_i]) \eta_L(w)^\omega
		\not=
		\eta_L(u)^\omega \eta_L(v[i\leftarrow (w)_i]) \eta_L(w)^\omega.
		\label{E X PSPACE}
	\end{align}
	We do not have enough space to store the words $u,v,w$.
We guess the letters of 
$u,v,w$ simultaneously from left to right. Writing
\begin{align}
v'=v[i\leftarrow u_i],
\qquad
v''=v[i\leftarrow w_i],
\end{align}
we maintain, after reading $j$ letters, the four matrices
\begin{align}
\mu(\alpha_j(u)),\quad
\mu(\alpha_j(v')),\quad
	\mu(\alpha_j(v'')),\quad\text{and}\quad
\mu(\alpha_j(w)).
\label{328}
\end{align}
All four products are initialized to the identity matrix. 
At position $\ell$, the first and fourth products are multiplied by 
$t_{u_\ell}$ and $t_{w_\ell}$, respectively. 
The two middle products are both multiplied by $t_{v_\ell}$ when 
$\ell\ne i$, and by $t_{u_i}$ and $t_{w_i}$, respectively, 
when $\ell=i$. We retain only the current products and the position 
counter, using polynomial space. At $j=k$, 
these products yield the matrices
\begin{align}
	M_u=\mu(u), \quad
	M_{v'}=\mu(v'), \quad
	M_{v''}=\mu(v''), \quad\text{and}\quad
	M_{w}=\mu(w).\label{E Matrices}
\end{align}

Recall that $|S(L)|\leq |T|\leq B$.
We first compute in polynomial space the matrices
\begin{align}
    t=M_u^\omega M_{v'}M_w^\omega,
    \qquad
    t'=M_u^\omega M_{v''}M_w^\omega.
\end{align}
Indeed, idempotent powers can be computed by successively
multiplying by the respective matrix until the current
power is idempotent, which occurs among its first $B$
positive powers. Since $\beta$ preserves idempotent powers,
it remains to decide whether $\beta(t)\neq\beta(t')$.

For every matrix $s\in\mathbb{B}^{Q\times Q}$ and $q\in Q$,
write $s(q,F)=\bigvee_{f\in F}s(q,f)$, and put
$T^1=T\cup\{I\}$, where $I$ is the identity matrix.
By definition of the syntactic morphism, we have
\begin{align}
    \beta(t)=\beta(t')
    \quad\Longleftrightarrow\quad
    \forall t_p,t_q\in T^1:
    (t_p t t_q)(q_0,F)=(t_p t't_q)(q_0,F).
    \end{align}
To verify this condition, we enumerate in polynomial space
all pairs
$(t_p,t_q)\in(\bbB^{Q\times Q})^2$
and first verify whether $t_p,t_q\in T^1$.
Membership in $T^1$ can be tested by checking whether the
given matrix is $I$, or by guessing on the fly, letter by
letter, a nonempty word of length at most $B$ whose image
under $\mu$ is the given matrix. 
If either membership test fails, we continue with the next
pair. Otherwise, we check whether
\begin{align}
    (t_p t t_q)(q_0,F)=(t_p t't_q)(q_0,F).
\end{align}
Thus both equality and inequality of $\beta(t)$ and
$\beta(t')$ can be decided in polynomial space.

Finally, since $|S(L)|\leq B$, we may restrict the witness
search to $k\in[1,3B^3]$. Guessing $k$, $i$, and the letters
of $u,v,w$ as above, followed by the test
$\beta(t)\neq\beta(t')$, gives a nondeterministic
polynomial-space procedure for $\eta_L\notin\mathbf{X}$.
\end{proof}

\subsection{Lower Bound}

\begin{samepage}
\begin{proposition}\label{P sublinear PSPACE lower}
Whenever $\calC$ is either the class of regular languages
		of almost logarithmic or linear circuit complexity
		as stated by Theorem~\ref{T Trichotomy},
		the $\calC$-membership problem 
		is $\mathbf{PSPACE}$-hard under logarithmic space
		reductions.
\end{proposition}
\end{samepage}
\begin{proof}
It is folklore that universality of NFAs is $\textbf{PSPACE}$-hard
	under logarithmic space reductions: we refer
	to Theorem~3.13 in~\cite{EsparzaBlondin2023}, 
	where for a fixed linear bounded automaton $\calB$ and some input $x$
	one computes in logarithmic space~\footnote{Technically speaking, a polynomial time reduction is provided in 
	Theorem~3.13 in~\cite{EsparzaBlondin2023}, yet it is clear
	that the reduction can also be carried out in logarithmic space.}
	an NFA $\calA$ over
	some alphabet $\Sigma$ with $|\Sigma|\geq 2$ such that there is a word
	$w\in\Sigma^+$ with $|w|\geq 2$	for which:
	\begin{align}
		L(\calA)=\begin{cases}
\Sigma^+\setminus\{w\} &\text{if $\calB$ accepts $x$}\\
			\Sigma^+&\text{otherwise}
			\label{E univ}
		\end{cases}
	\end{align}
	Let $\Gamma=\Sigma\uplus\{\$\}$ for some fresh symbol $\$\not\in\Sigma$.
	Let 
	\begin{align}
		L'=(L(\calA)\$\cup\{\$\})^+.
	\end{align}
	Clearly, one can compute from $\calA$ in logarithmic space some
	NFA $\calA'$ such that $L(\calA')=L'$.

	Let us first consider the case when $\calC$ is the class
	of regular languages of linear circuit complexity.
	
	Since $L'\in\calC$, if and only if, $\opc(L')\not\in o(n)$
	by Theorem~\ref{T Trichotomy} and since 
	$\mathbf{PSPACE}$ is closed under complement
	it suffices to show that 
	$L(\calA)=\Sigma^+$ if, and only if, 
	$L'\in\SIZE(o(n))$.

	For the only-if direction let us assume $L(\calA)=\Sigma^+$. Then 
	$L'=(\Sigma^+\$\cup\{\$\})^+=\Gamma^*\$$. Since a circuit
of constant size over the input alphabet $\Gamma$ 
	can simply check if the last input letter
	is $\$$, we clearly have $L'\in\SIZE(O(1))\subset \SIZE(o(n))$.

	For the if direction let us assume 
	$L(\calA)\not=\Sigma^+$. Then $L(\calA)=\Sigma^+\setminus\{w\}$
	by (\ref{E univ}).
	Let $w=w_0\dots w_{\ell-1}$ for some $\ell\geq 2$. 
	Since $|\Sigma|\geq 2$, for each $i\in[0,\ell-1]$ we can 
	fix some letter $\overline{w_i}\in\Sigma$
	such that $\overline{w_i}\not=w_i$.
	Let $k$ be an exponent of $S(L')$. Then 
	\begin{align}
		(\overline{w_0}w_1\dots w_{\ell-1}\$)^k
		(\overline{w_0}w_1\dots w_{\ell-1}\$)^k
		(w_0\overline{w_1}\dots \overline{w_{\ell-1}}\$)^k
		\in\ L'
	\end{align}
	whereas 
	\begin{align}
		(\overline{w_0}w_1\dots w_{\ell-1}\$)^k
		w_0w_1\dots w_{\ell-1}\$\
		(\overline{w_0}w_1\dots w_{\ell-1}\$)^{k-1}
		(w_0\overline{w_1}\dots \overline{w_{\ell-1}}\$)^k
		\not\in\ L'.
	\end{align}
	Hence, $\eta_{L'}$ is $(k(\ell+1),1)$-distinguishable,
	so $\eta_{L'}\not\in\bfX_{k(\ell+1)}$, in particular $\eta_{L'}
	\not\in \bfX$.
	By Points (3) and (4) of Theorem~\ref{T log main}, we obtain
	$L'\not\in\SIZE(o(n))$.

	It remains to consider the case when $\calC$ is
	the class of regular languages of almost logarithmic circuit complexity.

	Recall that $L'\subseteq\Gamma^+$. Let $a,b,c$ be fresh letters
	that, without loss of generality, do not appear in $\Gamma$
	and recall the language $L_1=(a+b)^*c(a+b)c(a+b)^*\in\calC$
	from Example~\ref{E almost log}.
	In particular we have $L_1\subseteq\{a,b,c\}^+$ and 
	therefore $L'\cap L_1=\emptyset$.
	Let
	\begin{align}
		L''=\underbrace{(L(\calA)\$\cup\{\$\})^+}_{L'}\cup L_1.
	\end{align}
An NFA recognizing $L''$ can also be constructed from
$\mathcal{A}$ in logarithmic space. Since $L'$ and $L_1$
are defined over disjoint alphabets, restricting a circuit
for $L''$ to either alphabet yields a circuit for the
corresponding language. Conversely, circuits for $L'$ and
$L_1$ can be combined using constantly many additional
gates, including the respective alphabet checks.
This has the following implications.

	Firstly, in case $L'\in\SIZE(O(1))$, then $L''\in\cal\calC$
	since $L_1\in\calC$.
	Secondly, if $L'\not\in\SIZE(o(n))$, then
	$L''\not\in\SIZE(o(n))$.

	We claim that $L(\calA)=\Sigma^+$ if, and only if, 
	$L''$ is in $\calC$.

	For the only-if direction let us assume $L(\calA)=\Sigma^+$. 
	Then 
	$L'\in\SIZE(O(1))$ as already argued above
	and due to $L_1\in\calC$, we obtain $L''\in\calC$, as required.

	For the if direction let us assume that  $L(\calA)\not=\Sigma^+$.
	Then $L'\not\in\SIZE(o(n))$ as argued above,
	so $L''\not\in\calC$.
\end{proof}

%!TEX root = main.tex

\section{Conclusions and Open Problems}\label{S Conclusios}

We established a trichotomy for the circuit complexity $\opc(L)$ of regular
languages $L$ into constant, almost logarithmic (i.e. 
$\opc(L)\in\Omega(\log n)\cap O(\log n\log\log n)$) and linear.

Our main technical contribution is the dichotomy theorem (Theorem~\ref{T log main})
between the regular languages of sublinear and linear circuit complexity.
We have shown that the regular languages of sublinear circuit complexity
are contained in both $\SIZEDEPTH(O(\log^2 n),O(1))$ and
$\SIZEDEPTH(O(\log n\log\log n),O((\log\log n)^2))$.
The theorem moreover provides various characterizations:
 algebraically in terms of the pseudovariety of stamps $\bfX$, 
combinatorially in terms of the congruences $\simeq_k$ on finite words, 
language-theoretically in terms of modular disjoint monomials, 
and logically in terms of the one-variable 
fragment of first-order logic with bounded Lindström quantifiers.

Combined with established results, our characterization yields
the following corollaries.
We have shown that the class of regular languages of
sublinear circuit complexity is contained 
in the one associated with $\QDA$.
Consequently, they are definable in $\rmFO^2[\Sigma,<,\bmod]$
and in the $\Sigma_2$ and $\Pi_2$ fragments of $\rmFO[\Sigma,<,\bmod]$,
and are in particular contained in $\WLACO$.

For languages given by NFAs, we proved that deciding membership
in each of the three classes is $\mathbf{PSPACE}$-complete under
logarithmic-space reductions.

We have derived an analogous characterization of the regular neutral
letter languages of sublinear circuit complexity, involving
the well-studied pseudovariety of finite monoids $\bfM\mathbf{Nil}$.

Several questions remain open.

\paragraph{Closing the logarithmic gap.}
Our upper and lower bounds for the almost logarithmic class differ by
a factor of $\log\log n$. Can this factor be removed? More precisely,
does every regular language in $\SIZE(o(n))$ belong to
$\SIZE(O(\log n))$?
A concrete approach is to improve the circuits for the promise
problem \textsl{Occ$_t$}: given a word containing exactly $t$ occurrences
of $1$, compute their positions in binary. An $O(\log n)$-size
construction for every fixed $t\geq 2$ would, through Point~(8) of
Theorem~\ref{T log main}, yield the desired bound for all regular
languages of sublinear circuit complexity. The case $t=1$ in fact already
admits such a construction of depth one.

\paragraph{Size and depth tradeoffs.}
Our constructions give two different bounds: circuits of size
$O(\log^2 n)$ and constant depth, and circuits of size
$O(\log n\log\log n)$ and depth $O((\log\log n)^2)$.
Can the latter depth be reduced to $O(\log\log n)$, or even to
a constant, while preserving the size bound?
The stronger question is whether every regular language of
sublinear circuit complexity admits circuits of size $O(\log n)$
and constant depth. Such a result would follow from corresponding
circuits for \textsl{Occ$_t$}, for every fixed $t$.

\paragraph{Other characterizations of $\bfX$.}
We have shown that the sublinear class does not have any characterization
in terms of the (isomorphism type of the) stable semigroup of the syntactic morphism,
in contrast to the regular languages in $\ACO$ or $\WLACO$.
The identities defining $\bfX$ characterize sublinear circuit
complexity, but leave open a structural description of this class.
Can $\bfX$ nevertheless be described using suitable algebraic
operations on simpler $lm$-pseudovarieties of stamps?
For regular languages with a neutral letter, we have shown that
sublinear circuit complexity is equivalent to having an aperiodic
syntactic monoid in which every idempotent commutes with every
element, that is, a syntactic monoid belonging to $\mathbf{MNil}$.
 It would be interesting to understand how this characterization relates to
the modular position information present in the general case.

\paragraph{Restrictions and extensions.}
More broadly, can one obtain a finer classification of the regular
languages in $\mathrm{AC}^0$ according to their circuit size under
a constant-depth restriction? For each fixed depth bound $d$, which
asymptotic size bounds occur, and can the resulting classes be
characterized algebraically? One may also ask whether analogous
size classifications hold when the circuit basis is extended by
$\mathrm{MOD}_m$ gates, for a fixed $m\geq 2$, or by majority gates,
both with and without a constant-depth restriction.

\bibliographystyle{alphaurl}
\bibliography{references}

@article{mnil,
	author = {Straubing, Howard},
	date = {1982/12/01},
	doi = {10.1007/BF02572753},
	id = {Straubing1982},
	isbn = {1432-2137},
	journal = {Semigroup Forum},
	number = {1},
	pages = {25--38},
	title = {The variety generated by finite nilpotent monoids},
	url = {https://doi.org/10.1007/BF02572753},
	volume = {24},
	year = {1982}}

@incollection{Kouckyhandbook,
  author       = {Michal Kouck{\'{y}}},
  editor       = {Jean{-}{\'{E}}ric Pin},
  title        = {Circuit complexity of regular languages},
  booktitle    = {Handbook of Automata Theory},
  pages        = {493--523},
  publisher    = {European Mathematical Society Publishing House, Z{\"{u}}rich,
                  Switzerland},
  year         = {2021},
  url          = {https://doi.org/10.4171/Automata-1/14},
  doi          = {10.4171/AUTOMATA-1/14}
}

@article{GrosshansMS22,
  author       = {Nathan Grosshans and
                  Pierre McKenzie and
                  Luc Segoufin},
  title        = {{Tameness and the power of programs over monoids in {DA}}},
  journal      = {Log. Methods Comput. Sci.},
  volume       = {18},
  number       = {3},
  year         = {2022},
  url          = {https://doi.org/10.46298/lmcs-18(3:14)2022},
  doi          = {10.46298/LMCS-18(3:14)2022}}

@article{CadilhacP22,
  author       = {Micha{\"{e}}l Cadilhac and
                  Charles Paperman},
  title        = {{The regular languages of wire linear AC\({}^{\mbox{0}}\)}},
  journal      = {Acta Informatica},
  volume       = {59},
  number       = {4},
  pages        = {321--336},
  year         = {2022},
  url          = {https://doi.org/10.1007/s00236-022-00432-2},
  doi          = {10.1007/S00236-022-00432-2}
}

@INPROCEEDINGS{Chaubard,
  author={Chaubard, L. and Pin, J. and Straubing, H.},
  booktitle={21st Annual IEEE Symposium on Logic in Computer Science (LICS'06)}, 
  title={{First Order Formulas with Modular Predicates}},
  year={2006},
  volume={},
  number={},
  pages={211-220},
  doi={10.1109/LICS.2006.24}}

@article{VitanyiM84,
  author       = {Paul M. B. Vit{\'{a}}nyi and
                  Lambert G. L. T. Meertens},
  title        = {Big omega versus the wild functions},
  journal      = {Bull. {EATCS}},
  volume       = {22},
  pages        = {14--19},
  year         = {1984},
  bibsource    = {dblp computer science bibliography, https://dblp.org}
}

@InProceedings{Papermanfo2,
  author =	{Dartois, Luc and Paperman, Charles},
  title =	{{Two-variable first order logic with modular predicates over words}},
  booktitle =	{STACS 2013},
  pages =	{329--340},
  series =	{LIPIcs},
  ISBN =	{978-3-939897-50-7},
  ISSN =	{1868-8969},
  year =	{2013},
  volume =	{20},
  address =	{Dagstuhl, Germany},
  URL =		{https://drops.dagstuhl.de/entities/document/10.4230/LIPIcs.STACS.2013.329}}

@InProceedings{Straubing02,
author="Straubing, Howard",
editor="Rajsbaum, Sergio",
title="{On Logical Descriptions of Regular Languages}",
booktitle="LATIN 2002: Theoretical Informatics",
year="2002",
publisher="Springer Berlin Heidelberg",
address="Berlin, Heidelberg",
pages="528--538",
isbn="978-3-540-45995-8"
}

@article{StraubingPin10,
author = {Pin, Jean-\'{E}ric and Straubing, Howard},
journal = {RAIRO - Theoretical Informatics and Applications},
language = {eng},
number = {1},
pages = {239-262},
publisher = {EDP Sciences},
title = {{Some results on {C}-varieties}},
url = {http://eudml.org/doc/92759},
volume = {39},
year = {2005},
}

@article{Kunc10,
author = {Kunc, Michal},
journal = {RAIRO - Theoretical Informatics and Applications},
language = {eng},
number = {3},
pages = {243-254},
publisher = {EDP Sciences},
title = {{Equational description of pseudovarieties of homomorphisms}},
url = {http://eudml.org/doc/92722},
volume = {37},
year = {2003},
}

@article{Friedman86,
  author       = {Joel Friedman},
  title        = {{Constructing O(n log n) Size Monotone Formulae for the k-th Threshold
                  Function of n Boolean Variables}},
  journal      = {{SIAM} J. Comput.},
  volume       = {15},
  number       = {3},
  pages        = {641--654},
  year         = {1986},
  url          = {https://doi.org/10.1137/0215047},
  doi          = {10.1137/0215047}
}

@article{KW15,
  author       = {Manfred Kufleitner and
                  Tobias Walter},
  title        = {{One quantifier alternation in first-order logic with modular predicates}},
  journal      = {{RAIRO} Theor. Informatics Appl.},
  volume       = {49},
  number       = {1},
  pages        = {1--22},
  year         = {2015},
  url          = {https://doi.org/10.1051/ita/2014024},
  doi          = {10.1051/ITA/2014024}
}

@manual{Pin2025Mathematical,
  title        = {{Mathematical Foundations of Automata Theory}},
  author       = {Pin, Jean--{\'E}ric},
  year         = {2025},
  note         = {Lecture notes, MPRI, Universit{\'e} Paris and IRIF},
  url          = {https://www.irif.fr/~jep/PDF/MPRI/MPRI.pdf}
}

@article{Schutz65,
  author  = {Sch{\"u}tzenberger, Marcel-Paul},
  title   = {{On Finite Monoids Having Only Trivial Subgroups}},
  journal = {Information and Control},
  volume  = {8},
  year    = {1965},
  pages   = {190--194}
}

@article{LogicMeetsAlgebra,
    title      = {{Logic Meets Algebra: the Case of Regular Languages}},
    author     = {Pascal Tesson and Denis Therien},
    url        = {https://lmcs.episciences.org/2226},
    doi        = {10.2168/LMCS-3(1:4)2007},
    journal    = {Logical Methods in Computer Science},
    issn       = {1860-5974},
    volume     = {Volume 3, Issue 1},
    eid        = 4,
    year       = {2007},
}

@article{BridgesTessonTherien,
  author       = {Pascal Tesson and
                  Denis Th{\'{e}}rien},
  title        = {Bridges between Algebraic Automata Theory and Complexity Theory},
  journal      = {Bull. {EATCS}},
  volume       = {88},
  pages        = {37--64},
  year         = {2006}
}

@book{McNPap71,
  author    = {McNaughton, Robert and Papert, Seymour},
  title     = {{Counter-Free Automata}},
  publisher = {The MIT Press},
  address   = {Cambridge, MA},
  year      = {1971}
}

@article{McKenzieNC1,
	author = {McKenzie, Pierre and P{\'e}ladeau, Pierre and Therien, Denis},
	date = {1991/12/01},
	doi = {10.1007/BF01212963},
	isbn = {1420-8954},
	journal = {computational complexity},
	number = {4},
	pages = {330--359},
	title = {{$\mathrm{NC}^1$: The automata-theoretic viewpoint}},
	url = {https://doi.org/10.1007/BF01212963},
	volume = {1},
	year = {1991}}

@article{ActionCvariety,
title = {{Actions, wreath products of {C}-varieties and concatenation product}},
journal = {Theoretical Computer Science},
volume = {356},
number = {1},
pages = {73-89},
year = {2006},
note = {In honour of Professor Christian Choffrut on the occasion of his 60th birthday},
author = {Laura Chaubard and Jean-\'{E}ric Pin and Howard Straubing},
issn = {0304-3975}
}

@article{PeladeauTCS,
title = {{Finite semigroup varieties defined by programs}},
journal = {Theoretical Computer Science},
volume = {180},
number = {1},
pages = {325-339},
year = {1997},
issn = {0304-3975},
doi = {https://doi.org/10.1016/S0304-3975(96)00297-6},
author = {Pierre Péladeau and Howard Straubing and Denis Therien},
}

@InProceedings{NathanJ,
author="Grosshans, Nathan",
title="{The Power of Programs over Monoids in {$\bf{J}$}}",
booktitle="Language and Automata Theory and Applications",
year="2020",
publisher="Springer International Publishing",
address="Cham",
pages="315--327"
}

@book{Straubingbook,
  author    = {Straubing, Howard},
  title     = {{Finite Automata, Formal Logic, and Circuit Complexity}},
  publisher = {Birkh{\"a}user},
  address   = {Boston},
  year      = {1994}
}

@article{BCST92,
  author       = {David A. Mix Barrington and
                  Kevin J. Compton and
                  Howard Straubing and
                  Denis Th{\'{e}}rien},
  title        = {{Regular Languages in NC{\({^1}\)}}},
  journal      = {J. Comput. Syst. Sci.},
  volume       = {44},
  number       = {3},
  pages        = {478--499},
  year         = {1992},
  url          = {https://doi.org/10.1016/0022-0000(92)90014-A},
  doi          = {10.1016/0022-0000(92)90014-A}
}

@article{Barrington1989,
  author  = {David A. Barrington},
  title   = {{Bounded-Width Polynomial-Size Branching Programs Recognize Exactly Those Languages in {NC}$^1$}},
  journal = {Journal of Computer and System Sciences},
  volume  = {38},
  number  = {1},
  pages   = {150--164},
  year    = {1989}
}

@inproceedings{AaronsonGS19,
  author       = {Scott Aaronson and
                  Daniel Grier and
                  Luke Schaeffer},
  editor       = {David Zuckerman},
  title        = {{A Quantum Query Complexity Trichotomy for Regular Languages}},
  booktitle    = {60th {IEEE} Annual Symposium on Foundations of Computer Science, {FOCS}
                  2019, Baltimore, Maryland, USA, November 9-12, 2019},
  pages        = {942--965},
  publisher    = {{IEEE} Computer Society},
  year         = {2019},
  url          = {https://doi.org/10.1109/FOCS.2019.00061}
}

@article{FurstSaxeSipser1984,
  author  = {Merrick Furst and James B. Saxe and Michael Sipser},
  title   = {{Parity, Circuits, and the Polynomial-Time Hierarchy}},
  journal = {Mathematical Systems Theory},
  volume  = {17},
  number  = {1},
  pages   = {13--27},
  year    = {1984},
  doi     = {10.1007/BF01744431}
}

@article{ChandraFortuneLipton1985,
  author  = {Ashok K. Chandra and Steven Fortune and Richard J. Lipton},
  title   = {{Unbounded Fan-In Circuits and Associative Functions}},
  journal = {Journal of Computer and System Sciences},
  volume   = {30},
  number   = {2},
  pages    = {222--234},
  year     = {1985},
  doi      = {10.1016/0022-0000(85)90015-7}
}

@article{Koucky2009,
  author  = {Michal Kouck{\'y}},
  title   = {{Circuit Complexity of Regular Languages}},
  journal = {Theory of Computing Systems},
  volume   = {45},
  number   = {4},
  pages    = {865--879},
  year     = {2009},
  doi      = {10.1007/s00224-009-9180-z}
}

@article{ALP26,
  author       = {Antoine Amarilli and
                  S{\'{e}}bastien Labb{\'{e}} and
                  Charles Paperman},
  title        = {{Out-of-Order Membership to Regular Languages}},
  journal      = {CoRR},
  volume       = {abs/2602.13100},
  year         = {2026},
  url          = {https://doi.org/10.48550/arXiv.2602.13100},
  doi          = {10.48550/ARXIV.2602.13100},
  eprinttype   = {arXiv},
  eprint       = {2602.13100}
}

@article{Hromkovic85a,
  author       = {Juraj Hromkovic},
  title        = {{Linear Lower Bounds on Unbounded Fan-In Boolean Circuits}},
  journal      = {Inf. Process. Lett.},
  volume       = {21},
  number       = {2},
  pages        = {71--74},
  year         = {1985},
  url          = {https://doi.org/10.1016/0020-0190(85)90035-3},
  doi          = {10.1016/0020-0190(85)90035-3}
}

@article{BarringtonTherien1988,
  author  = {Barrington, David A. Mix and Th{\'e}rien, Denis},
  title   = {{Finite Monoids and the Fine Structure of {$\mathrm{NC}^1$}}},
  journal = {Journal of the ACM},
  volume  = {35},
  number  = {4},
  pages   = {941--952},
  year    = {1988},
  doi     = {10.1145/48014.63138}
}

@article{Ganardi2025SlidingWindow,
  author  = {Ganardi, Moses and Hucke, Danny and Lohrey, Markus and Mamouras, Konstantinos and Starikovskaya, Tatiana},
  title   = {{Regular Languages in the Sliding Window Model}},
  journal = {TheoretiCS},
  volume  = {4},
  number  = {8},
  pages   = {1--75},
  year    = {2025},
  doi     = {10.46298/theoretics.25.8}
}

@article{TessonTherien2005,
  author  = {Pascal Tesson and Denis Th{\'e}rien},
  title   = {{Complete Classifications for the Communication Complexity of Regular Languages}},
  journal = {Theory of Computing Systems},
  volume  = {38},
  number  = {2},
  pages   = {135--159},
  year    = {2005},
  doi     = {10.1007/s00224-004-1190-2}
}

@book{EsparzaBlondin2023,
  author    = {Javier Esparza and Michael Blondin},
  title     = {{Automata Theory: An Algorithmic Approach}},
  publisher = {MIT Press},
  year      = {2023},
  isbn      = {978-0-262-04863-7}
}

@article{GM26arxiva,
  author       = {Stefan G{\"{o}}ller and
                  Amaldev Manuel},
  title        = {Rational Reductions and Regular Languages of Constant Circuit Complexity},
  journal      = {CoRR},
  volume       = {abs/2609.18484},
  year         = {2026},
  url          = {https://doi.org/10.48550/arXiv.2609.18484},
  doi          = {10.48550/ARXIV.2609.18484},
  eprinttype   = {arXiv},
  eprint       = {2609.18484}
}

@article{Spielman96,
  author       = {Daniel A. Spielman},
  title        = {Linear-time encodable and decodable error-correcting codes},
  journal      = {{IEEE} Trans. Inf. Theory},
  volume       = {42},
  number       = {6},
  pages        = {1723--1731},
  year         = {1996},
  url          = {https://doi.org/10.1109/18.556668},
  doi          = {10.1109/18.556668}
}

@misc{khoulani2026semigroups,
  author  = {Haneen Khoulani},
  title   = {Web Application for Computing Semigroups},
  year    = {2026},
  url     = {https://github.com/HaneenKl/Bachelor-Project},
  note    = {Bachelor project, University of Kassel}
}

\end{document}